\documentclass[prd,aps,showpacs,nofootinbib]{revtex4}

\usepackage{mathrsfs,amssymb,amsmath,epsfig,amsbsy,amsfonts}
\usepackage[usenames,dvipsnames]{color}
\usepackage[bookmarks]{hyperref}
\usepackage[svgnames]{xcolor}
\usepackage{subfigure}
\usepackage{cancel} 
\usepackage{times}
\usepackage{multirow}

\newcommand{\Real}{\mathbb{R}}

\newcommand{\ve}[1]{{\bf #1}}

\newcommand{\Torus}{\mathbb{T}}

\begin{document}

\title{Axisymmetric, stationary collisionless gas clouds trapped in a Newtonian potential}

\author{Carlos Gabarrete and Olivier Sarbach}
\affiliation{Instituto de F\'isica y Matem\'aticas,
Universidad Michoacana de San Nicol\'as de Hidalgo,
Edificio C-3, Ciudad Universitaria, 58040 Morelia, Michoac\'an, M\'exico.}

\begin{abstract}
The properties of an axisymmetric, stationary gas cloud surrounding a massive central object are discussed. It is assumed that the gravitational field is dominated by the central object which is modeled by a nonrelativistic rotationally-symmetric potential. Further, we assume that the gas consists of collisionless, identical massive particles that follow bound orbits in this potential. Several models for the one-particle distribution function are considered and the essential formulae that describe the relevant macroscopical observables, such as the particle and energy densities, pressure tensor, and the kinetic temperature are derived. The asymptotic decay of the solutions at infinity is discussed and we specify configurations with finite total mass, energy and (zero or non-zero) angular momentum. Finally, our configurations are compared to their hydrodynamic analogs. In an accompanying paper, the equivalent general relativistic problem is discussed, where the central object consists of a Schwarzschild black hole.
\end{abstract}

\date{\today}


\maketitle

\section{Introduction}

The main goal of this work is to initiate a systematic investigation for the description of steady-state kinetic gas configurations surrounding massive compact objects. From a theoretical point of view the motivation for this problem stems from the interest in understanding the behavior of a kinetic gas cloud which is trapped in a strong gravitational field background. This has many potential astrophysical applications, including the modeling of low-density hot accretion disks around compact stars or black holes, the description of distributions of stars around supermassive black holes or studying the behavior of dark matter surrounding a black hole. In particular, the recent breakthrough observations by the Event Horizon Telescope Collaboration (EHTC)~\cite{EHTC,EHTCI,EHTCXII} showing the first image of the shadows of the supermassive black holes M87$^*$ and Sgr A$^*$ provide a strong motivation for a thorough understanding of the behavior of a hot plasma in the vicinity of a strong gravitational field beyond the hydrodynamic approximation.

In this article, we start with a rather simple and idealized model which allows for a complete analytic description. However, despite its simplicity, it has many interesting features, as we will see. Furthermore, it serves as a starting point and benchmark for more complicated models. The models described in this article are based on the following assumptions. First, we assume that the gas is collisionless, that is, we completely neglect collisions between the individual gas particles. Second, we assume that the gas consists of identical, massive and uncharged particles. Third, we assume the gravitational field is dominated by the potential of the central massive object, such that the self-gravity of the gas can be neglected. Fourth, we assume that the central object is spherically symmetric, leading to a central gravitational potential. Fifth, in this article we further assume that the gravitational potential and the kinetic gas can be treated non-relativistically; the fully relativistic case will be treated in the accompanying paper~\cite{cGoS2022c}. Sixth, we focus on axisymmetric steady-states in which each gas particle follows a bound trajectory in the central gravitational potential generated by the central object.

The physical meaning of these assumptions are the following. The collisionless approximation is based on the assumption that the mean free path is large compared to the length scale over which the spatial gradients of the macroscopic quantities vary. This is most probably a reasonable assumption in all the aforementioned astrophysical scenarios. Our next assumption, namely that the gas consists of identical massive and uncharged particles makes sense for the modeling of distribution of stars and dark matter; however it is unrealistic for the description of low-density hot plasma disks since in this case the electromagnetic field is expected to play a prominent role. For this scenario, a model involving at least two species (protons and electrons) based on the Vlasov-Maxwell equations should be considered. The third assumption is justified as long as the central object is much more massive than the gas cloud. The fourth and fifth assumptions are well-founded if the gas configuration lies sufficiently far from the central object and the gas temperature is small compared to the particle's rest energy. The sixth assumption is partially based on the expectation that at sufficiently late times the gas is well-described by a distribution function (DF) depending only on integrals of motion (and hence describing a steady-state) and whose support lies in the region of phase space describing bound orbits. This is indeed expected to hold due to the fact that particles following unbound trajectories either disperse or fall into the central object in finite time~\cite{pRoS17a} and due to the phase space mixing phenomena~\cite{pRoS2020,pRoS18,pDeJmAeMdN17}.

There has been much related work describing axisymmetric steady-state gas configurations without a massive central object, in which case the self-gravity cannot be neglected. In the non-relativistic limit, Shapiro and Teukolsky~\cite{sSsT1992} numerically constructed axisymmetric static solutions of the Vlasov-Poisson system to model equilibrium stellar systems. Later, these solutions were generalized to the general relativistic case~\cite{sSsT1993a,sSsT1993b}. More recently, Rein and Andr\'easson constructed axially symmetric disk solutions of the Vlasov-Poisson system~\cite{hAgR2015} to model the rotation curves of disk galaxies without introducing dark matter. Further (analytic and numerical) models describing axisymmetric solutions of the Vlasov-Poisson system are discussed in section 4.4 of the book by Binney and Tremaine~\cite{BinneyTremaine-Book}, see also~\cite{gRyG2003}. For mathematical and numerical results regarding the stationary, axisymmetric Einstein-Vlasov system we refer the reader to the review article by Andr\'easson~\cite{hA11}. Recently, in~\cite{eAhAaL16,eAhAaL19}, axisymmetric and stationary solutions of the Vlasov-Poisson and Einstein-Vlasov systems were constructed numerically, based on a variety of ans\"atze for the DF leading to configurations with toroidal, disk-like, spindle-like and composite structures. Axially symmetric, stationary solutions of the Einstein-Vlasov and Einstein-Vlasov-Maxwell systems with and without rotation were constructed in~\cite{hAmKgR11,hAmKgR14,mTh2019} by deforming a spherically symmetric, static solution of the Vlasov-Poisson system and using the implicit function theorem. For related astrophysical work describing tori of axisymmetric collisionless plasmas around compact objects based on a quasi-stationary approximation, see~\cite{cCzS13,cCetAl2013}.

We mention in passing that a very similar model to the one considered in this work has been used to describe accretion phenomena of a collisionless gas into a central object, both in the non-relativistic~\cite{ZelNovik-Book,Shapiro-Book,aGetal2021} and the relativistic~\cite{pRoS17a,pRoS17b,pMoA2021a,pMoA2021b,aCpM2020,aGetal2021} regimes. The main difference between the accretion model and the one considered in the present article resides in the fact that in the former case, unbounded trajectories are relevant, whereas here we focus on bound trajectories.

Returning to our model, we now describe its main features and properties. We use a four-parametric polytropic ansatz similar to Refs.~\cite{sSsT1992,eAhAaL16,eAhAaL19} for which the one-particle DF depends only on the energy and the azimuthal angular momentum of the particles through a power law with certain cut-offs. Our ansatz allows for the description of both rotating and non-rotating stationary and axially symmetric configurations with finite total mass, energy and angular momentum. For the case of H\'enon's isochrone potential (see, e.g.~\cite{jB04,BinneyTremaine-Book}), which includes the Kepler potential as a special case, we compare configurations with the same total mass with each other. To this purpose we perform a detailed analytic derivation of the macroscopic observables corresponding to the DF, namely: the particle and the energy densities, the mean particle velocity, and the components of the pressure tensor. From these observables we define an anisotropy parameter and a kinetic temperature whose properties are analyzed. Finally, we compare our configuration's particle density and kinetic temperature with those of an analogous hydrodynamic model. We show that although the kinetic configurations are more compact than the fluid models, their normalized temperature profiles may be very similar to each other.

As mentioned before, relativistic generalizations of the models discussed in this article are provided in an accompanying paper~\cite{cGoS2022c}. See also Ref.~\cite{cGoS2022a} for a very similar disk model based on a DF depending on the energy and the inclination angle (instead of the energy and azimuthal component of the angular momentum).

This work is organized as follows: in section~\ref{Sec:Basics} we summarize the most relevant properties of a stationary, collisionless gas configuration trapped in a central potential and discuss the simple model of a spherical polytrope. In section~\ref{Sec:StationaryAxisymmetric} we present our stationary and axisymmetric models, and we derive explicit formulae for the macroscopic observables and the total mass, energy and angular momentum associated with our models. In section~\ref{Sub:Examples} we discuss the behavior of the macroscopic observables including the morphology of the resulting gas configurations. Further, in this section, we compare our kinetic models with the corresponding fluid models. Conclusions are drawn in section~\ref{Sec:Conclusions}. Technical details including a list of relevant integrals, properties of H\'enon's isochrone model, the derivation of the expressions for the total mass, energy and angular momentum and a short review of the circular fluid models can be found in appendices~\ref{App:Integrals}--\ref{App:ClassicalPolishDoughnuts}.

\section{Review of basic equations and polytropic model}
\label{Sec:Basics}

In this section, we first summarize the most relevant equations describing the properties of a stationary, collisionless gas configuration which is trapped in a central Newtonian potential $\Phi(r)$. Next, we consider the particular example of the polytropic model in which the one-particle DF depends only on the energy of the particles through a power law, thus yielding a spherically symmetric configuration. This model will be generalized to describe axisymmetric configurations in the subsequent section.

\subsection{Basic equations describing a collisionless, stationary gas configuration in a central potential}

In the following, we assume the potential $\Phi: (0,\infty)\to \Real$ to be a smooth increasing function satisfying $\Phi(r)\to 0$ as $r\to \infty$ with the additional property that $r^3\Phi': (0,\infty)\to \Real$ is an increasing function from $0$ to $\infty$. This implies that for each non-zero value of the total angular momentum $L\neq 0$ the effective potential
\begin{equation}
V_L(r) := m\Phi(r) + \frac{L^2}{2m r^2},\qquad r > 0,
\label{Eq:EffectivePotentialNewton}
\end{equation}
has a unique global minimum whose location $r = r_0$ is determined by the equation $r_0^3\Phi'(r_0) = L^2/m^2$ and corresponds to a stable circular trajectory with minimum energy $E_0(L) = V_L(r_0)$. For $L > 0$ any trajectory with energy $E_0(L) < E < 0$ is bounded and its radial motion is periodic with period $T_r(E,L) = \frac{\partial A(E,L)}{\partial E}$, where $A(E,L)$ denotes the area function (see for instance~\cite{Arnold-Book})
\begin{equation}
A(E,L) := \oint p_r dr = 2\int\limits_{r_1(E,L)}^{r_2(E,L)} \sqrt{2m(E - V_L(r))} dr
\label{Eq:AreaFunction}
\end{equation}
with $r_1(E,L) < r_2(E,L)$ referring to the turning points.

In this article, we consider collisionless gas configurations in which individual gas particles follow bound orbits. These are described by a one-particle DF $f$ whose support lies in the region $E < 0$. As long as the potential $\Phi(r)$ satisfies the non-degeneracy condition $\det[ D^2 A(E,L)] \neq 0$ for almost all $(E,L)$, it has been shown in~\cite{pRoS2020} that an arbitrary initial configuration relaxes in time to a stationary configuration which can be described by a one-particle DF $f$ depending only on the integrals of motion, that is, only on the energy and the angular momentum,
\begin{equation}
E = \mathcal{H}(\ve{x},\ve{p}) := \frac{|\ve{p}|^2}{2m} + m\Phi(r),\qquad
\ve{L} = \ve{x}\wedge \ve{p}.
\label{Eq:Energy}
\end{equation}
We shall be particularly interested in the properties of the resulting macroscopic observables describing the gas, namely (see, for instance~\cite{Huang-Book}):
\begin{eqnarray}
n(\ve{x}) := \int f(\ve{x},\ve{p}) d^3 p &&\hbox{(particle density)},
\label{Eq:ParticleDensity}\\
\varepsilon(\ve{x}) := \int \mathcal{H}(\ve{x},\ve{p}) f(\ve{x},\ve{p}) d^3 p &&\hbox{(energy density)},
\label{Eq:EnergyDensity}\\
\ve{u}(\ve{x}) := \int \frac{\ve{p}}{m n(\ve{x})} f(\ve{x},\ve{p}) d^3 p 
&&\hbox{(mean particle velocity)},
\label{Eq:MeanParticleVelocity}\\
P_{ij}(\ve{x}) := \frac{1}{m}\int (p_i - m u_i)(p_j - m u_j)  f(\ve{x},\ve{p}) d^3 p 
&&\hbox{(pressure tensor)}.
\label{Eq:PressureTensor}
\end{eqnarray}
As a consequence of the collisionless Boltzmann equation and the assumption that $f$ is time-independent, it follows that these quantities satisfy the conservation laws~\cite{Huang-Book}
\begin{equation}
\nabla \cdot (n \ve{u} ) = 0,\qquad
(\ve{u} \cdot \nabla) u_i = -\partial^j P_{ij} - mn\partial_i\Phi.
\label{Eq:ConservationLaws}
\end{equation}
If interested in thermodynamical considerations, one further considers the quantities
\begin{eqnarray}
S(\ve{x}) := -k_B\int f(\ve{x},\ve{p})\log(A f(\ve{x},\ve{p})) d^3 p &&\hbox{(entropy density)},
\label{Eq:EntropyDensity}\\
\ve{S}(\ve{x}) := -k_B\int \frac{\ve{p}}{m} f(\ve{x},\ve{p})\log(A f(\ve{x},\ve{p})) d^3 p 
&&\hbox{(entropy flux)},
\label{Eq:EntropyFlux}
\end{eqnarray}
where $k_B$ denotes Boltzmann's constant and $A$ is an arbitrary positive constant which makes sure that $A f$ is dimensionless.\footnote{Note that a rescaling $A\mapsto \lambda A$ of $A$ by a positive factor $\lambda$ induces the transformations $S\mapsto S - (\log\lambda) k_B n$, $\ve{S}\mapsto \ve{S} - \log(\lambda) k_B n\ve{u}$, such that Eq.~(\ref{Eq:EntropyConservation}) remains invariant by this rescaling, due to the conservation law~(\ref{Eq:ConservationLaws}).} Since there are no collisions, the entropy flux is divergence-free, i.e.
\begin{equation}
\nabla\cdot\ve{S} = 0,
\label{Eq:EntropyConservation}
\end{equation}
and $S(\ve{x})$ is constant in time. Finally, we define the ``kinetic temperature" $T$ at position $\ve{x}$ through~\cite{BinneyTremaine-Book}
\begin{equation}
k_B T(\ve{x}) = \frac{\delta^{ij} P_{ij}(\ve{x})}{3n(\ve{x})}
 = \int \frac{|\ve{p} - m\ve{u}|^2}{3 m n(\ve{x})} f(\ve{x},\ve{p}) d^3 p,
\label{Eq:Temperature}
\end{equation}
which is motivated by the ideal gas equation $P = n k_B T$ with $P = \delta^{ij} P_{ij}/3$ the isotropic pressure. Note, however, that the configurations analyzed in this article are not in thermodynamic equilibrium, since collisions are completely neglected. Therefore, $T(\ve{x})$ should be interpreted as a measure for the molecular mean square velocity rather than the property associated with a thermal state. An example illustrating this characteristic will be provided shortly.

Eqs.~(\ref{Eq:Energy}--\ref{Eq:PressureTensor}) and the definition of the temperature in Eq.~(\ref{Eq:Temperature}) imply the general relation
\begin{equation}
\varepsilon = n\left[ \frac{m}{2} |\ve{u}|^2 + \frac{3}{2} k_B T + m\Phi \right].
\label{Eq:varepsilon}
\end{equation}

\subsection{Spherical polytropes}
\label{SubSec:Polytropes}

To provide a simple but useful example, we first discuss the polytropes (see, for instance, section~4.3.3 in Ref.~\cite{BinneyTremaine-Book}) for which
\begin{equation}
f(\ve{x},\ve{p}) = F_{\text{poly}}(E) := \alpha\left(-\frac{E}{E_0} \right)_+^{k-\frac{3}{2}},
\label{Eq:Polytrope}
\end{equation}
with some positive constants $E_0$ (with units of energy) and $\alpha$ (with units of time$^3$ $\times$ mass$^{-3}$ $\times$ length$^{-6}$) and the polytropic index $k > 1/2$. Here and in the following, the notation $F_+$ refers to the positive part of $F$, that is $F_+ = F$ if $F > 0$ and $F_+ = 0$ otherwise. The observables~(\ref{Eq:ParticleDensity}--\ref{Eq:PressureTensor}) and (\ref{Eq:EntropyDensity},\ref{Eq:EntropyFlux}) can easily be computed by means of the ``partition function"
\begin{equation}
Z(\alpha,k,\ve{x}) := \alpha\int \left(-\frac{E}{E_0} \right)_+^{k-\frac{3}{2}} d^3 p,
\end{equation}
according to the following relations:
\begin{equation}
n(\ve{x}) = Z(\alpha,k,\ve{x}),\qquad
\varepsilon(\ve{x}) = -E_0 Z(\alpha,k+1,\ve{x}),\qquad
S(\ve{x}) = -k_B n(\ve{x})\left[ \log(\alpha A) 
 + \left( k - \frac{3}{2} \right)\frac{\partial}{\partial k}\log Z(\alpha,k,\ve{x}) \right],
\end{equation}
whereas $\ve{u} = \ve{S} = 0$ and $P_{ij} = P\delta_{ij}$ as a direct consequence of the fact that the DF only depends on the energy. Computing $Z(\alpha,k,\ve{x})$ explicitly and taking into account the relation~(\ref{Eq:varepsilon}) yields
\begin{eqnarray}
&& n(\ve{x}) = c_k \psi(r)^k,\qquad
c_k := \alpha(2\pi m E_0)^{\frac{3}{2}}\frac{\Gamma\left( k - \frac{1}{2} \right)}{\Gamma(k+1)},
\label{Eq:ParticleDensity1}\\
&& k_B T(\ve{x}) = \frac{P(\ve{x})}{n(\ve{x})} = \frac{E_0}{k+1}\psi(r),\qquad
\varepsilon(\ve{x}) = -\left( k - \frac{1}{2} \right) P(\ve{x}),
\label{Eq:EnergyDensity1}\\
&& \frac{S(\ve{x})}{n(\ve{x})} = -k_B\left\{ \log\left[ \alpha A \psi(r)^{k-\frac{3}{2}}\right] 
 + \left(k-\frac{3}{2}\right)\left[ \Psi_{\textrm{di}}\left( k - \frac{1}{2} \right) - \Psi_{\textrm{di}}\left( k + 1 \right)\right] \right\},
\end{eqnarray}
where $\Psi_{\textrm{di}}(k) := \frac{\partial}{\partial k}\log\Gamma(k)$, $k > 0$, denotes the digamma function (see, e.g.~\cite[Eq. 5.2.2]{DLMF}) and for convenience we have introduced the dimensionless quantity
\begin{equation}
\psi(r) := -\frac{m\Phi(r)}{E_0}.
\label{Eq:psi}
\end{equation}
Note that the polytropic relation $P = K n^\gamma$ holds, with adiabatic index $\gamma = 1 + 1/k$ and constant $K = E_0 c_k^{-1/k}/(k+1)$.  Hence, the configuration described by the DF~(\ref{Eq:Polytrope}) gives rise to a static, isotropic perfect fluid configuration with polytropic equation of state. Remarkably, the radial profiles of the particle density, pressure and energy density are given by simple powers of the dimensionless potential $\psi(r)$; in particular they are monotonously decreasing for all $k >1/2$ since $\psi(r)$ is, with their decay properties at $r\to \infty$ controlled by the ones of $\psi(r)$. A self-gravitating configuration can be obtained by requiring the fulfillment of Poisson's equation, which leads to the famous Lane-Emden equation (see~\cite{Chandrasekhar-BookStellar} and references therein). However, we shall not go further into this direction since our work focuses on the case where the gravitational potential is dominated and shaped by a central object.

We conclude this section by remarking that the above configurations do not describe thermal equilibria. Indeed, by varying the DF~(\ref{Eq:Polytrope}) with respect to $\alpha$ and $k$, one obtains
\begin{equation}
d\left( \frac{\varepsilon}{n} \right) - T d\left( \frac{S}{n} \right) + P d\left( \frac{1}{n} \right)
 = \frac{P}{n} \left\{ -\frac{3}{2}\frac{1}{k+1} + \left(k-\frac{3}{2}\right)\left[ \Psi_{\textrm{di}}'\left( k - \frac{1}{2} \right) - \Psi_{\textrm{di}}'\left( k + 1 \right)\right]
\right\} dk.
\end{equation}
It can be verified that the right-hand side does not vanish for any $k > 1/2$, such that the Gibbs relation does not hold. Consequently, the gas configurations described by the  DF~(\ref{Eq:Polytrope}) are not in local thermodynamical equilibrium, despite of the fact that they describe steady-states. Of course, there is not contradiction since collisions are completely neglected in our model, implying that there is no entropy production driving the gas towards thermal equilibrium.

\section{Stationary, axisymmetric models}
\label{Sec:StationaryAxisymmetric}

In this section, we consider steady-state configurations which are axisymmetric. Assuming that mixing has taken place\footnote{As discussed in~\cite{pRoS2020} mixing takes place in H\'enon's isochrone potential used later in this work, as long as the parameter $b$ characterizing this potential is positive.} such that the DF only depends on the integrals of motion $E$ and $\ve{L}$, and assuming without loss of generality that the axis of symmetry coincides with the $z$-axis, this implies that the one-particle DF only depends on $E$, $L := |\ve{L}|$ and $L_z$, such that
\begin{equation}
f(\ve{x},\ve{p}) = F( E,L,L_z).
\label{Eq:EquilibriumDistributionFunction}
\end{equation}
Here, $F$ is a function which vanishes if its arguments $(E,L,L_z)$ lie outside the admissible range describing bound trajectories, see Eq.~(\ref{Eq:Admissible}) below. We start our investigation with some general bounds in section~\ref{SubSec:Bounds}, which provide sufficient conditions for the ansatz~(\ref{Eq:EquilibriumDistributionFunction}) to describe a gas configuration with finite total mass and angular momentum. Next, in Section~\ref{SubSec:TheELzModels} we discuss the $(E,L_z)$-models in which the DF is a product between the polytropic function $F_{\text{poly}}(E)$ and some function $I(L_z)$ of the azimuthal angular momentum $L_z$. Finally, in Section~\ref{SubSec:TotalMassAngularMomentum} we derive general expressions for the total mass, energy and angular momentum for the $(E,L_z)$-models and compute them explicitly for the case of H\'enon's gravitational potential which interpolates between the potential belonging to a point mass and the one belonging to a constant density sphere~\cite{BinneyTremaine-Book}.

\subsection{General bounds on the observables}
\label{SubSec:Bounds}

Before specifying our models and deriving the corresponding expressions for the macroscopic observables, we derive some elementary bounds that can be inferred from the results in the previous subsection. For this, we assume that $F$ satisfies the bound
\begin{equation}
0\leq F(E,L,L_z)\leq F_{\text{poly}}(E),
\label{Eq:FBound}
\end{equation}
for all admissible values $(E,L,L_z)$ of the constants of motion leading to bound trajectories, where here $F_{\text{poly}}(E)$ is any function of the form given in Eq.~(\ref{Eq:Polytrope}) for \emph{some fixed} positive constants $\alpha$, $E_0$ and $k > 1/2$. In principle, one might think that the upper bound~(\ref{Eq:FBound}) on $F$ which is uniform in $(L,L_z)$ is rather restrictive. However, this is not the case as one can see by noticing that the admissible range of values for the constants of motion leading to bound trajectories is described by the inequalities
\begin{equation}
E_{\text{min}} := m\Phi(0) < E < 0,\qquad
0\leq L\leq L_{\text{ub}}(E),\qquad
|L_z|\leq L,
\label{Eq:Admissible}
\end{equation}
with $L_{\text{ub}}(E)$ denoting the critical angular momentum such that the minimum of the effective potential $V_L(r)$ is precisely $E$, i.e. such that $E_0(L_{\text{ub}}) = E$. Therefore, given $E \in (E_{\text{min}}, 0)$, the admissible set of $(L,L_z)$ satisfying Eq.~(\ref{Eq:Admissible}) is compact, implying that Eq.~(\ref{Eq:FBound}) is really a bound on the energy-dependency of the DF.

Due to the monotony properties of the integral, Eq.~(\ref{Eq:FBound}) implies that the particle density associated with $F$ is bounded according to
\begin{equation}
0\leq n(\ve{x}) \leq c_k\psi(r)^k,
\qquad \ve{x}\in \Real^3,
\label{Eq:nBound}
\end{equation}
with $c_k$ as in Eq.~(\ref{Eq:ParticleDensity1}). In particular, the decay of $n(\ve{x})$ for $r\to \infty$ is controlled by the decay of the dimensionless gravitational potential $\psi(r)$. For the typical $1/r$-decay of the potential, it follows that the particle density decays at least as fast as $1/r^k$ which leads to a finite mass configuration provided that $k > 3$.

From the bound~(\ref{Eq:nBound}) and the fact that $|\mathcal{H}| = -m\Phi - |\ve{p}|^2/(2m)\leq E_0\psi$, one also obtains
\begin{equation}
|\varepsilon(\ve{x})| \leq E_0 c_k\psi(r)^{k+1},
\qquad \ve{x}\in \Real^3,
\label{Eq:EBound}
\end{equation}
showing that the energy density's magnitue decays at least as fast as $\psi(r)^{k+1}$.

Similarly, we can estimate the magnitude of the mean velocity $|\ve{u}|$ according to
\begin{equation}
|\ve{u}(\ve{x})| \leq \frac{1}{m n(\ve{x})} \int |\ve{p}| F_{\text{poly}}(E) d^3 p
 = \frac{32\pi m \alpha}{n(\ve{x})} \frac{E_0^2}{4k^2-1}\psi(r)^{k+\frac{1}{2}},
\qquad \ve{x}\in \Real^3,
\label{Eq:uBound}
\end{equation}
which shows that $|n(\ve{x}) \ve{u}(\ve{x})|$ decays at least as fast as $1/r^{k+1/2}$ if $\psi(r)$ decays like $1/r$. 

Next, the kinetic energy density can be bounded according to
\begin{equation}
0\leq \int \frac{|\ve{p}|^2}{2m} f(\ve{x},\ve{p}) d^3 p
 \leq \int  \frac{|\ve{p}|^2}{2m} F_{\text{poly}}(E) d^3 p = \frac{3E_0 c_k}{2(k+1)} \psi(r)^{k+1},
\label{Eq:TBound}
\end{equation}
which implies that the individual components of the pressure tensor can be estimated according to
\begin{equation}
| P_{ij}(\ve{x}) | \leq \frac{1}{m}\int | \ve{p} - m\ve{u} |^2 f(\ve{x},\ve{p}) d^3 p
 = \frac{1}{m}\int |\ve{p}|^2 f(\ve{x},\ve{p}) d^3 p - m n(\ve{x}) |\ve{u}(\ve{x})|^2
  \leq \frac{3E_0 c_k}{k+1} \psi(r)^{k+1},
\label{Eq:PBound}
\end{equation}
where we have used the bound~(\ref{Eq:TBound}) in the last step.

Summarizing, any stationary, axisymmetric DF satisfying the estimates~(\ref{Eq:FBound}) has the property that the quantities $n(\ve{x})$, $|\varepsilon(\ve{x})|$, $|P_{ij}(\ve{x})|$ and $n(\ve{x}) |\ve{u}(\ve{x})|$ are bounded by powers of the dimensionless gravitational potential $\psi(r)$. In particular, any such configuration for which $k > 3$ and $\psi(r)$ is regular at $r = 0$ and decays as $1/r$ for $r\to \infty$ has finite total mass and energy, as follows from the bounds~(\ref{Eq:nBound},\ref{Eq:EBound}). Furthermore, taking into account the bound~(\ref{Eq:uBound}) one can also conclude that the slightly stronger bound $k > 7/2$ leads to configurations with finite total angular momentum. The results obtained in this subsection will be useful to describe the range of validity of the free parameters in the models introduced in the next subsection.

\subsection{The $(E,L_z)$-models}
\label{SubSec:TheELzModels}

For the following, we restrict our attention to DFs of the form~(\ref{Eq:EquilibriumDistributionFunction}) which are independent of $L$ and which factorize, such that
\begin{equation}
F(E,L,L_z) = F_0(E) I(L_z),
\label{Eq:FFactorizationAnsatz}
\end{equation}
with two functions $F_0(E)$ and $I(L_z)$. The independence of $L$ can be motivated by the observation that if the self-gravity of the gas configuration was taken into account, then the gravitational potential would in general no longer be spherically symmetric, and thus $L$ would cease to be an integral of motion, whereas an ansatz in which the DF depends on $(E,L_z)$ is fully compatible with a self-gravitating axisymmetric configuration. Ans\"atze of the form~(\ref{Eq:FFactorizationAnsatz}) have been presented in the context of Newtonian~\cite{aM1982,sSsT1992} and relativistic~\cite{sSsT1993a,sSsT1993b} self-gravitating, stationary and axisymmetric stellar systems. See also Refs.~\cite{eAhAaL16,eAhAaL19} for more recent work on the numerical construction of self-gravitating disk, spindle and torus configurations in the relativistic case. For the following, we consider an ansatz of the form~(\ref{Eq:FFactorizationAnsatz}) with $F_0(E) = F_{\text{poly}}(E)$ as in Eq.~(\ref{Eq:Polytrope}) and $I(L_z) = I^{(\text{even},\text{rot})}_{\text{poly}}(L_z)$ given by one of the following two models:
\begin{equation}
I^{(\text{even})}_{\text{poly}}(L_z) := \left( \frac{|L_z|}{L_0} - 1 \right)_+^l,
\label{Eq:PolytropeLzEven}
\end{equation}
or
\begin{equation}
I^{(\text{rot})}_{\text{poly}}(L_z) := 2\left( \frac{L_z}{L_0} - 1 \right)_+^l, 
\label{Eq:PolytropeLzRot}
\end{equation}
with parameters $L_0 > 0$ and $l = 0,1,2,\ldots$. The parameter $L_0 > 0$ describes a lower-bound for the (magnitude of the) azimuthal angular momentum. In particular, since $|L_z|\leq L$, it implies that radial and almost radial trajectories are not populated, meaning that the gas configuration must vanish close to the center. On the other hand, since $|L_z|\leq L\leq L_{\text{ub}}(E)$ for bound orbits, one can show that $|I(L_z)|$ can be bounded by a power of $1/|E|$ for small $|E|$. For example, for  typical $1/r$-decay at infinity, such that $\Phi(r) \approx -GM/r$ for large values of $r$, one finds the Keplerian expressions $r_0\approx L^2/(GM m^2)$ and $L_{\text{ub}}(E) \approx GM m^2/\sqrt{-2m E}$, which means that for both ans\"atze~(\ref{Eq:PolytropeLzEven}) and (\ref{Eq:PolytropeLzRot}), $|I(Lz)|$ can be bounded from above by a constant times $(-E/E_0)^{-l/2}$, and all the bounds obtained in the previous subsection hold provided $k$ is replaced with $k-l/2$. In particular, one obtains finite mass configurations if $k > 3 + l/2$. Finally, we note that the main difference between the two ans\"atze~(\ref{Eq:PolytropeLzEven}) and (\ref{Eq:PolytropeLzRot}) lies in the absolute value of $L_z$, which implies that $I^{(\text{even})}_{\text{poly}}$ is an even function of the azimuthal angular momentum, such that for every orbit with $L_z > L_0$ there is a corresponding orbit rotating in the opposite direction with reversed sign $-L_z$ of the azimuthal angular momentum, whereas $I^{(\text{rot})}_{\text{poly}}$ only considers orbits rotating with positive $L_z$. The factor $2$ in Eq.~(\ref{Eq:PolytropeLzRot}) is introduced such that both configurations described by Eqs.~(\ref{Eq:PolytropeLzEven},\ref{Eq:PolytropeLzRot}) have the same density profile and total mass (see below).

After having specified our particular ans\"atze for the DF $F(E,L,L_z)$, we compute the associated observables~(\ref{Eq:ParticleDensity},\ref{Eq:EnergyDensity},\ref{Eq:MeanParticleVelocity},\ref{Eq:PressureTensor}). To this purpose, we transform the Cartesian components of the momentum $\ve{p}$ to its spherical components $(p_r,p_\vartheta,p_\varphi)$ and next to the constants of motion $(E,L,L_z)$, which can be written as
\begin{equation}
E = \frac{p_r^2}{2m} + V_L(r),\qquad
L^2 = p_\vartheta^2 + \frac{p_\varphi^2}{\sin^2\vartheta},\qquad
L_z = p_\varphi.
\label{Eq:IntegralsOfMotion}
\end{equation}
Consequently, the volume form transforms according to
\begin{equation}
d^3 p = \frac{dp_r dp_\vartheta dp_\varphi}{r^2\sin\vartheta}
 = \frac{m L}{|p_\vartheta| |p_r|} \frac{dE dL dL_z}{r^2\sin\vartheta}.
\label{Eq:JacobianTransformation}
\end{equation}
At this point, it is important to realize that the variable substitution $(p_r,p_\vartheta,p_\varphi)\mapsto (E,L,L_z)$ is not one-to-one but four-to-one, which is due to the possible sign choices for
\begin{equation}
p_r = \pm \sqrt{2m(E - V_L(r))},\qquad
p_\vartheta = \pm \sqrt{L^2 - L_z^2\sin^{-2}\vartheta},
\end{equation}
when the spherical components of $\ve{p}$ are reconstructed from the integrals of motion. These sign choices imply that there are four corresponding contributions when rewriting Eqs.~(\ref{Eq:ParticleDensity},\ref{Eq:EnergyDensity},\ref{Eq:MeanParticleVelocity},\ref{Eq:PressureTensor}) as integrals over $(E,L,L_z)$. The domain of integration is restricted by the requirement that (for given values of $r$ and $\vartheta$) one must have $V_L(r)\leq E$ and $L_z^2\sin^{-2}\vartheta \leq L^2$, which gives\footnote{Note that the conditions in Eq.~(\ref{Eq:AdmissibleFixedrtheta}) automatically imply that $(E,L,L_z)$ lies in the admissible range described by the inequalities~(\ref{Eq:Admissible}) since $L\sin\vartheta\leq L$ and since, by definition, $L_{\text{max}}(E,r)\leq L_{\text{ub}}(E)$.} 
\begin{equation}
E_{\text{min}}(r) < E < 0,\qquad
0\leq L < L_{\text{max}}(E,r) ,\qquad
|L_z|\leq L\sin\vartheta,
\label{Eq:AdmissibleFixedrtheta}
\end{equation}
with $E_{\text{min}}(r) := m\Phi(r)$ and $L_{\text{max}}(E,r) := r\sqrt{2m(E - m\Phi(r))}$.

For an arbitrary DF of the form~(\ref{Eq:EquilibriumDistributionFunction}) this yields the following expression for the particle density
\begin{equation}
n(\ve{x}) = \frac{4m}{r^2\sin\vartheta}\int\limits_{E_{\text{min}}(r)}^{0} dE\int\limits_0^{L_{\text{max}}(E,r)} dL L\int\limits_{-L\sin\vartheta}^{+L\sin\vartheta} dL_z
\frac{F(E,L,L_z)}{\sqrt{L^2 - L_z^2\sin^{-2}\vartheta}\sqrt{2m(E - V_L(r))}},
\end{equation}
and similar expressions for $\varepsilon(\ve{x})$, $\ve{u}(\ve{x})$ and $P_{ij}(\ve{x})$ (taking into account the correct signs of $p_r$ and $p_\vartheta$ in each of the four contributions). For the particular product ansatz~(\ref{Eq:FFactorizationAnsatz}) considered in this subsection, one can interchange the order of the integrals over $L$ and $L_z$ and explicitly compute the resulting integral over $L$ (see Appendix~\ref{App:Integrals} for details), which gives
\begin{equation}
n(\ve{x}) = \frac{2\pi m}{r\sin\vartheta}\int\limits_{E_{\text{min}}(r)}^{0} dE F_0(E)
\int\limits_{-L_{\text{max}}(E,r)\sin\vartheta}^{+L_{\text{max}}(E,r)\sin\vartheta} dL_z I(L_z),
\end{equation}
and similarly for the other observables. Specializing to the specific functions~(\ref{Eq:Polytrope}) for $F_0$ and Eqs.~(\ref{Eq:PolytropeLzEven},\ref{Eq:PolytropeLzRot}) for $I(L_z)$ these integrals can be computed explicitly. Introducing the cylindrical radius $R := r\sin\vartheta$ and the dimensionless variable (depending on $r$ and $\vartheta$)
\begin{equation}
\eta := \frac{R}{L_0}\sqrt{2m E_0\psi} ,
\label{Eq:eta}
\end{equation}
(with the dimensionless potential $\psi$ depending on $r$ defined as in Eq.~(\ref{Eq:psi})) one obtains\footnote{
In the limit $\eta\to \infty$ one finds $\displaystyle J_{s,l}(\eta) \to \frac{1}{2} B\left(s+1,\frac{l+3}{2}\right)$. In particular, taking $l=0$ and $L_0\to 0$ yields
$$
n^{(\text{rot})} = 2\pi \alpha(2m E_0)^{\frac{3}{2}} B\left( k-\frac{1}{2},\frac{3}{2} \right)\psi^k,
\qquad
\varepsilon^{(\text{rot})} = -2\pi \alpha(2m E_0)^{\frac{3}{2}} B\left( k+\frac{1}{2},\frac{3}{2} \right)E_0\psi^{k+1},
$$
which coincides with the expression in Eqs.~(\ref{Eq:ParticleDensity1},\ref{Eq:EnergyDensity1}), as expected.} 
\begin{eqnarray}
n^{(\text{rot})} &=& 4\pi \alpha(2m E_0)^{\frac{3}{2}}\frac{\eta^l}{l+1}\psi^{k} J_{k-\frac{3}{2},l}(\eta),
\label{Eq:nrot}\\
\varepsilon^{(\text{rot})} &=& -4\pi \alpha(2m E_0)^{\frac{3}{2}} E_0
\frac{\eta^l}{l+1}\psi^{k + 1} J_{k-\frac{1}{2},l}(\eta),
\label{Eq:epsrot}\\
\ve{u}^{(\text{rot})} &=& \sqrt{\frac{2E_0}{m}\psi}\left[ \frac{l+1}{l+2}
\frac{J_{k-\frac{3}{2},l+1}(\eta)}{J_{k-\frac{3}{2},l}(\eta)} + \frac{1}{\eta} \right]\ve{e}_\varphi,
\label{Eq:urot}
\end{eqnarray}
with $\ve{e}_{\varphi} := (-\sin\varphi,\cos\varphi,0)$. Herein, $J_{s,l}$ refer to the integrals (defined for $s,l\geq 0$)
\begin{equation}
J_{s,l}(\eta) = \frac{1}{\eta^{2s+l+3}}\int\limits_1^\eta (\eta^2-\xi^2)^s(\xi-1)_+^{l+1}\xi d\xi,
\qquad \eta > 0,
\label{Eq:Jsl}
\end{equation}
which vanish if $\eta\leq 1$.
Using integration by parts, the variable substitution $\xi = 1 + x$ and~\cite[Eq. 3.197.8]{iGiR2007} these integrals can be written as
\begin{equation}
J_{s,l}(\eta) 
 = \frac{1}{2} B(s+1,l+2) {}_{2}F_{1}\left( -(s+1), l+1; s+l+3; -\frac{\eta-1}{\eta+1} \right)
 \left( 1 + \frac{1}{\eta} \right)^{s+1}\left( 1 - \frac{1}{\eta} \right)_+^{s+l+2},
\end{equation}
where $B(a,b)$ denotes the Beta function and ${}_{2}F_{1}(a,b;c;z)$ Gauss' hypergeometric function (see~\cite{DLMF}). Taking into account the behavior of the integrand under the transformation $L_z\mapsto -L_z$, the observables belonging to the even ansatz~(\ref{Eq:PolytropeLzEven}) are given by
\begin{equation}
n^{(\text{even})}(\ve{x}) = n^{(\text{rot})}(\ve{x}),\qquad
\varepsilon^{(\text{even})}(\ve{x}) = \varepsilon^{(\text{rot})}(\ve{x}),\qquad
\ve{u}^{(\text{even})}(\ve{x}) = \ve{0}.
\label{Eq:neven}
\end{equation}
The pressure tensor can be computed from Eq.~(\ref{Eq:PressureTensor}) and results to be diagonal:
\begin{equation}
{\bf P} = P_{\hat{r}} {\bf e}_r\otimes {\bf e}_r + P_{\hat{\vartheta}} {\bf e}_{\vartheta}\otimes {\bf e}_{\vartheta}
 + P_{\hat{\varphi}} {\bf e}_\varphi\otimes {\bf e}_{\varphi},
\label{Eq:PressureTensorDecomposition}
\end{equation}
with $\ve{e}_r := (\cos\varphi\sin\vartheta,\sin\varphi\sin\vartheta,\cos\vartheta)$ and $\ve{e}_\vartheta := \partial\ve{e}_r/\partial\vartheta$. Due to the identity~(\ref{Eq:A2}) in Appendix~\ref{App:Integrals} the principle pressures $P_{\hat{r}}$ and $P_{\hat{\vartheta}}$ turn out to be equal, and for the ansatz~(\ref{Eq:PolytropeLzRot}) one finds
\begin{eqnarray}
P_{\hat{r}}^{(\text{rot})} &=& 8\pi \alpha(2mE_0)^{\frac{3}{2}} E_0 \frac{\eta^l \psi^{k+1}}{l+1} \left[ 
 \frac{1}{\eta}\frac{J_{k-\frac{3}{2},l+1}(\eta)}{l+2} + \frac{J_{k-\frac{3}{2},l+2}(\eta)}{l+3} \right],
\label{Eq:P1}\\
P_{\hat{\varphi}}^{(\text{rot})} &=& 8\pi \alpha(2m E_0)^{\frac{3}{2}} E_0\frac{\eta^l \psi^{k+1}}{l+1} \left[\frac{1}{\eta^2} J_{k-\frac{3}{2},l}(\eta) + \frac{2}{\eta}\frac{l+1}{l+2} J_{k-\frac{3}{2},l+1}(\eta) + \frac{l+1}{l+3} J_{k-\frac{3}{2},l+2}(\eta) \right]- m n^{(\text{rot})}\left| {\bf u}^{\text{(rot)}} \right|^2.
\label{Eq:P3}
\end{eqnarray}
For the interpretation of our results in the next section, it is convenient to introduce the isotropic pressure, defined by $P := (P_{\hat{r}} + P_{\hat{\vartheta}} + P_{\hat{\varphi}})/3$, and the corresponding kinetic temperature $T$, such that $P = n k_B T$. Using the identity
\begin{equation}
 \left( \frac{1}{\eta^2} - 1 \right) J_{s,l}(\eta) + \frac{2}{\eta} J_{s,l+1}(\eta)
 + J_{s,l+2}(\eta) = -J_{s+1,l}(\eta),
\qquad s,l \geq 0.
\label{Eq:PIso}
\end{equation}
one finds
\begin{equation}
P^{(\text{rot})} = \frac{8\pi \alpha}{3} (2mE_0)^{\frac{3}{2}} E_0 \frac{\eta^l \psi^{k+1}}{l+1} 
 \left[ J_{k-\frac{3}{2},l}(\eta) - J_{k-\frac{1}{2},l}(\eta)  \right] 
 - \frac{1}{3} m n^{(\text{rot})}\left| {\bf u}^{\text{(rot)}} \right|^2,
\label{Eq:IsotropicPressureRot}
\end{equation}
and it can be checked that the relation~(\ref{Eq:varepsilon}) holds. The pressure variables belonging to the even ansatz~(\ref{Eq:PolytropeLzEven}) are given by the same expressions as in Eqs.~(\ref{Eq:P1},\ref{Eq:P3},\ref{Eq:IsotropicPressureRot}) where one replaces ${\bf u}^{\text{(rot)}}$ with zero. In the next subsection, we derive explicit expressions for the total mass, energy and angular momentum associated with the configurations described by the models~(\ref{Eq:FFactorizationAnsatz},\ref{Eq:PolytropeLzEven},\ref{Eq:PolytropeLzRot}).

We conclude this subsection by noting that the support of the particle and energy densities, as well as the pressure components is delimited by the condition $\eta\geq 1$, or
\begin{equation}
r\sqrt{\psi(r)}\sin\vartheta \geq  \frac{L_0}{\sqrt{2m E_0}}.
\label{Eq:SupportNewton}
\end{equation}
This boundary is generated by parabolic-type orbits of maximal energy ($E=0$) whose orbital plane has an inclination angle $\pi/2 - \vartheta$ with respect to the equatorial plane and whose total angular momentum $L = L_0/\sin\vartheta$ is minimal. Examples of the resulting surface will be presented in the next section. For the moment, we note that for the Kepler potential, for which $\psi$ is proportional to $1/r$, this surface has the form
\begin{equation}
z^2\leq R^2\left( \frac{R^2}{R_0^2} - 1 \right)
\end{equation}
for some constant $R_0 > 0$ corresponding to the radius of the inner edge of the disk.

\subsection{Total mass, energy and angular momentum}
\label{SubSec:TotalMassAngularMomentum}

In this subsection we derive useful formulae that allow one to compute the total mass $M_{\text{gas}}$ as well as the total energy $E_{\text{gas}}$ and angular momentum $\ve{J}_{\text{gas}}$ of the stationary, axisymmetric gas configurations described by DFs of the form~(\ref{Eq:EquilibriumDistributionFunction}) and the ans\"atze~(\ref{Eq:FFactorizationAnsatz},\ref{Eq:PolytropeLzEven},\ref{Eq:PolytropeLzRot}) in particular. In the latter case, we also compute explicit expressions for H\'enon's isochrone potential. Recall that for the polytropes discussed in subsection~\ref{SubSec:Polytropes} and their axisymmetric generalizations discussed in the previous subsection, these quantities are finite provided the gravitational potential $\Phi(\ve{x})$ decays to zero as fast as $1/r$ for $r\to \infty$ and $k > 7/2 + l/2$.

The computation of the total mass for a DF depending only on integrals of motion as in Eq.~(\ref{Eq:EquilibriumDistributionFunction}) is greatly simplified by introducing action-angle variables $(\ve{Q},\ve{J})$ (see for instance Refs.~\cite{Arnold-Book,BinneyTremaine-Book}) on the region of phase space corresponding to bound orbits (that is, those orbits lying in the admissible range defined by Eq.~(\ref{Eq:Admissible})). Using the fact that the transformation $(\ve{x},\ve{p})\mapsto (\ve{Q},\ve{J})$ is symplectic and hence volume-preserving, one obtains
\begin{equation}
M_{\text{gas}} = m\int n(\ve{x}) d^3 x = m\int f(\ve{x},\ve{p}) d^3 x d^3 p
 = m\int\limits_{\Omega_J}\int\limits_{\Torus^3} f(\ve{x},\ve{p}) d^3 Q d^3 J,
\label{Eq:TotalMass}
\end{equation}
where $\Omega_J\subset \Real^3$ denotes the domain over which the action variables $\ve{J}$ vary and $\Torus^3$ the three-torus parametrized by the angles $\ve{Q}$. A DF of the form given in Eq.~(\ref{Eq:EquilibriumDistributionFunction}) is independent of the angle variables, and hence in this case the integral over $\Torus^3$ is trivial and yields the factor $(2\pi)^3$. In order to perform the integral over the action variables, one uses the relation~\cite{BinneyTremaine-Book,pRoS2020} $\ve{J} = (J_1,J_2,J_3) = (A(E,L)/(2\pi),L,L_z)$, with $A(E,L)$ the area function defined in Eq.~(\ref{Eq:AreaFunction}), and rewrites it as an integral over the constants of motion $(E,L,L_z)$. Taking into account Eq.~(\ref{Eq:Admissible}) to formulate the appropriate integration limits yields
\begin{equation}
M_{\text{gas}} = 
4\pi^2 m\int\limits_{E_{\text{min}}}^0 dE 
\int\limits_0^{L_{\text{ub}}(E)} dL 
\int\limits_{-L}^L dL_z T_r(E,L) F(E,L,L_z), \qquad
T_r(E,L) = \frac{\partial A(E,L)}{\partial E},
\label{Eq:TotalMass1}
\end{equation}
which allows one to compute $M_{\text{gas}}$ if the period of the radial motion $T_r$ and the function $F$ are known. The expression for $E_{\text{gas}}$ is obtained from Eq.~(\ref{Eq:TotalMass1}) by inserting the factor $E/m$ inside the integral. Similarly, the total angular momentum of the gas configuration is given by
\begin{equation}
\ve{J}_{\text{gas}} = \int\int \ve{x}\wedge\ve{p} f(\ve{x},\ve{p}) d^3 x d^3 p = \int\limits_{\Omega_J}\int\limits_{\Torus^3} \ve{x}\wedge\ve{p} f(\ve{x},\ve{p}) d^3 Q d^3 J,
\label{Eq:JgasOriginal}
\end{equation}
which, for the stationary and axisymmetric configuration of the form~(\ref{Eq:EquilibriumDistributionFunction}) yields
\begin{equation}
\ve{J}_{\text{gas}} = 
4\pi^2\int\limits_{E_{\text{min}}}^0 dE 
\int\limits_0^{L_{\text{ub}}(E)} dL 
\int\limits_{-L}^L dL_z L_z T_r(E,L) F(E,L,L_z) \ve{e}_z
\label{Eq:TotalAngularMomentum1}
\end{equation}
with $\ve{e}_z = (0,0,1)$. Without specifying the function $F$ in more details, this is how far we can get.\footnote{Note that if $F(E,L,L_z) = F_0(E)$ is independent of $L$ and $L_z$ one can carry out the integrals over $L$ and $L_z$ explicitly. Using Eq.~(\ref{Eq:AreaFunction}) one obtains
\begin{equation}
M_{\text{gas}} = 16\pi^2 m^2\int_{E_{\text{min}}}^0 dE F_0(E) \int_0^{r_2(E,0)} dr r^2\sqrt{2m(E -m\Phi(r))}.
\nonumber
\end{equation}
Another interesting case is the one described by a DF of the form $F(E,L,L_z) = F_0(E) G(\beta)$ with $\beta = L_z/L$, for which
\begin{equation}
M_{\text{gas}} = \frac{1}{2}\int_{-1}^1 G(\beta) d\beta\times \left. M_{\text{gas}}  \right|_{G=1}.
\nonumber
\end{equation}
}
For the following, we compute $M_{\text{gas}}$, $E_{\text{gas}}$ and $\ve{J}_{\text{gas}}$ for the particular ans\"atze~(\ref{Eq:FFactorizationAnsatz},\ref{Eq:PolytropeLzEven},\ref{Eq:PolytropeLzRot}) of the previous subsection, assuming that the gravitational potential is given by H\'enon's isochrone model (see~\cite{jB04} and~\cite{BinneyTremaine-Book}),
\begin{equation}
\Phi(r) = \Phi_b^{\text{iso}}(r) := -\frac{GM}{b+\sqrt{b^2 + r^2}}, \qquad r > 0,
\label{Eq:IsochronePotential}
\end{equation}
which is parametrized in terms of the parameter $b > 0$ which has units of length. Note that in the limit $b\to 0$ this potential reduces to Kepler's potential; however for $b > 0$ the potential $\Phi_b^{\text{iso}}$ is finite and regular at the center. One of the interesting properties of this potential is that the period of the radial motion is independent of $L$ and given by the same expression as Kepler's potential, i.e.
\begin{equation}
T_r(E,L) = 2\pi G M \left( \frac{m}{-2E} \right)^{\frac{3}{2}}.
\label{Eq:RadialPeriod}
\end{equation}
The relevant limits of integration in this case are given by (see Appendix~\ref{App:EffectivePotential} for details)
\begin{equation}
E_{\text{min}} = m\Phi(0) = -\frac{GMm}{2b}, \qquad 
L_{\text{ub}}(E) = \sqrt{\frac{m}{2}}\left(\frac{GMm}{\sqrt{-E}}-2b\sqrt{-E} \right).
\label{Eq:LimitsOfIntegration}
\end{equation}
Introducing these expressions into Eqs.~(\ref{Eq:TotalMass1},\ref{Eq:TotalAngularMomentum1}) yields, after some calculations which are carried out in Appendix~\ref{App:TotalMass},
\begin{eqnarray}
M_{\text{gas}}^{(\text{rot})}
 &=& 16\pi^3\alpha L_0^3 m\left( \frac{G^2 M^2 m^3}{2E_0 L_0^2} \right)^{k - \frac{3}{2}}
 \frac{\Gamma(l+1)\Gamma(2k-l-6)}{\Gamma(2k-3)} 
 \frac{{}_{2}F_{1}\left(-(l+2), 2k-l-6, 2k-3, -\tanh^2\chi \right)}{(\cosh\chi)^{4k-2l-12}},
\label{Eq:Mgas}\\
E_{\text{gas}}^{(\text{rot})}
 &=& -16\pi^3\alpha L_0^3 E_0\left( \frac{G^2 M^2 m^3}{2E_0 L_0^2} \right)^{k - \frac{1}{2}}
 \frac{\Gamma(l+1)\Gamma(2k-l-4)}{\Gamma(2k-1)} 
 \frac{{}_{2}F_{1}\left(-(l+2), 2k-l-4, 2k-1, -\tanh^2\chi \right)}{(\cosh\chi)^{4k-2l-8}},
\label{Eq:Egas}\\
\ve{J}_{\text{gas}}^{(\text{rot})} 
 &=& L_0\frac{M_{\text{gas}}^{(\text{rot})}}{m}
 \left[ 1 + \frac{l+1}{2k-l-7}\cosh^2\chi \frac{{}_{2}F_{1}\left(-(l+3), 2k-l-7, 2k-3, -\tanh^2\chi \right)}{{}_{2}F_{1}\left(-(l+2), 2k-l-6, 2k-3, -\tanh^2\chi \right)}\right] \ve{e}_z,
\label{Eq:Jgas}
\end{eqnarray}
where the hyperbolic angle $\chi$ is defined by
\begin{equation}
\chi = \frac{1}{2}\sinh^{-1}\left( 2\frac{\sqrt{G M m^2 b}}{L_0} \right).
\end{equation}
Using the symmetries of the DF with respect to $L_z$, the corresponding quantities for the even ansatz~(\ref{Eq:PolytropeLzEven}) yield
\begin{equation}
M_{\text{gas}}^{(\text{even})} = M_{\text{gas}}^{(\text{rot})},\qquad
E_{\text{gas}}^{(\text{even})} = E_{\text{gas}}^{(\text{rot})},\qquad
\ve{J}_{\text{gas}}^{(\text{even})} = 0.
\label{Eq:EvenRotRelation}
\end{equation}
Note that in the Kepler limit $b\to 0$ one has $\chi\to 0$ and the last factors on the right-hand sides of Eqs.~(\ref{Eq:Mgas}) and (\ref{Eq:Egas}) converge to one, while $\ve{J}_{\text{gas}}^{(\text{rot})}/M_{\text{gas}}^{(\text{rot})} \to (L_0/m) (2k-6)/(2k-l-7)$. Another interesting limit consists in taking $L_0\to 0$ and $l = 0$, in which case one obtains\footnote{For this, the identities~\cite[Eq. 15.4.26]{DLMF}
${}_{2}F_{1}\left(a,b;c;-1\right) = {}_{2}F_{1}\left(b,a;c;-1\right) = \Gamma(1+a-b)\Gamma(1+a/2)/(\Gamma(1+a)\Gamma(1-b+a/2))$ are useful.
}
\begin{eqnarray}
\left.\lim_{L_0\to 0} M_{\text{gas}}^{(\text{rot})}\right|_{l=0} 
&=& 8\pi^3\alpha m\frac{(2m b^2 E_0)^{\frac{3}{2}}}{(k-1)(k-2)(k-3)} 
\left( \frac{G M m}{2 b E_0} \right)^k,
\label{Eq:MgasZeroAngular}\\
\left.\lim_{L_0\to 0} E_{\text{gas}}^{(\text{rot})}\right|_{l=0} 
&=& -8\pi^3\alpha E_0\frac{(2m b^2 E_0)^{\frac{3}{2}}}{k(k-1)(k-2)} 
\left( \frac{G M m}{2 b E_0} \right)^{k+1},
\label{Eq:EgasZeroAngular}\\
\left.\lim_{L_0\to 0} \ve{J}_{\text{gas}}^{(\text{rot})}\right|_{l=0} 
&=& 128\pi^3\alpha \frac{(2m b^2 E_0)^2}{(2k-1)(2k-3)(2k-5)(2k-7)} 
\left( \frac{G M m}{2 b E_0} \right)^{k+\frac{1}{2}} \ve{e}_z,
\label{Eq:JgasZeroAngular}
\end{eqnarray}
where the corresponding expressions for the even ansatz can be obtained from the general relations in Eq.~(\ref{Eq:EvenRotRelation}). In the even case, one can check that these expressions give the total mass and energy of the spherical polytropes discussed in subsection~\ref{SubSec:Polytropes}. Note that the Kepler limit $b\to 0$ of the expressions in Eqs.~(\ref{Eq:MgasZeroAngular},\ref{Eq:EgasZeroAngular},\ref{Eq:JgasZeroAngular}) are ill-defined. This is due to the fact that the Kepler potential diverges at the center, implying that particles with low angular momentum can have arbitrarily low energy.

The qualitative behavior of $M_{\text{gas}}$, $E_{\text{gas}}$ and $\ve{J}_{\text{gas}}$ as functions of the parameters $\lambda_0,\kappa,k,l$ will be analyzed in more detail in the next section.

\section{Properties of the macroscopic observables}
\label{Sub:Examples}

In this section we analyze the behavior of the macroscopic observables derived in the previous section. For definiteness, we focus on the particle density, the mean velocity, the kinetic temperature, and the anisotropy parameter (defined below), assuming that the central gravitational potential is given by H\'enon's isochrone potential, see Eq.~(\ref{Eq:IsochronePotential}). In general, these quantities depend on the parameters $\alpha > 0$, $E_0 > 0$, $L_0 > 0$, $l\geq 0$, $k > 7/2 + l/2$ and $M > 0$, $b\geq 0$ appearing in the DF and H\'enon's potential. In the following, we compare configurations of equal total mass, which fixes the parameter $\alpha$ (see Eq.~(\ref{Eq:Mgas})), and we introduce the length scale
\begin{equation}
\Lambda := \frac{G M m}{E_0},
\label{Eq:LengthScaleKinetic}
\end{equation}
in order to express our results in terms of the dimensionless quantities:
\begin{equation}
\xi := \frac{r}{\Lambda},\qquad
\kappa := \frac{b}{\Lambda},\qquad
\lambda := \frac{L}{\sqrt{m\Lambda^2 E_0}},\qquad
\lambda_0 := \frac{L_0}{\sqrt{m\Lambda^2 E_0}},
\label{Eq:DimensionlessQuantitiesKinetic}
\end{equation}
which leaves the following dimensionless parameters: $k$, $l$, $\lambda_0$ and $\kappa$.

\subsection{Normalized particle density and morphology}

\begin{figure}[h!]
\centering
\subfigure{\includegraphics[scale=0.3]{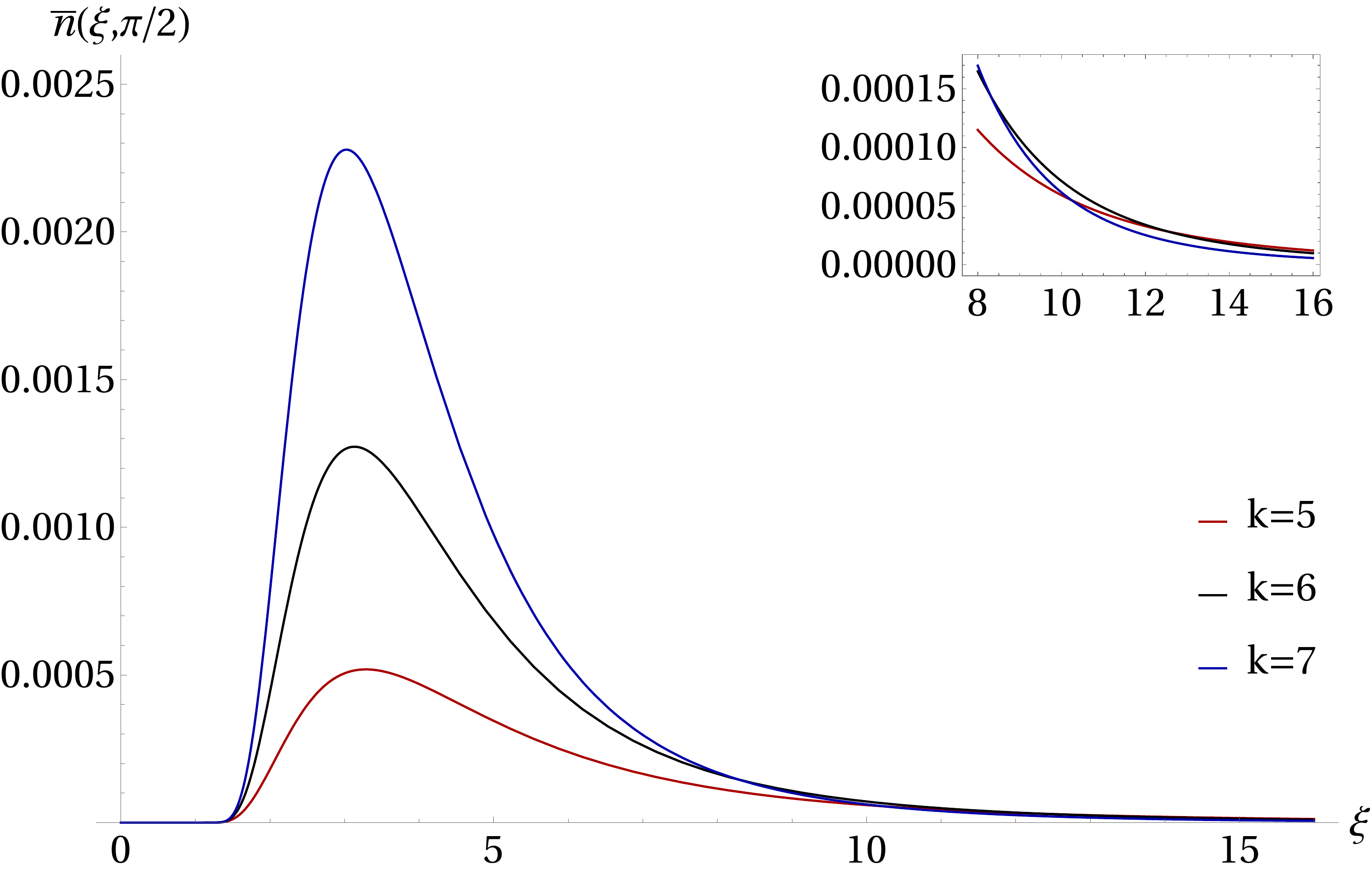}} 
\subfigure{\includegraphics[scale=0.28]{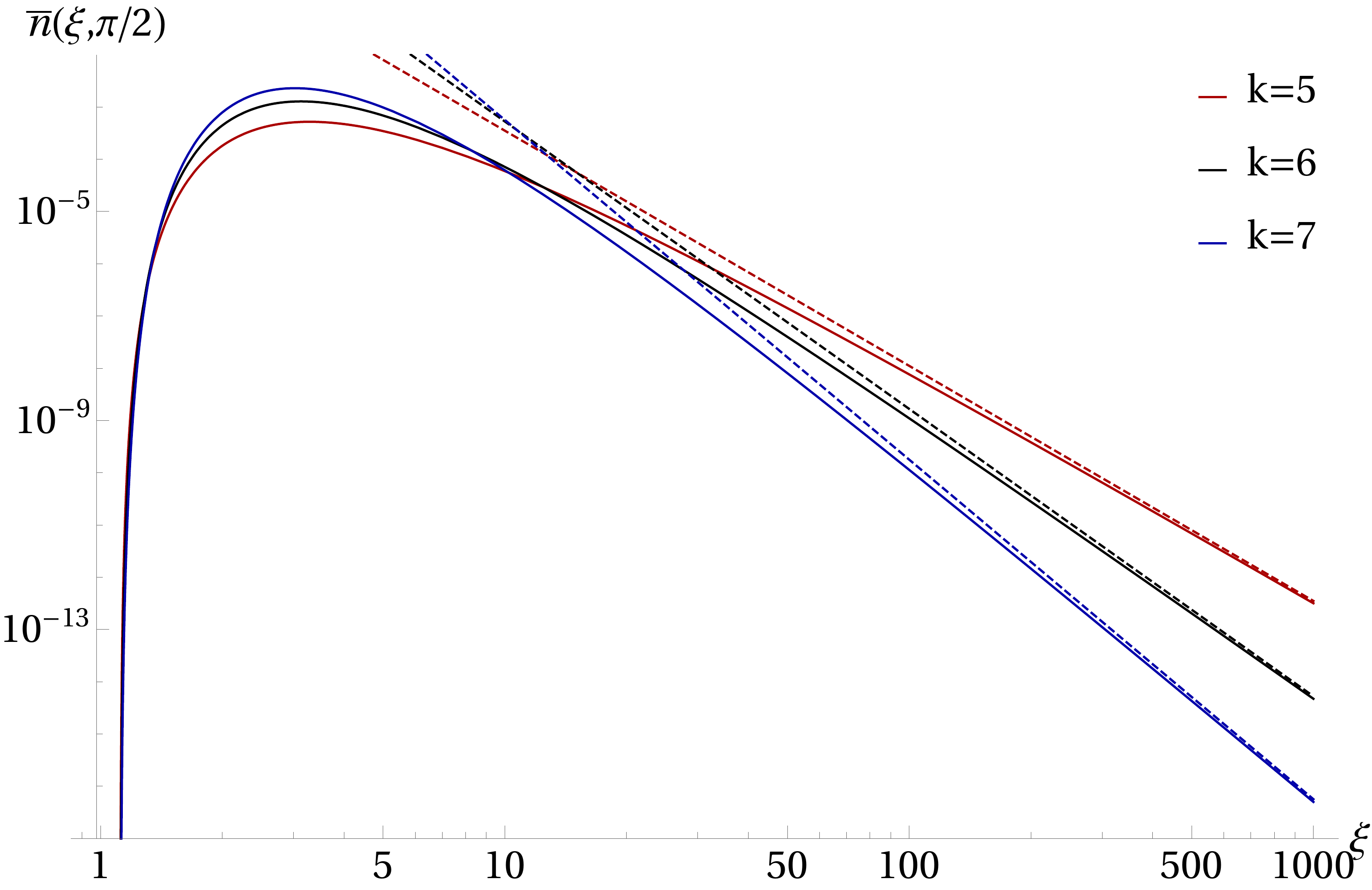}} 
\caption{Normalized particle density $\bar{n}$ vs radius on the equatorial plane for $l=1$ and $k=5,6,7$. The remaining parameters have been chosen to be $\kappa=1$ and $\lambda_0=1$. The left panel is a linear plot showing the range $0\leq\xi \leq 16$, with the inset zooming into the region $8\leq\xi\leq 16$ which makes visible the intersection between the different curves. This intersection is expected, since the higher concentration of particles near the central object has to be compensated by a faster decay in the density due to the fact that we compare configurations with equal masses. The right panel shows the same quantity in a log-log plot, making visible the different inverse power law decay behavior at large $\xi$. The dashed lines correspond to the inverse power law behavior found in Eq.~(\ref{Eq:AsymptoticBehavior}).}
\label{Fig:ParticleDensity1}
\end{figure}

\begin{figure}[h!]
\centering 
	\subfigure{\includegraphics[scale=0.30]{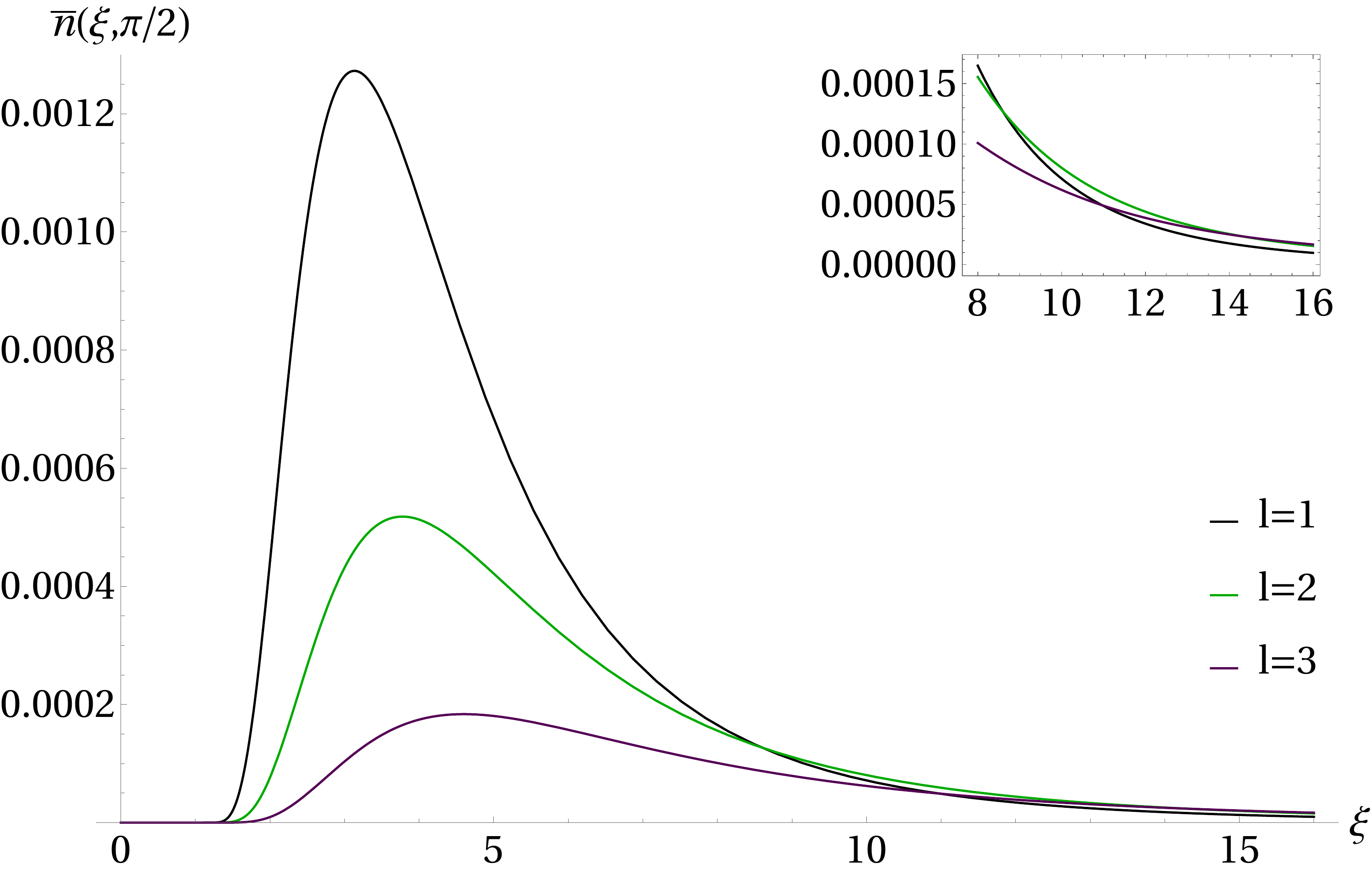}} 
	\subfigure{\includegraphics[scale=0.28]{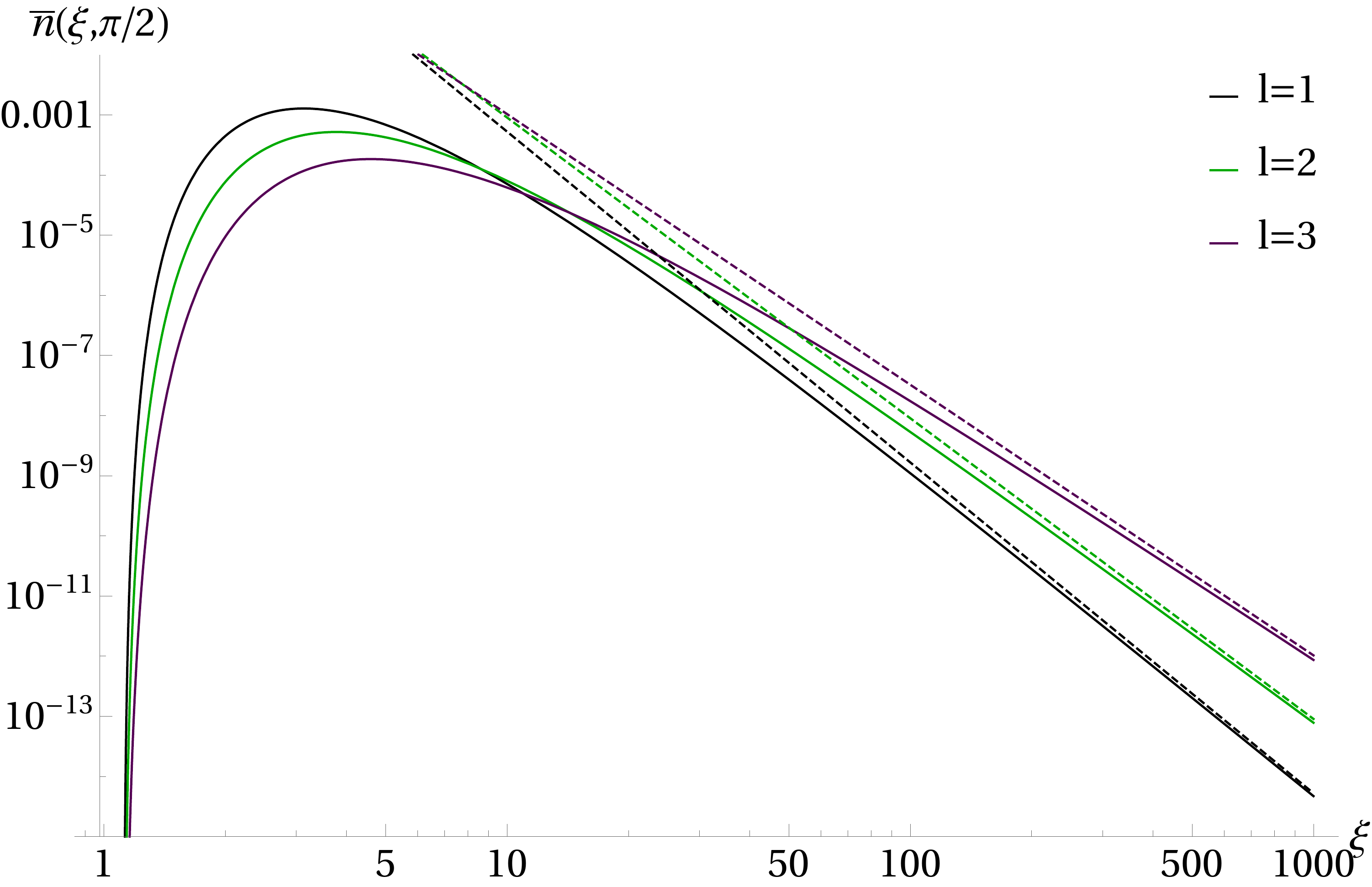}} 
\caption{Same as in previous plot, except that here $k=6$ and $l=1,2,3$. Note that the maximum of $\bar{n}$ moves to the right as $l$ increases.}
\label{Fig:ParticleDensity2}
\end{figure}

We start with the analysis of the dimensionless particle density per unit total mass, given by the function (see Eqs.~(\ref{Eq:nrot},~\ref{Eq:Mgas}))
\begin{equation}
\bar{n}(\xi,\vartheta) 
 := \Lambda^3 n^{\text{(rot)}}\frac{m}{M^{\text{(rot)}}_{\text{gas}}} 
 = \frac{2^{k-2}}{\pi^2} \frac{\Gamma(2k-3)}{\Gamma(l+2)\Gamma(2k-l-6)} 
\frac{\eta^l \psi^k J_{k-\frac{3}{2},l}(\eta) \lambda_0^{2k-6}(\cosh^2\chi)^{2k-l-6}}{{}_{2}F_{1}\left( -(l+2),2k-l-6,2k-3,-\tanh^2\chi\right)},
\label{Eq:nrotDimensionless}
\end{equation}
where in terms of the dimensionless quantities defined in Eq.~(\ref{Eq:DimensionlessQuantitiesKinetic}),
\begin{equation}
\chi = \frac{1}{2}\sinh^{-1}\left( \frac{2\sqrt{\kappa}}{\lambda_0} \right),\qquad
\eta = \frac{\xi\sin\vartheta}{\lambda_0}\sqrt{2\psi},\qquad
\psi = \frac{1}{\kappa + \sqrt{\kappa^2 + \xi^2}}.
\label{Eq:chietaDimensionless}
\end{equation}
Note that $\psi$ is a function of the dimensionless radius $\xi$, $\eta$ a function of $(\xi,\vartheta)$, while $\chi$ is constant. We show in figures~\ref{Fig:ParticleDensity1} and~\ref{Fig:ParticleDensity2} the behavior of the normalized particle density~(\ref{Eq:nrotDimensionless}) in the equatorial plane ($\vartheta=\pi/2$) for different values of $k$ and $l$ and fixed values of $\lambda_0 = \kappa = 1$. First, we observe from figure~\ref{Fig:ParticleDensity1} that, for fixed $l=1$, the configurations become more concentrated near their inner edge (determined by $\eta = 1$ corresponding to $\xi = \sqrt{5}/2$ when $\vartheta = \pi/2$) and exhibit a faster asymptotic decay as $k$ increases. In this sense, they become more compact with growing $k$. Next, from the plots in figure~\ref{Fig:ParticleDensity2}, we observe that, for fixed $k=6$, the bulk of the particles moves away from the central region as $l$ increases. This is, of course, expected, since $l$ controls the exponent of the $L_z$-dependency of the DF, implying that the higher the value of $l$, the more populated are the configurations with high angular momentum.

As is visible from the plots in the right panels of figures~\ref{Fig:ParticleDensity1} and~\ref{Fig:ParticleDensity2}, the normalized density $\bar{n}$ exhibits an inverse power-law behavior for large values of $\xi$. Using Eqs.~(\ref{Eq:nrotDimensionless},\ref{Eq:chietaDimensionless}) one finds, for large values of $\xi$:
\begin{equation}
\bar{n}\sim \sigma\frac{\sin^l\vartheta}{\xi^{k-\frac{l}{2}}},
\label{Eq:AsymptoticBehavior}
\end{equation}
where the constant $\sigma$ is given by
\begin{equation}
\sigma = \frac{(\sqrt{2})^{2k+l-6}}{\pi^2}\frac{\Gamma(2k-3)\Gamma(k-1/2)}{\Gamma(2k-l-6)\Gamma(k+l+3/2)} \frac{{}_{2}F_{1}\left(-(k-1/2), l+1, k+l+3/2, -1 \right)}{{}_{2}F_{1}\left( -(l+2),2k-l-6,2k-3,-\tanh^2\chi\right)} \left(\lambda_0\cosh^2\chi\right)^{2k-l-6}.
\label{Eq:Slopen}
\end{equation}
For comparison, the asymptotic behavior~(\ref{Eq:AsymptoticBehavior}) is also shown in the right panels of figures~\ref{Fig:ParticleDensity1} and~\ref{Fig:ParticleDensity2}.

\begin{figure}[h!]
\centering
\subfigure{\includegraphics[scale=0.325]{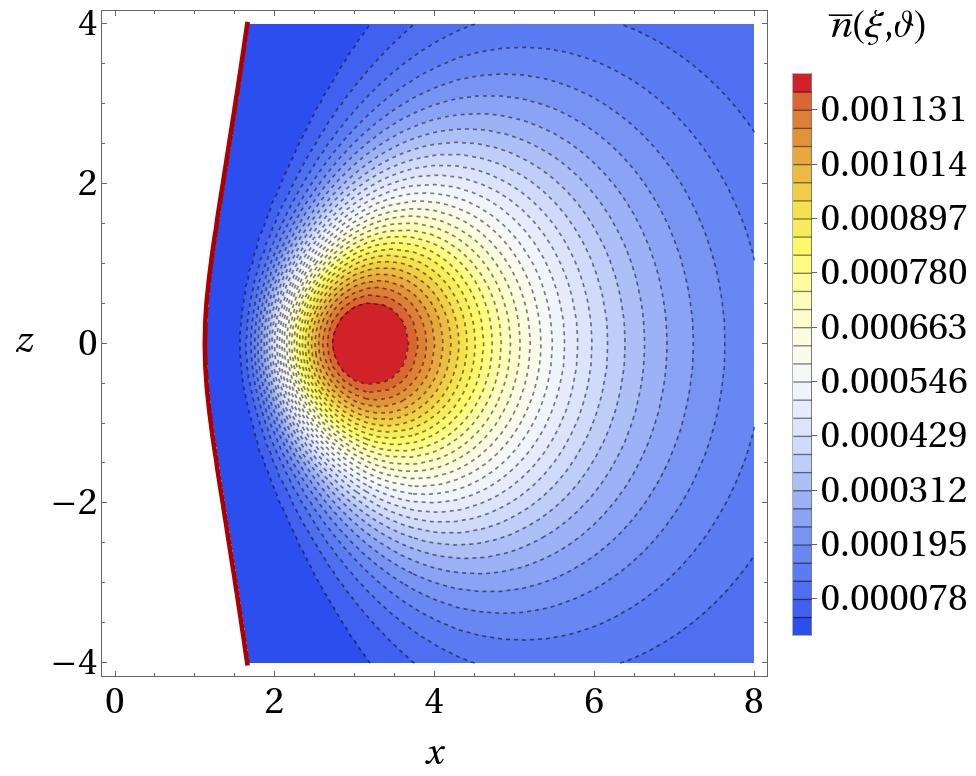}}
\subfigure{\includegraphics[scale=0.325]{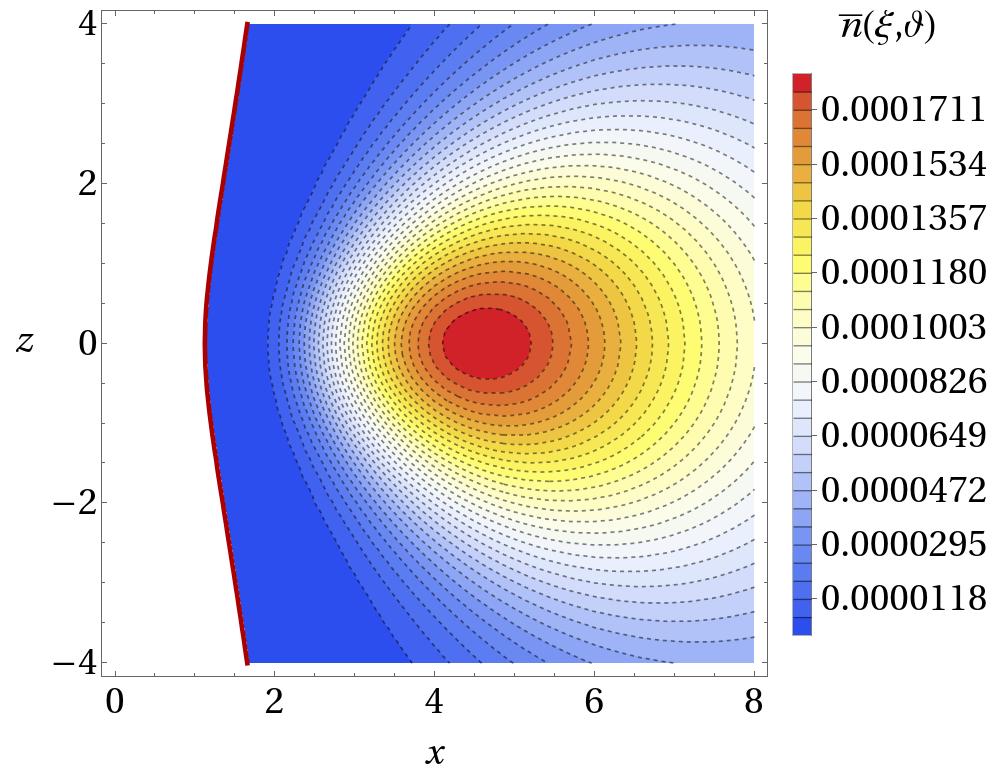}}
\subfigure{\includegraphics[scale=0.325]{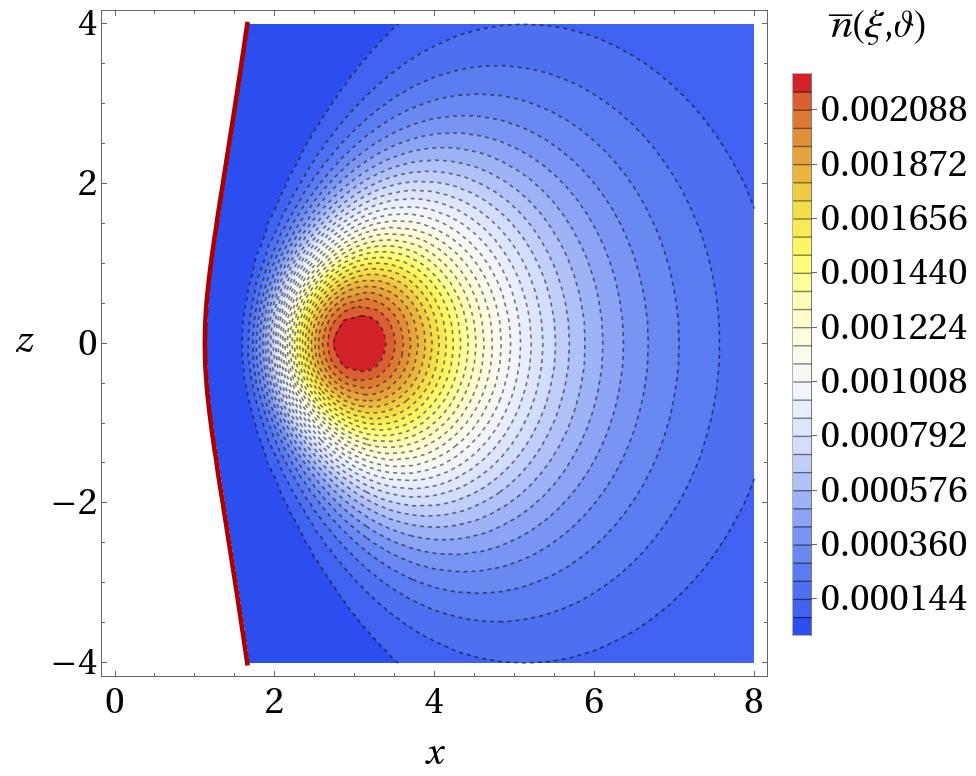}}
\subfigure{\includegraphics[scale=0.325]{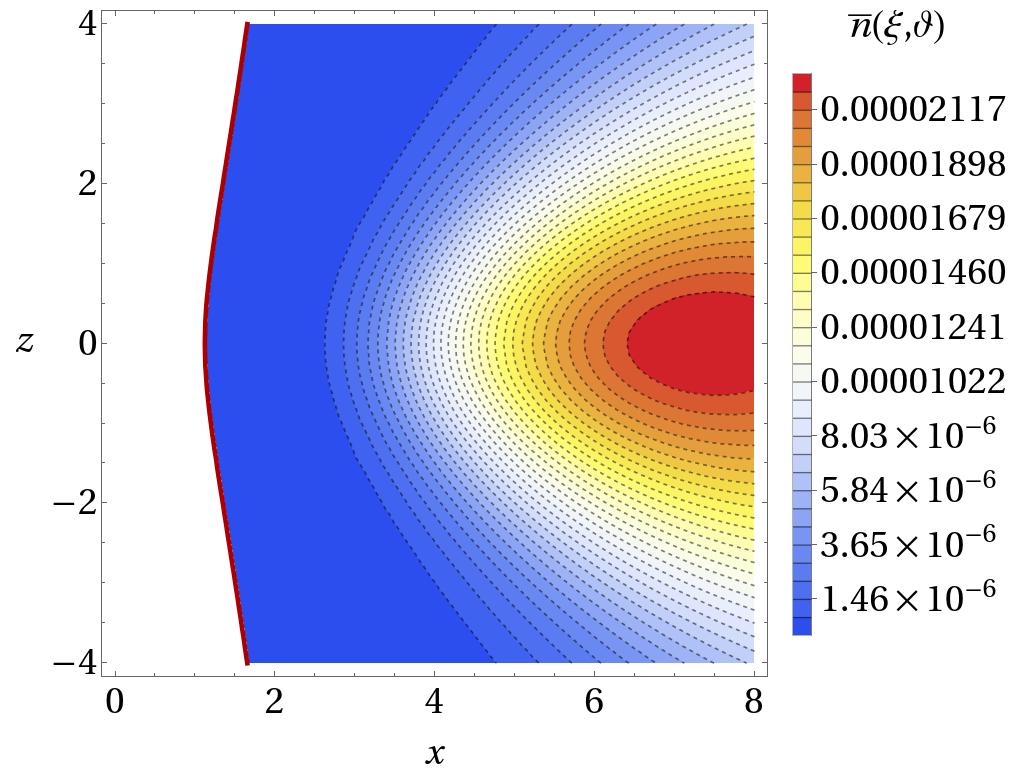}}
\caption{Contour plot for the normalized particle density $\bar{n}$ in the $xz$-plane for $\lambda_0 = \kappa = 1$ and different values of $k$ and $l$. Top left panel: $(k,l) = (6,1)$. Top right panel: $(k,l) = (6,3)$. Bottom left panel: $(k,l) = (7,1)$. Bottom right panel: $(k,l) = (7,6)$. Note that as the value of $l$ increases the disk becomes slimmer and the maximum of the density moves away from the central object, as expected. The thick red line indicates the location of the inner boundary of the configuration which is determined by $\eta\geq 1$.}
\label{Fig:KineticDescription} 
\end{figure}

In order to analyze the dependency of the particle density with respect to the polar angle, we show in figure~\ref{Fig:KineticDescription} contour plots of $\bar{n}$ in the $xz$-plane for the parameter values $k=5$, $l=1,2,3$, and $\lambda_0 = \kappa = 1$. As is visible from these plots, the particle density is everywhere regular and the configurations become slimmer as $l$ increases, which is compatible with the polar dependency of the asymptotic behavior~(\ref{Eq:AsymptoticBehavior}).

\subsection{Mean particle velocity}

\begin{figure}[h!]
\centerline
	\subfigure{\includegraphics[scale=0.29]{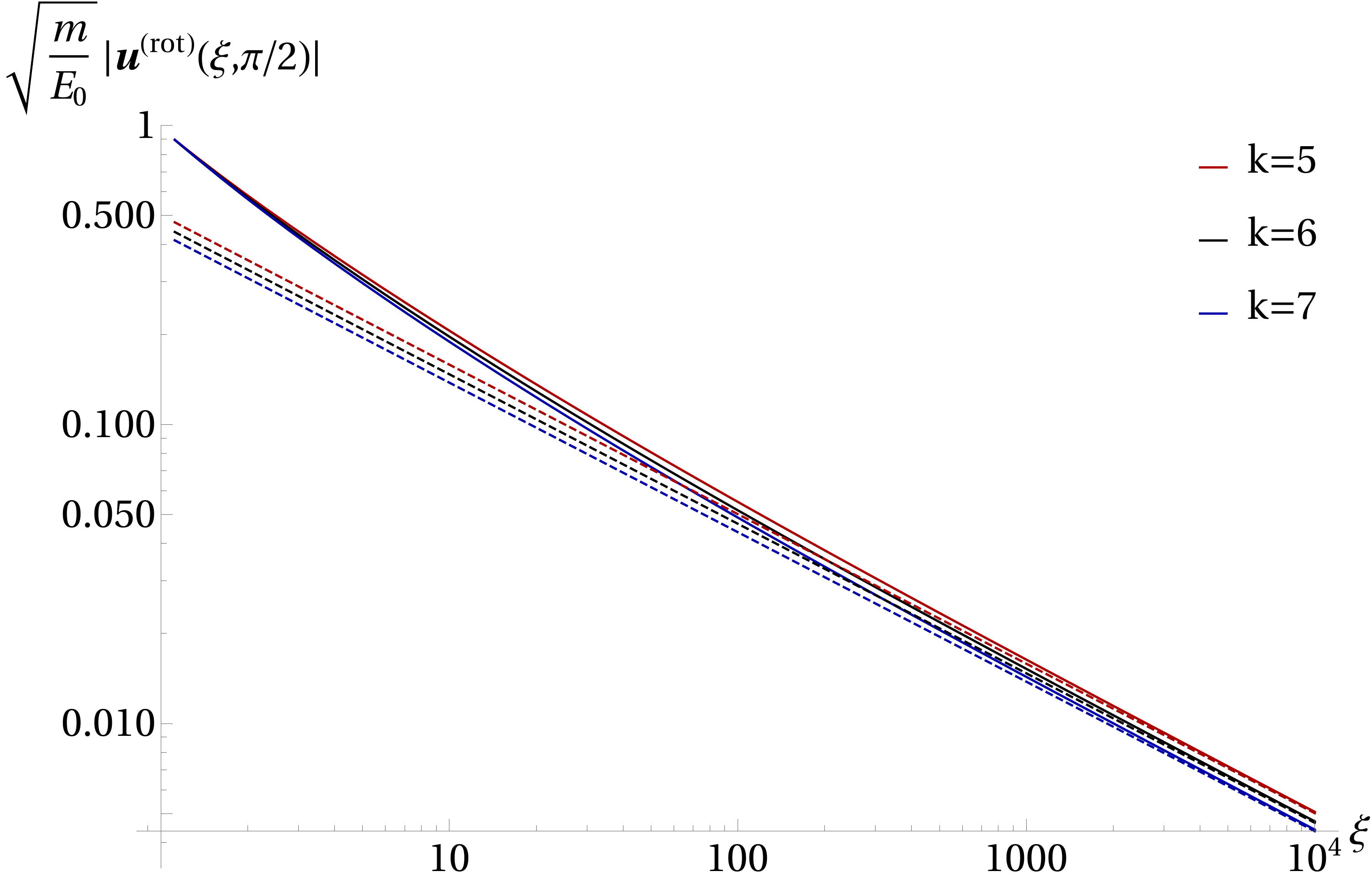}} 
	\subfigure{\includegraphics[scale=0.29]{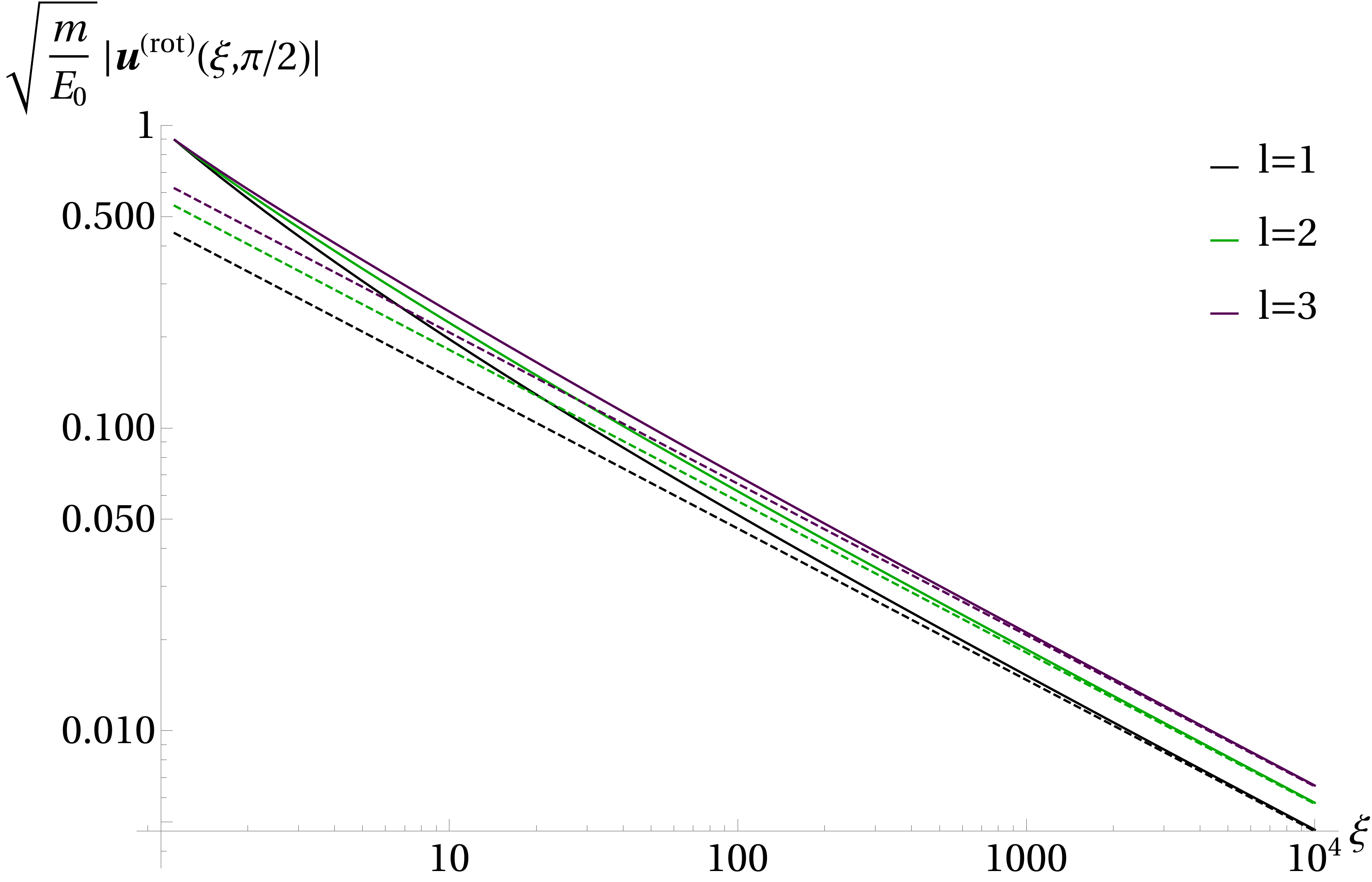}} 
\caption{Log-log plot for the magnitude of the mean velocity vs radius for some values of $l$ and $k$ and fixed parameter values $\lambda_0 = \kappa=1$. From the left panel we see that for $l=1$ this magnitude decreases as $k$ increases, while the right panel shows that, as expected, this magnitude increases as $l$ increases. The dashed lines show the asymptotic behavior predicted by Eq.~(\ref{Eq:Slopeu}).}
\label{Fig:MeanVelocity1}
\end{figure}

Next, we analyze the behavior of the mean particle velocity for rotating configurations, see Eq.~(\ref{Eq:urot}) (recall that this velocity vanishes for the even ansatz). For definiteness, we restrict ourselves to the equatorial plane. In figure~\ref{Fig:MeanVelocity1} we show the magnitude of the velocity as a function of the dimensionless radius $\xi$ for different values of $k$ and $l$ and fixed $\kappa = \lambda_0 = 1$. We observe that this magnitude has a maximum at the inner edge ($\xi = \sqrt{5}/2$) and decays monotonically as $\xi$ increases.
At the inner edge one finds
\begin{equation}
\lim\limits_{\eta\to 1^+} \sqrt{\frac{m}{E_0}} |\ve{u}^{(\text{rot})}(\ve{x})| 
 = \left. \sqrt{2\psi} \right|_{\xi = \sqrt{5}/2},
\end{equation}
which is independent of $k$ and $l$. For large $\xi$ one finds the following behavior:
\begin{equation}
\sqrt{\frac{m}{E_0}}|\ve{u}^{(\text{rot})}| \sim \frac{\tau}{\sqrt{\xi}},\qquad
\tau
 = \sqrt{2}\frac{\Gamma\left(\frac{l}{2}+1\right)\Gamma\left(k+\frac{l}{2}+1\right)}{\Gamma\left(\frac{l}{2}+\frac{1}{2}\right)\Gamma\left(k+\frac{l}{2}+\frac{3}{2}\right)},
\label{Eq:Slopeu}
\end{equation}
which is (up to the factor $\tau$ which is of order unity) the expression for the tangential velocity for circular orbits in the Kepler potential. Note that $\tau$ is independent of the polar angle $\vartheta$ and the parameters $\lambda_0$ and $\kappa$. 

\subsection{Kinetic temperature}

\begin{figure}[h!]
\centerline
	\subfigure{\includegraphics[scale=0.28]{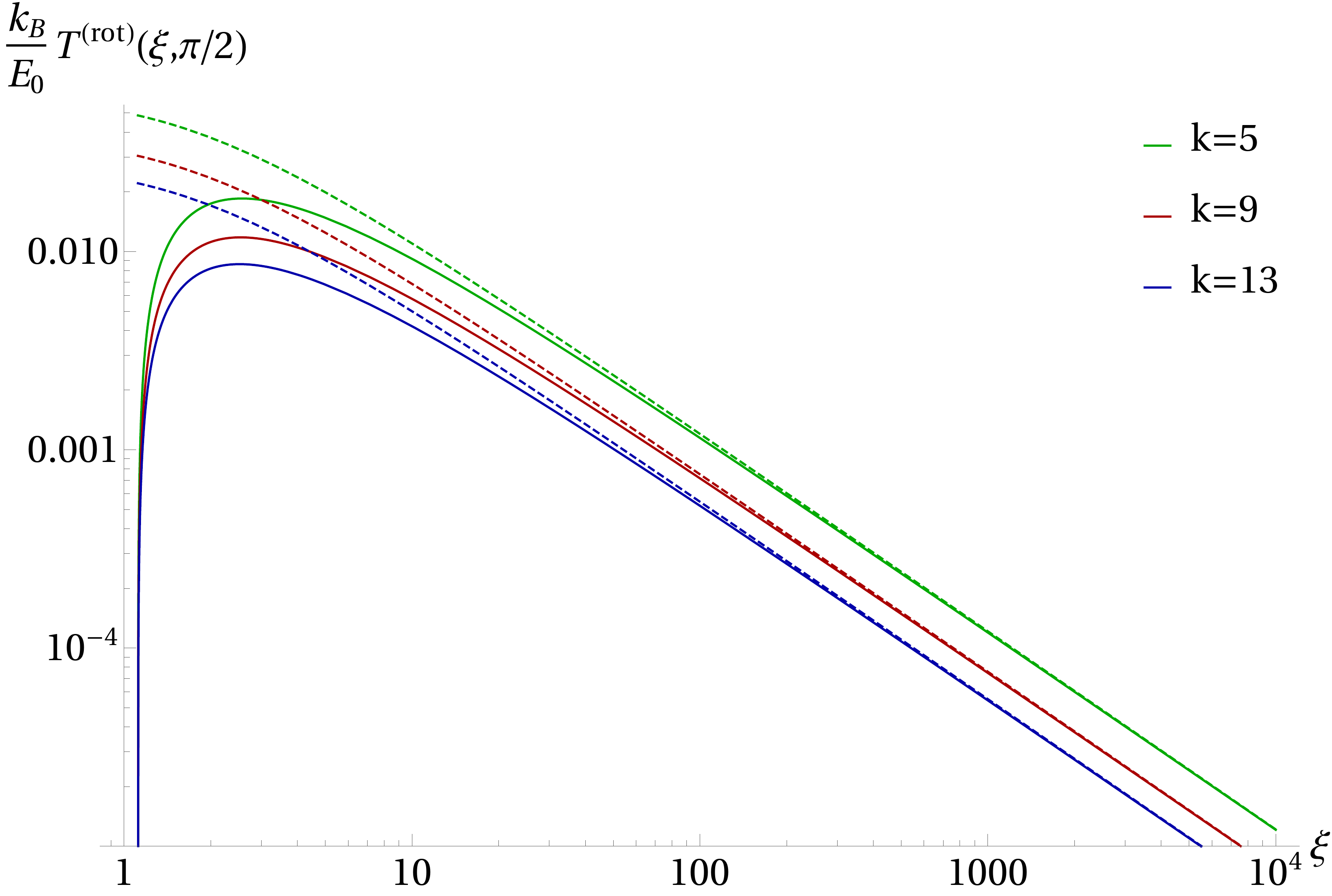}}
	\subfigure{\includegraphics[scale=0.28]{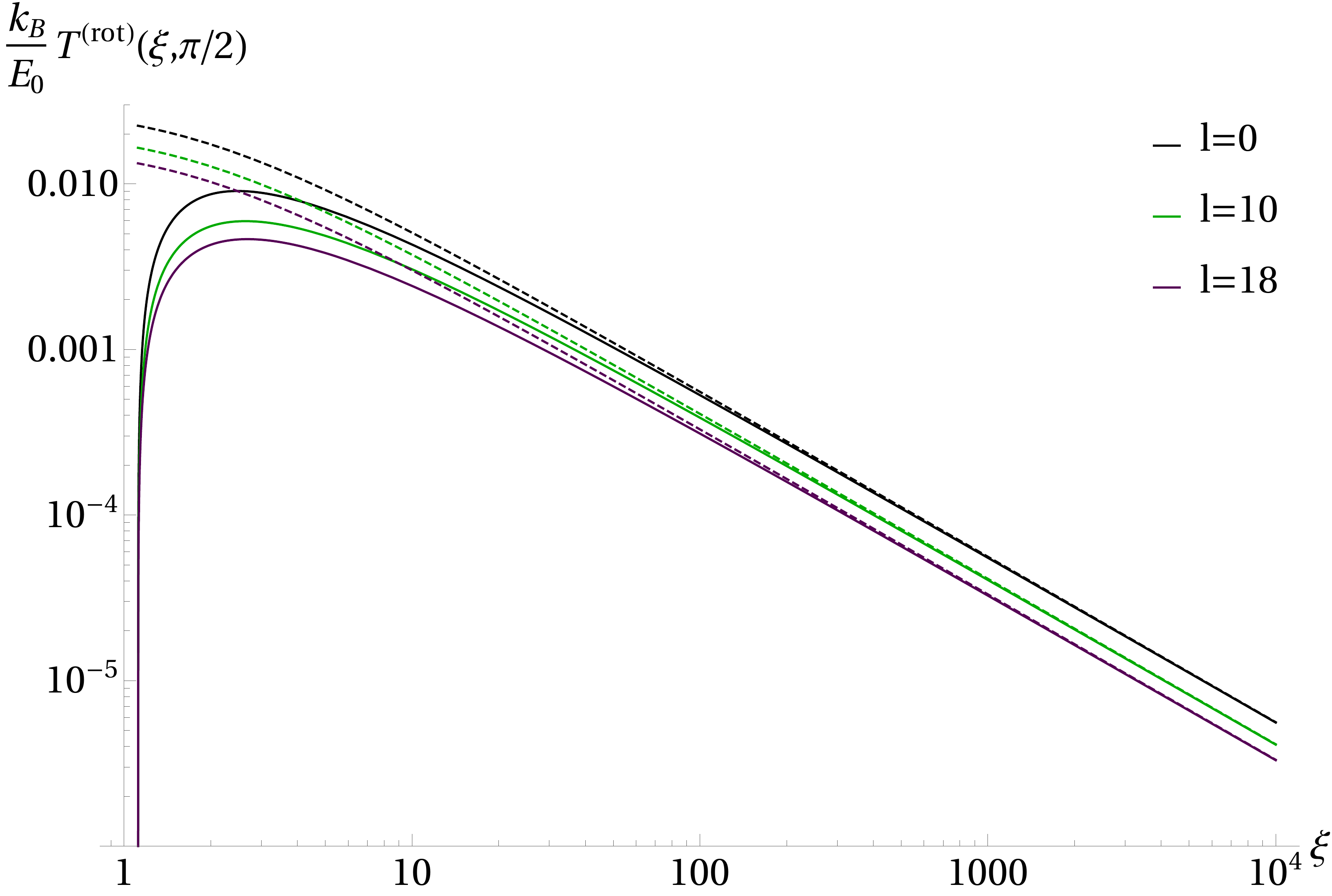}} 
\caption{Log-log plot for the kinetic temperature of the rotating model (see Eq.~(\ref{Eq:KineticTemperatureRot})) vs radius in the equatorial plane for $l=1$, $k=5,9,13$ (left panel) and $k=13$ and $l=0,10,18$ (right panel) and $\lambda_0 = \kappa =1$. The  corresponding asymptotic limits from Eq.~(\ref{Eq:TrotAsymptotic}) are shown in dashed lines.}
\label{Fig:TemperatureRot}
\end{figure}

\begin{figure}[h!]
\centerline
	\subfigure{\includegraphics[scale=0.28]{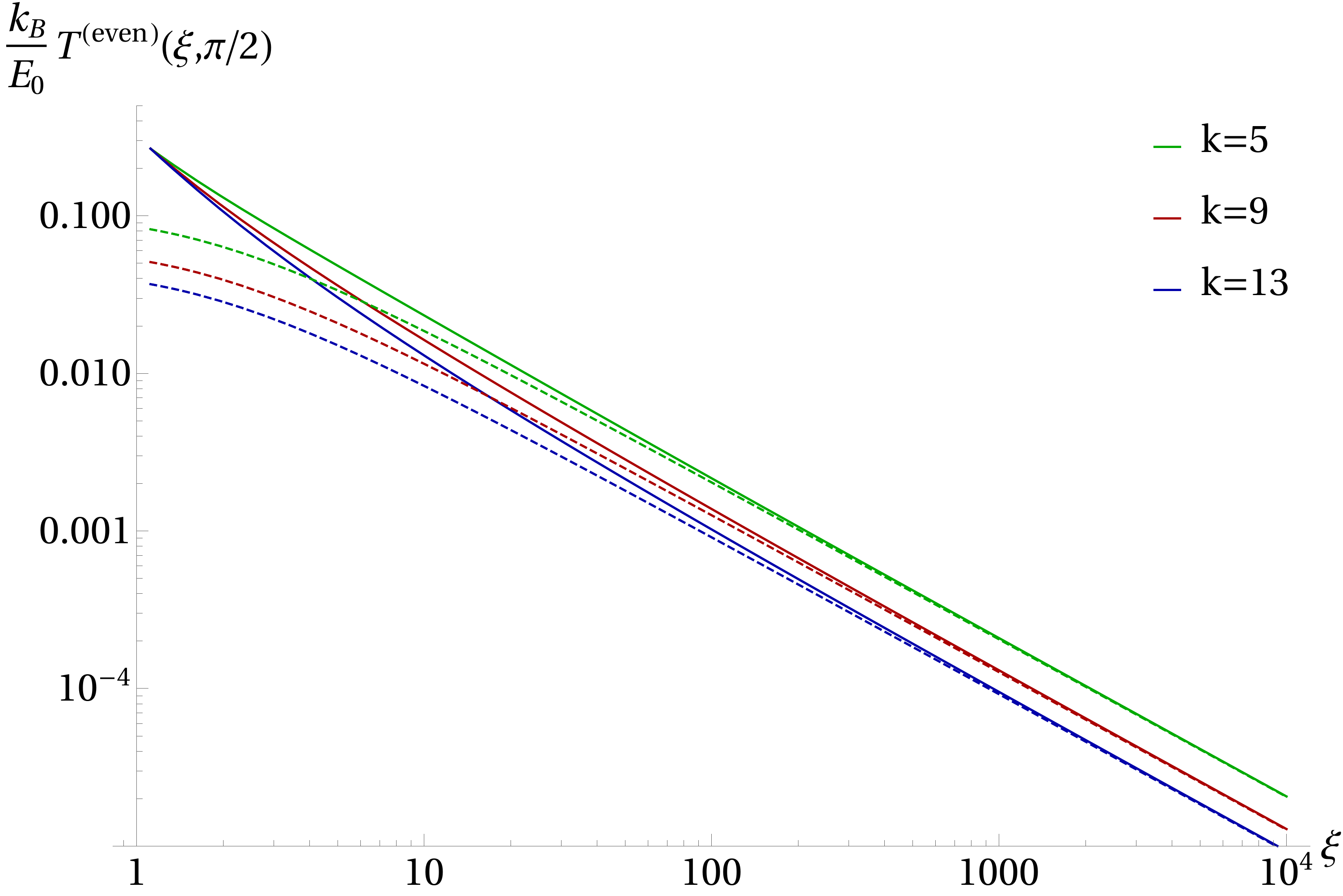}}
	\subfigure{\includegraphics[scale=0.28]{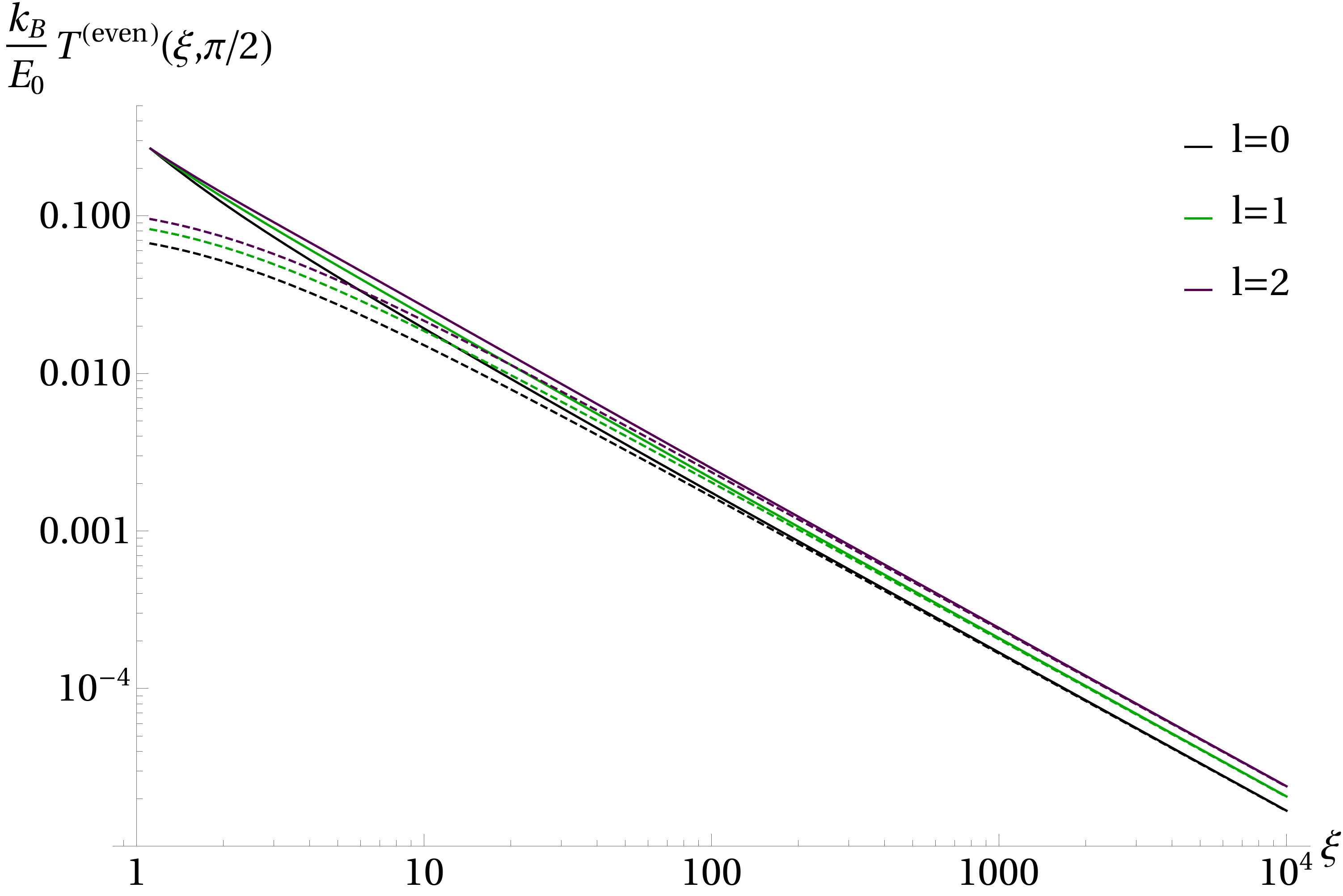}} 
\caption{Same as previous plot for the even model (see Eq.~(\ref{Eq:KineticTemperatureEven}))}
\label{Fig:TemperatureEven}
\end{figure}

Next, we analyze the behavior of the kinetic temperature, which is given by the ratio between the pressure~(\ref{Eq:IsotropicPressureRot}) and the particle density~(\ref{Eq:nrot}). For the ans\"atze~(\ref{Eq:PolytropeLzRot},\ref{Eq:PolytropeLzEven}) this yields
\begin{eqnarray}
\label{Eq:KineticTemperatureRot}
T^{(\text{rot})}(\ve{x}) &=& \frac{2}{3}\frac{E_0\psi}{k_B} \left[ 1 - \frac{ J_{k-\frac{1}{2},l}(\eta)}{ J_{k-\frac{3}{2},l}(\eta)} - \left(\frac{l+1}{l+2}\frac{ J_{k-\frac{3}{2},l+1}(\eta)}{ J_{k-\frac{3}{2},l}(\eta)} + \frac{1}{\eta}\right)^2 \right],\\
\label{Eq:KineticTemperatureEven}
T^{(\text{even})}(\ve{x}) &=& \frac{2}{3}\frac{E_0\psi}{k_B} \left[ 1 - \frac{ J_{k-\frac{1}{2},l}(\eta)}{ J_{k-\frac{3}{2},l}(\eta)} \right].
\end{eqnarray}
In figures~\ref{Fig:TemperatureRot} and~\ref{Fig:TemperatureEven} we show the temperature profiles on the equatorial plane for these two models. As before, we fix $\lambda_0 = \kappa = 1$ and vary the parameters $k$ and $l$. The asymptotic values in the limits $\eta\to 1$ and $\eta\to \infty$ for the kinetic temperature in both models are given by
\begin{eqnarray}
&& \lim\limits_{\eta\rightarrow 1^+} T^{(\text{rot})} = 0,\qquad 
\lim\limits_{\eta\rightarrow\infty} T^{(\text{rot})} 
 = \frac{2}{3}\frac{E_0\psi}{k_B}\left\{ \frac{l+3}{2k+l+2} - \left[\frac{\Gamma\left(\frac{l}{2}+1\right)\Gamma\left(k+\frac{l}{2}+1 \right)}{\Gamma\left(\frac{l}{2}+\frac{1}{2} \right)\Gamma\left(k+\frac{l}{2}+\frac{3}{2} \right)}\right]^2\right\}, 
\label{Eq:TrotAsymptotic}\\
&& \lim\limits_{\eta\rightarrow 1^+} T^{(\text{even})} = \frac{2}{3}\frac{E_0\psi}{k_B},
\qquad
\lim\limits_{\eta\rightarrow\infty} T^{(\text{even})} 
 = \frac{2}{3}\frac{E_0\psi}{k_B} \frac{l+3}{2k+l+2}.
\end{eqnarray}
From the plots and these expressions, we observe that the temperature profile for the rotating model is continuous, $T^{(\text{rot})}$ being zero at the inner boundary of the disk. The temperature has a maximum inside the disk configuration, as is visible from figure~\ref{Fig:TemperatureRot}, and decays as $1/\xi$ for large radii. In contrast to this, the temperature profile for the even model in the equatorial plane is monotonously decaying for all values of $\xi$, the maximum temperature being located at the inner edge of the disk. As is also visible from the plots, the temperature decreases as $k$ increases, which can be understood by remarking that the lower energy levels are more populated when $k$ is high.

\subsection{Pressure anisotropy}

\begin{figure}[h!]
\centerline
	\subfigure{\includegraphics[scale=0.29]{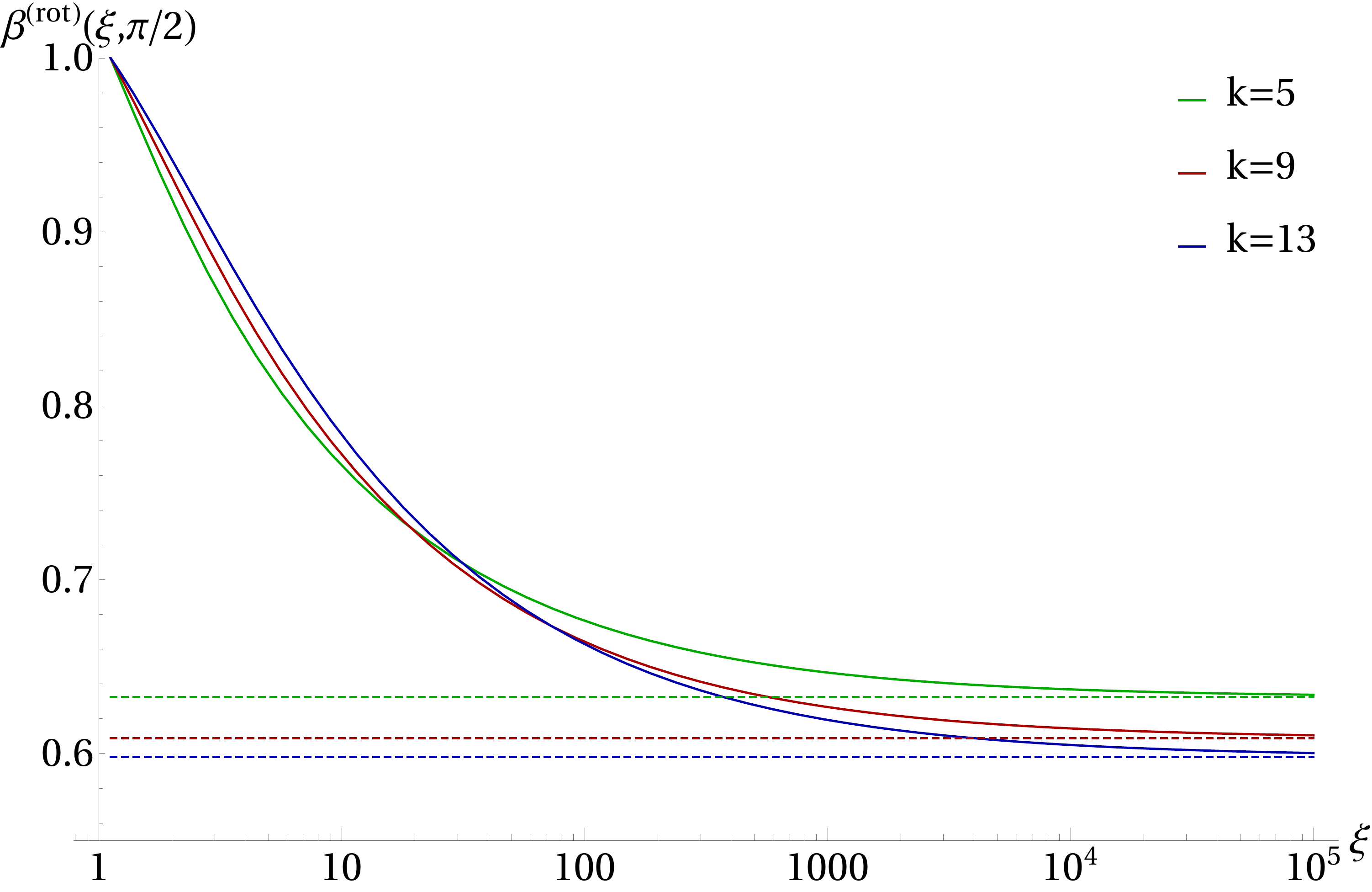}} 
	\subfigure{\includegraphics[scale=0.29]{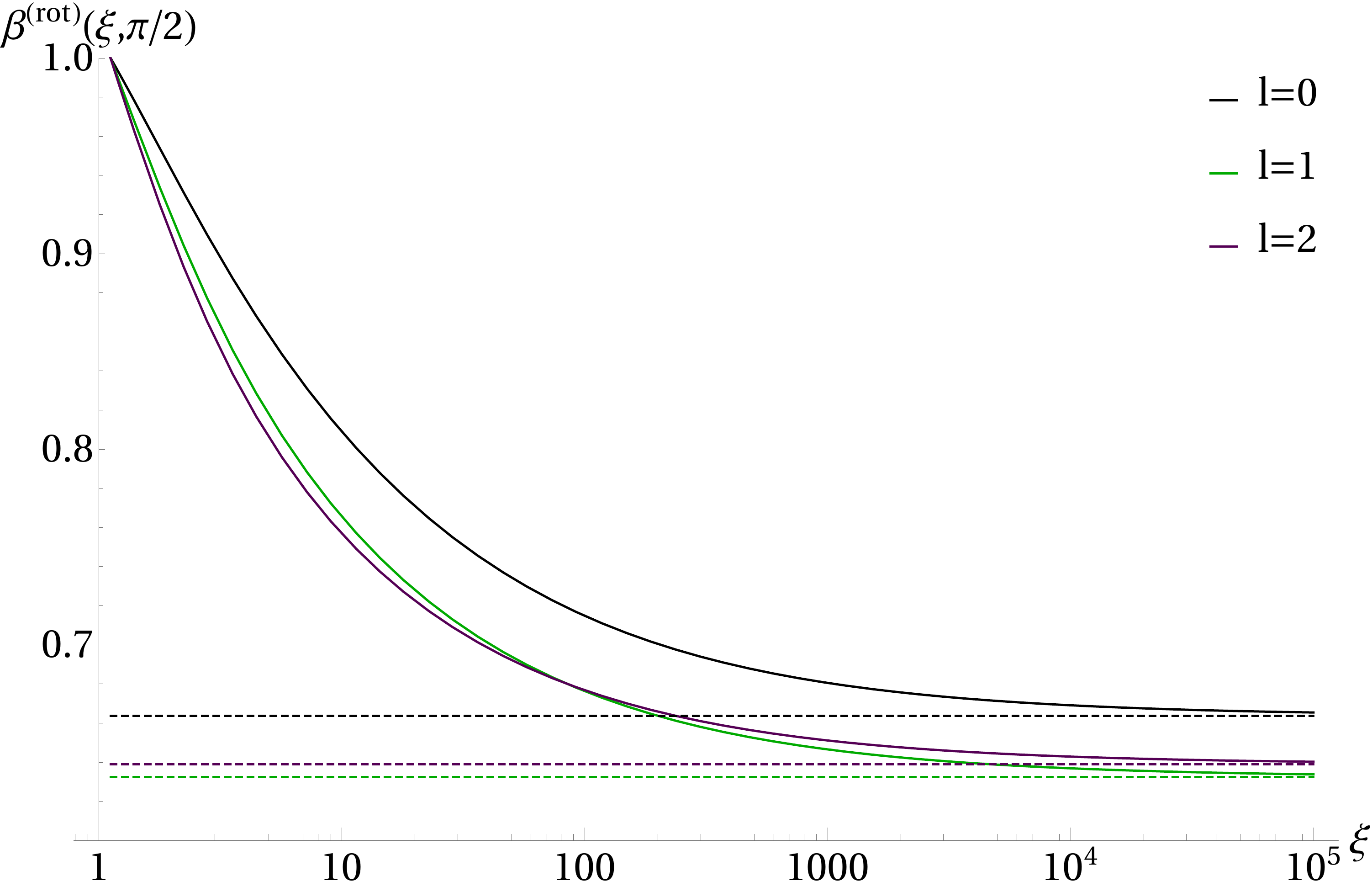}}
\caption{Anisotropy parameter $\beta$ vs $\xi$ for different values of $k$ and $l$ and fixed $\lambda_0 = \kappa = 1$. Note that we use a logarithmic scale in $\xi$. The dashed lines refer to the asymptotic limits computed in Eq.~(\ref{Eq:BetaRotEtaInf}). In the left panel $l=1$ and $k = 5,9,13$ while in the right panel $k=5$ is fixed and $l$ varies over $0,1,2$. In all cases, we observe that $\beta\to 1$ when $\xi$ approaches the inner edge's radius, while for large $\xi$, $\beta$ decreases and approaches a constant positive value, indicating that the radial pressure always dominates the azimuthal one.}
\label{Fig:AnisotropyRot}
\end{figure}

\begin{figure}[h!]
\centerline
	\subfigure{\includegraphics[scale=0.29]{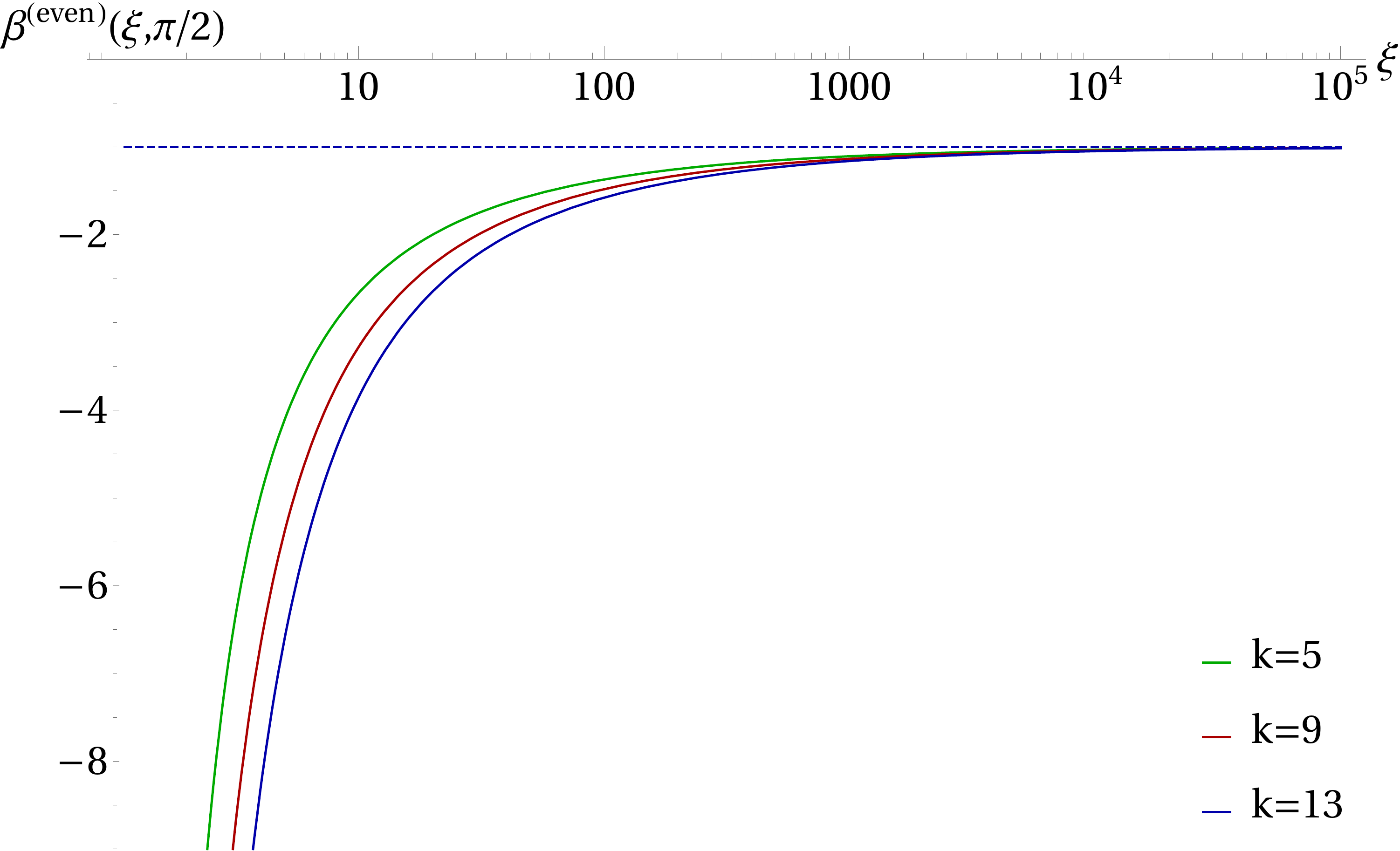}}
	\subfigure{\includegraphics[scale=0.29]{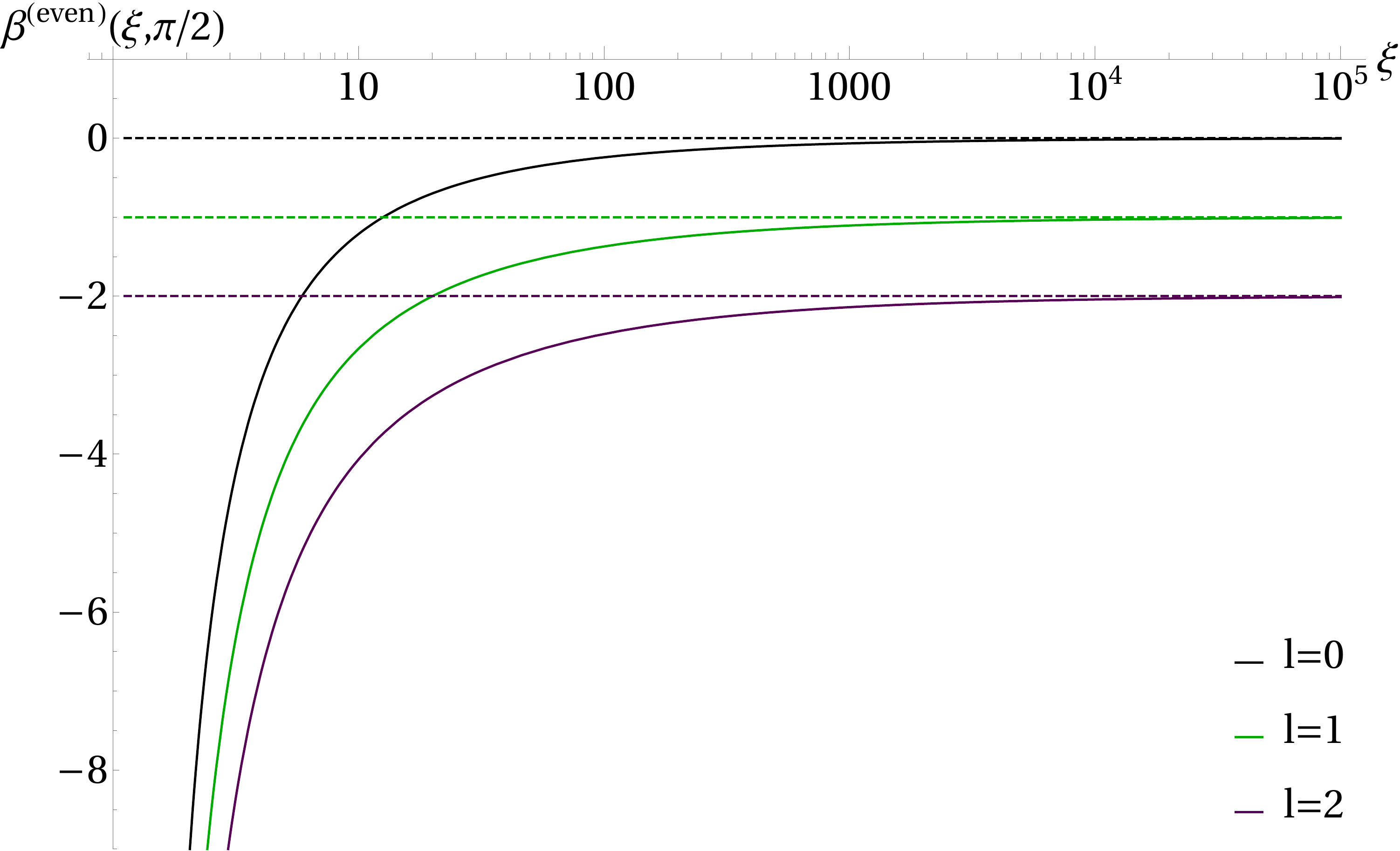}}
\caption{Same as previous plot for the even model. In this case, the anisotropy parameter $\beta$ diverges to $-\infty$ as $\xi$ approaches the inner edge's radius, while in the limit $\xi\to \infty$, $\beta$ approaches a negative value, indicating that the azimuthal pressure dominates in this case.}
\label{Fig:AnisotropyEven}
\end{figure}

In this section we analyze the pressure anisotropy of our configurations. To this purpose, recall that the principle pressures $P_{\hat{r}}$, $P_{\hat{\vartheta}}$ and $P_{\hat{\varphi}}$ defined in Eqs.~(\ref{Eq:PressureTensorDecomposition},\ref{Eq:P1},\ref{Eq:P3}) satisfy $P_{\hat{r}} = P_{\hat{\vartheta}}$. Therefore, in analogy to what is being defined in galactic dynamics~\cite{BinneyTremaine-Book,lCh2018} we introduce the anisotropy parameter
\begin{equation}
\beta := 1 - \frac{P_{\hat{\varphi}}}{P_{\hat{r}}},
\label{Eq:AnisotropyParameter}
\end{equation}
which is positive for a configuration in which the radial pressure dominates the azimuthal one and negative otherwise, while $\beta = 0$ for an isotropic configuration. For the rotating configurations~(\ref{Eq:PolytropeLzRot}) one obtains
\begin{eqnarray}
\beta^{(\text{rot})}(\ve{x}) 
&=& 1 - \left[\frac{l+1}{l+3} J_{k-\frac{3}{2},l+2}(\eta) - \left(\frac{l+1}{l+2}\right)^2 \frac{J_{k-\frac{3}{2},l+1}(\eta)^2}{J_{k-\frac{3}{2},l}(\eta)} \right] \left[\frac{1}{\eta}\frac{J_{k-\frac{3}{2},l+1}(\eta)}{l+2} + \frac{J_{k-\frac{3}{2},l+2}(\eta)}{l+3} \right]^{-1},
\label{Eq:BetaRot}
\end{eqnarray}
while for the even model~(\ref{Eq:PolytropeLzEven}) the anisotropy parameter yields
\begin{equation}
\beta^{(\text{even})}(\ve{x}) 
= 1 - \left[\frac{1}{\eta^2}J_{k-\frac{3}{2},l}(\eta) + \frac{2}{\eta}\frac{l+1}{l+2} J_{k-\frac{3}{2},l+1}(\eta) + \frac{l+1}{l+3} J_{k-\frac{3}{2},l+2}(\eta)\right] \left[\frac{1}{\eta}\frac{J_{k-\frac{3}{2},l+1}(\eta)}{l+2} + \frac{J_{k-\frac{3}{2},l+2}(\eta)}{l+3} \right]^{-1}.
\label{Eq:BetaEven}
\end{equation}
Note that these expressions only depend on $(\xi,\vartheta)$ through the parameter $\eta$. The behavior of $\beta$ on the equatorial plane is shown in figures~\ref{Fig:AnisotropyRot} and \ref{Fig:AnisotropyEven} for both models for different values of $k$ and $l$ and fixed $\lambda_0=\kappa=1$.

The asymptotic limits of $\beta$ when $\eta\to 1$ and $\eta\to \infty$ are given by
\begin{equation}
\lim\limits_{\eta \rightarrow 1^+} \beta^{(\text{rot})} = 1, \qquad 
\lim\limits_{\eta \rightarrow \infty} \beta^{(\text{rot})} 
= -l+\frac{2\Gamma\left(k+\frac{l}{2}+1\right)\Gamma\left(k+\frac{l}{2}+2\right)\Gamma\left(\frac{l}{2}+1\right)^2}{\Gamma\left(k+\frac{l}{2}+\frac{3}{2}\right)^2 \Gamma\left(\frac{l}{2}+\frac{1}{2}\right)^2},
\label{Eq:BetaRotEtaInf}
\end{equation}
and
\begin{equation}
\lim\limits_{\eta \rightarrow 1^+} \beta^{(\text{even})} = -\infty, 
\qquad 
\lim\limits_{\eta \rightarrow \infty} \beta^{(\text{even})} = -l.
\label{Eq:BetaEvenEtaInf}
\end{equation}
From these plots and expressions, we see that in the rotating case, the anisotropy parameter is always positive, showing that the azimuthal pressure is always less than the radial pressure, the difference being largest at the inner edge of the disk. This fact can be partially understood by noticing that in this case, all particles orbit in the same direction, leading to a small azimuthal velocity dispersion. However, the dependency of $\beta^{(\text{rot})}$ on $k$ and $l$ seems to be intricate.

In contrast to this, the anisotropy parameter for the even configurations is always negative, meaning that the azimuthal pressure dominates in this case (which is expected since each particle has a pair rotating in the opposite direction). As one approaches the boundary of the disk ($\eta\to 1$), this difference becomes larger and larger such that $\beta^{(\text{even})}\to -\infty$. This is expected, since the boundary of the disk is generated by the turning points of parabolic-type orbits (see the discussion following Eq.~(\ref{Eq:SupportNewton})), such that $P_{\hat{r}}^{(\text{even})}\to 0$ as $\eta\to 1$; however $P_{\hat{\varphi}}^{(\text{even})}$ remains positive since these orbits are populated by pairs of particles rotating in the opposite direction. Notice also that as $k$ and $l$ increase, the anisotropy of the even configurations becomes more pronounced.

\subsection{Total mass, energy and angular momentum}

Next, we discuss some qualitative properties of the total mass, energy and angular momentum given by the expressions found in Eqs.~(\ref{Eq:Mgas},\ref{Eq:Egas},\ref{Eq:Jgas}). Recall that the total mass and energy of the even model coincide with those of the rotating one, whereas $\ve{J}_{\text{gas}}^{(\text{even})} = 0$. Therefore, it is sufficient to focus our attention on the rotating case.

\begin{figure}[h!]
\centering
\includegraphics[scale=0.325]{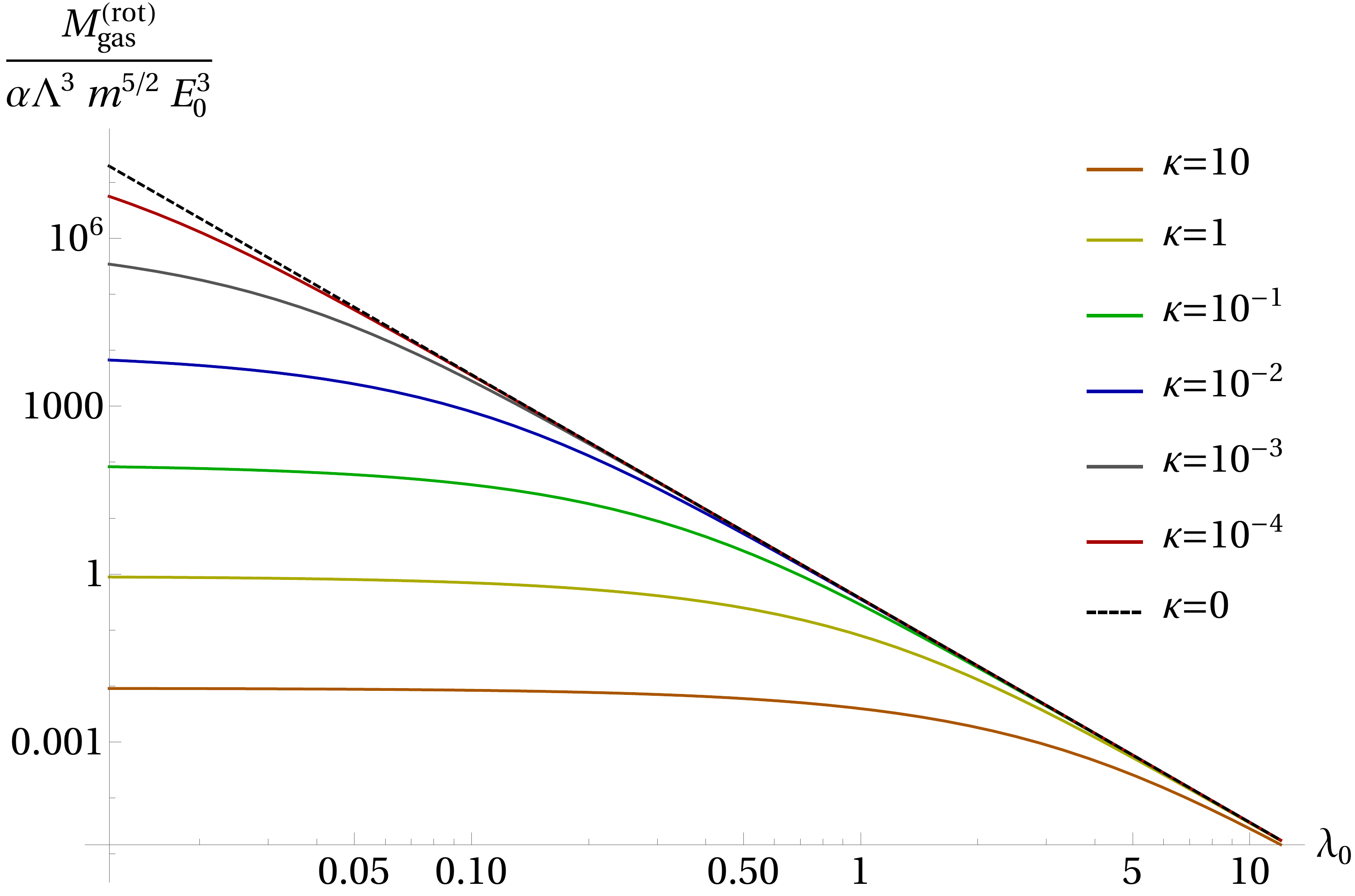}
\caption{Log-log plot for the total mass vs the cut-off value of the azimuthal angular momentum $\lambda_0$ for $k=5$, $l=0$ and different values of $\kappa$.}
\label{Fig:TotalMass}
\end{figure}

First, we analyze the behavior of the total mass as a function of the cut-off value $\lambda_0$ for the angular momentum and the parameter $\kappa$ in H\'enon's potential, fixing the values of $k$ and $l$. Figure~\ref{Fig:TotalMass} shows this behavior for the values $k=5$ and $l=0$.
As expected, when $\kappa$ approaches zero or $\lambda_0$ becomes very large, the total mass converges to the Kepler expression with the power-law dependency $\lambda_0^{-4}$, see Eq.~(\ref{Eq:Mgas}) with $\chi=0$ and $k=5$. When $\lambda_0 \to 0$, one recovers the limit computed in Eq.~(\ref{Eq:MgasZeroAngular}) corresponding to the spherical polytropes.

\begin{figure}[h!]
\centering
\includegraphics[scale=0.325]{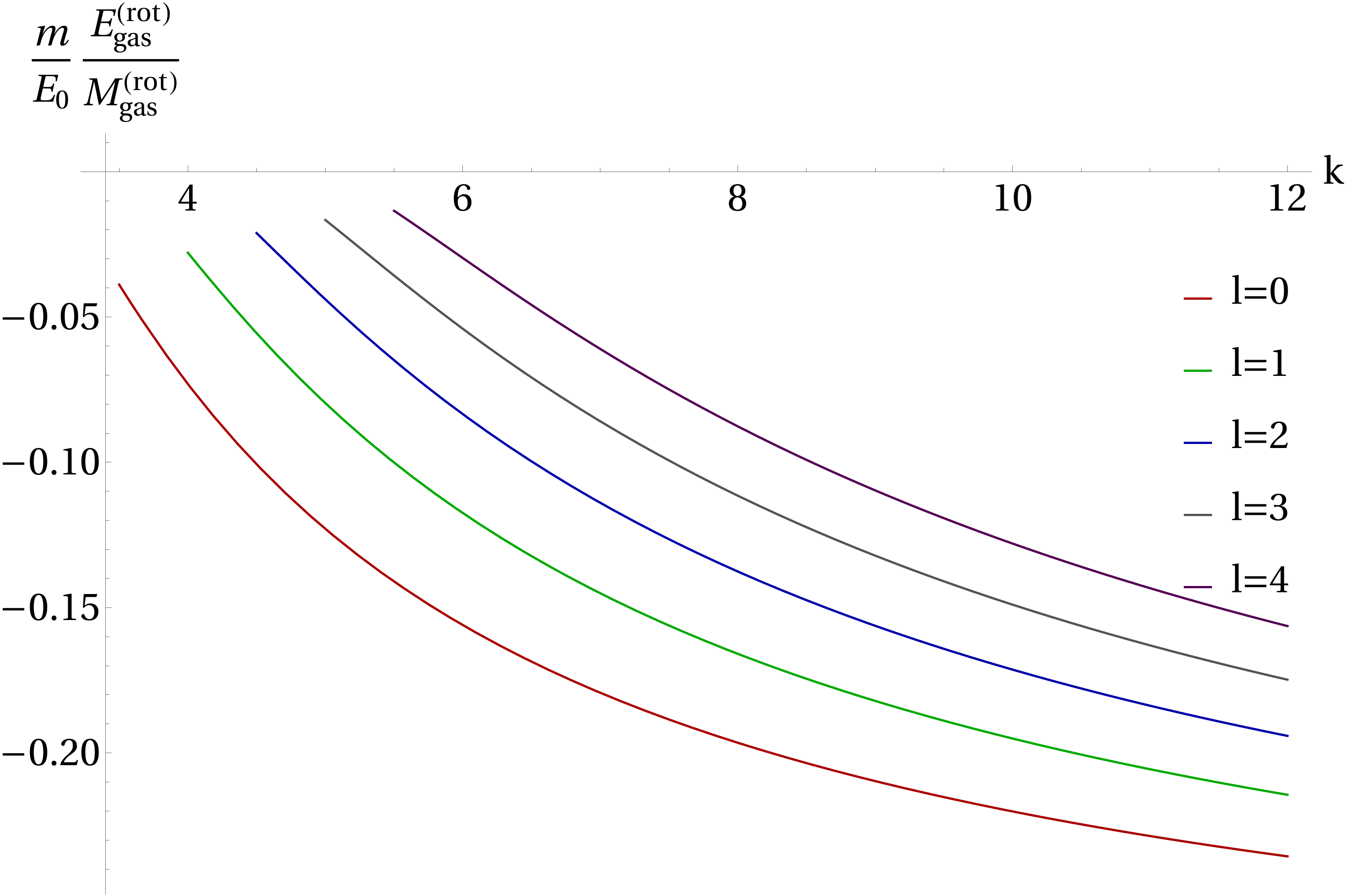}
\caption{Total energy per total mass vs $k$ for $l = 0,1,2,3,4$ and $\kappa=\lambda_0 = 1$.}
\label{Fig:TotalEnergy}
\end{figure}

Next, in figure~\ref{Fig:TotalEnergy} we show the ratio between the total energy and mass as a function of $k$ and $l$ for fixed values $\kappa = \lambda_0 = 1$.
As is visible from this plot, for fixed $l$ this ratio decreases for increasing $k$. This is expected since the lower energy levels become more populated as $k$ increases. In contrast, the ratio increases when $l$ becomes larger, which is also expected due to the fact that the orbits with larger angular momentum (which have larger energies) become more populated as $l$ increases. 

\begin{figure}[h!]
\centering
\includegraphics[scale=0.325]{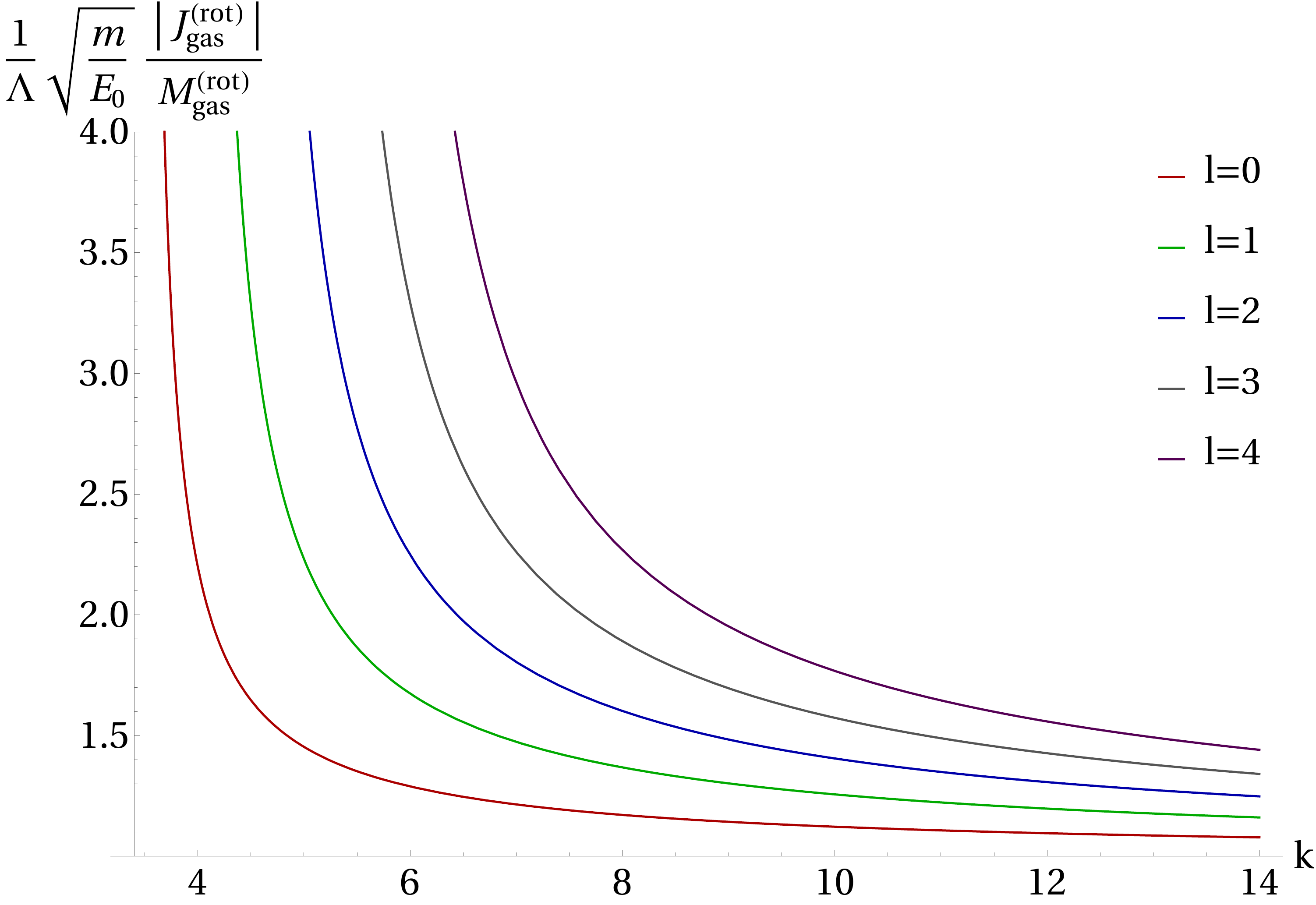}
\caption{Total angular momentum per total mass vs $k$ for different values of $l=0,1,2,3,4$ and fixed $\kappa=\lambda_0 = 1$.}
\label{Fig:TotalAngularMomentum}
\end{figure}

Finally, in figure~\ref{Fig:TotalAngularMomentum} we show the ratio between the total angular momentum and mass as a function of $k$ and $l$ for fixed values $\kappa = \lambda_0 = 1$.
As can be seen, this ratio decreases for increasing $k$ and fixed value of $l$, which is again expected since configurations with high $k$ have most of their particles lying on low-energy orbits for which the maximum angular momentum is low. In contrast, for fixed value of $k$ and increasing $l$, the ratio increases. Also, note that $|\ve{J}_{\text{gas}}^{(\text{rot})}|/M_{\text{gas}}^{(\text{rot})}$ diverges as $k\to (7+l)/2$ which means that the configurations can have arbitrary high angular momentum. This divergence is due to the fact that when $k = (7+l)/2$, the decay of the solution is not fast enough for the total angular momentum integral~(\ref{Eq:JgasOriginal}) to converge, while $M_{\text{gas}}^{(\text{rot})}$ is still finite.

\subsection{Comparison with fluid model}

We end this section with a comparison between the kinetic configurations based on the $(E,L_z)$-models and their hydrodynamic analogues whose properties are summarized in Appendix~\ref{App:ClassicalPolishDoughnuts}.

First, we compare the boundary surface of the configurations. As follows from Eq.~(\ref{Eq:EnthalpyFluid}), in the fluid case the gas is supported in the domain
\begin{equation}
\bar{h}_\infty + \psi(\xi) - \frac{\lambda_0^2}{2 \xi^2 \sin^2\vartheta} \geq 0,
\label{Eq:FluidBoundary}
\end{equation}
where $\bar{h}_\infty$ refers to the asymptotic value of the specific enthalpy (up to a constant factor), $\psi$ to the dimensionless isochrone potential defined in Eq.~(\ref{Eq:chietaDimensionless}) and $\xi$ and $\lambda_0$ to the dimensionless radius and constant azimuthal angular momentum per unit mass, see Eq.~(\ref{Eq:CDimensionless}). Interestingly, in the collisionless kinetic case, the condition $\eta\geq 1$ yields precisely the same domain if one sets $\bar{h}_\infty = 0$ and identifies $\xi$ and $\lambda_0$ with the corresponding quantities defined in Eq.~(\ref{Eq:DimensionlessQuantitiesKinetic}). This coincidence is due to the fact that as one approaches the boundary surface, the fluid elements' pressure gradient decreases to zero such that in this limit they move as point particles with zero energy and azimuthal angular momentum parameter equal to $\lambda_0$, as in the kinetic case. Therefore, taking $\bar{h}_\infty = 0$ for the following,\footnote{For the behavior of the boundary surface for solutions with $\bar{h}_\infty \neq 0$, see Appendix~\ref{App:ClassicalPolishDoughnuts}.} in both cases the minimum dimensionless radius at given $\vartheta$ is
\begin{equation}
\xi_{\text{min}}(\vartheta) = \frac{\lambda_0^2}{2\sin^2\vartheta}\sqrt{1 + \frac{4\kappa\sin^2\vartheta}{\lambda_0^2}}.
\label{Eq:XiMinimum}
\end{equation}

Whereas the boundary surface does not depend on whether one considers the fluid or the collisionless kinetic case, one clearly expects differences to occur in the interior region since the two matter models are rather different in nature. In the fluid description, the pressure is enforced to be isotropic and it is assumed that the specific azimuthal angular momentum is constant throughout the gas configuration. In contrast, in the kinetic model the gas is collisionless, anisotropic and the azimuthal angular momentum is not constant; rather it is distributed according to the functions given in Eqs.~(\ref{Eq:PolytropeLzEven}) or (\ref{Eq:PolytropeLzRot}). For the following, we focus on the latter distribution for which all particles rotate in the same direction with minimal azimuthal angular momentum $\lambda_0$ when performing the comparison with the fluid case, since in the former case the total angular momentum is zero.

To perform the comparison, for definiteness we restrict our attention to the profiles of the mass density and the temperature in the equatorial plane. In the kinetic case these profiles can be read off from the expressions~(\ref{Eq:nrot}) and~(\ref{Eq:KineticTemperatureRot}) with $\vartheta = \pi/2$. In the fluid case the mass density $\rho$ and temperature can be determined by using Eq.~(\ref{Eq:EnthalpyFluid}) again and assuming a polytropic equation of state $P = K\rho^\gamma$ with $K$ a constant and $\gamma > 1$ the adiabatic index. Integrating the first law of thermodynamics (with constant entropy) $dh = dP/\rho$ and taking into account the ideal gas equation $P = \rho k_B T/\bar{m}$ with $\bar{m}$ the averaged rest mass per particle, one obtains
\begin{equation}
h = \frac{\gamma}{\gamma-1}K\rho^{\gamma-1} = \frac{\gamma}{\gamma-1}\frac{P}{\rho}
 = \frac{\gamma}{\gamma-1}\frac{k_B T}{\bar{m}},
\label{Eq:hrhoT}
\end{equation}
which allows one to express $\rho$ and $T$ in terms of the specific enthalpy $h$. Note that taking $\bar{h}_\infty = 0$ as before, it follows from Eqs.~(\ref{Eq:hrhoT},\ref{Eq:EnthalpyFluid}) that the enthalpy and temperature decay as $\psi$ for large $\xi$. To proceed, we need a relation between the adiabatic index $\gamma$ and the parameters $k$ and $l$ characterizing the kinetic configuration. To this purpose, we note that in the fluid case,
\begin{equation}
\lim\limits_{\xi\to\infty}\frac{k_B T^{(\text{fluid})}(\xi,\vartheta)}{-\bar{m}\Phi(r)} 
 = 1 - \frac{1}{\gamma},
\end{equation}
whereas in the kinetic case one finds from Eq.~(\ref{Eq:TrotAsymptotic}) that
\begin{equation}
\lim\limits_{\xi\to\infty}\frac{k_B T^{(\text{kinetic})}(\xi,\vartheta)}{-m\Phi(r)} 
 = \frac{2}{3}\left\{ \frac{l+3}{2k+l+2} - \left[\frac{\Gamma\left(\frac{l}{2}+1\right)\Gamma\left(k+\frac{l}{2}+1 \right)}{\Gamma\left(\frac{l}{2}+\frac{1}{2} \right)\Gamma\left(k+\frac{l}{2}+\frac{3}{2} \right)}\right]^2\right\}.
\label{Eq:TempPsiLimit}
\end{equation}
This motivates the introduction of an effective adiabatic index $\gamma^{(\text{kinetic})}$ depending on $(k,l)$ (but not on $\kappa$ nor $\lambda_0$), which is defined in such a way that the right-hand side of Eq.~(\ref{Eq:TempPsiLimit}) is equal to $1 - 1/\gamma^{(\text{kinetic})}$. It follows from this definition that
\begin{equation}
1 < \gamma^{(\text{kinetic})} < \frac{6k + 3l + 6}{6k + l}.
\end{equation}
Some specific values for $\gamma^{(\text{kinetic})}$ are shown in table~\ref{Table:gammavskl}. Note that $\gamma^{(\text{kinetic})}\to 1$ as $k$ tends to infinity and $l$ remains finite, corresponding to the isothermal limit. Furthermore, the condition $k > (7+l)/2$ implies that $\gamma^{(\text{kinetic})} < 3/2$, although the maximum effective adiabatic index found from the table is for $l=0$ and $k=4$, for which
\begin{equation}
\gamma^{(\text{kinetic})} = \frac{297675\pi^2}{131072+238140\pi^2} \approx 1.1840.
\end{equation}
\begin{table}
\centering
\caption{Effective adiabatic index $\gamma^{(\text{kinetic})}$ for different values of $k$ and $l$ (five significant figures are shown).}
\label{Table:gammavskl}
\begin{tabular}{|c|c|c|c|c|c|}
\hline
 & $k=4$ & $k=5$ & $k=6$ & $k=7$ & $k=10$ \\ 
\hline
$l=0$ & $1.1840$ & $1.1492$ & $1.1254$ & $1.1082$ & $1.0766$ \\
\hline
$l=1$ & & $1.1382$ & $1.1181$ & $1.1031$ & $1.0746$ \\
\hline
$l=2$ & & $1.1267$ & $1.1097$ & $1.0968$ & $1.0714$ \\
\hline
$l=3$ & & & $1.1020$ & $1.0908$ & $1.0682$ \\
\hline
$l=4$ & & & $1.0951$ & $1.0853$ & $1.0651$ \\
\hline
$l=5$ & & & & $1.0804$ & $1.0623$ \\
\hline
$l=6$ & & & & $1.0760$ & $1.0596$ \\
\hline
$l=7$ & & & & & $1.0572$ \\
\hline
$l=8$ & & & & & $1.0549$ \\
\hline
\end{tabular}
\end{table} 
Therefore, for what follows we compare our kinetic gas configurations with a fluid model with adiabatic index $\gamma = \gamma^{(\text{kinetic})}\leq 1.1840$ (note that our prescription excludes the typical values of $4/3$ and $5/3$ describing a monoatomic gas in the ultra and non-relativistic limits, respectively).

\begin{figure}[h!]
\centering
\subfigure{\includegraphics[scale=0.28]{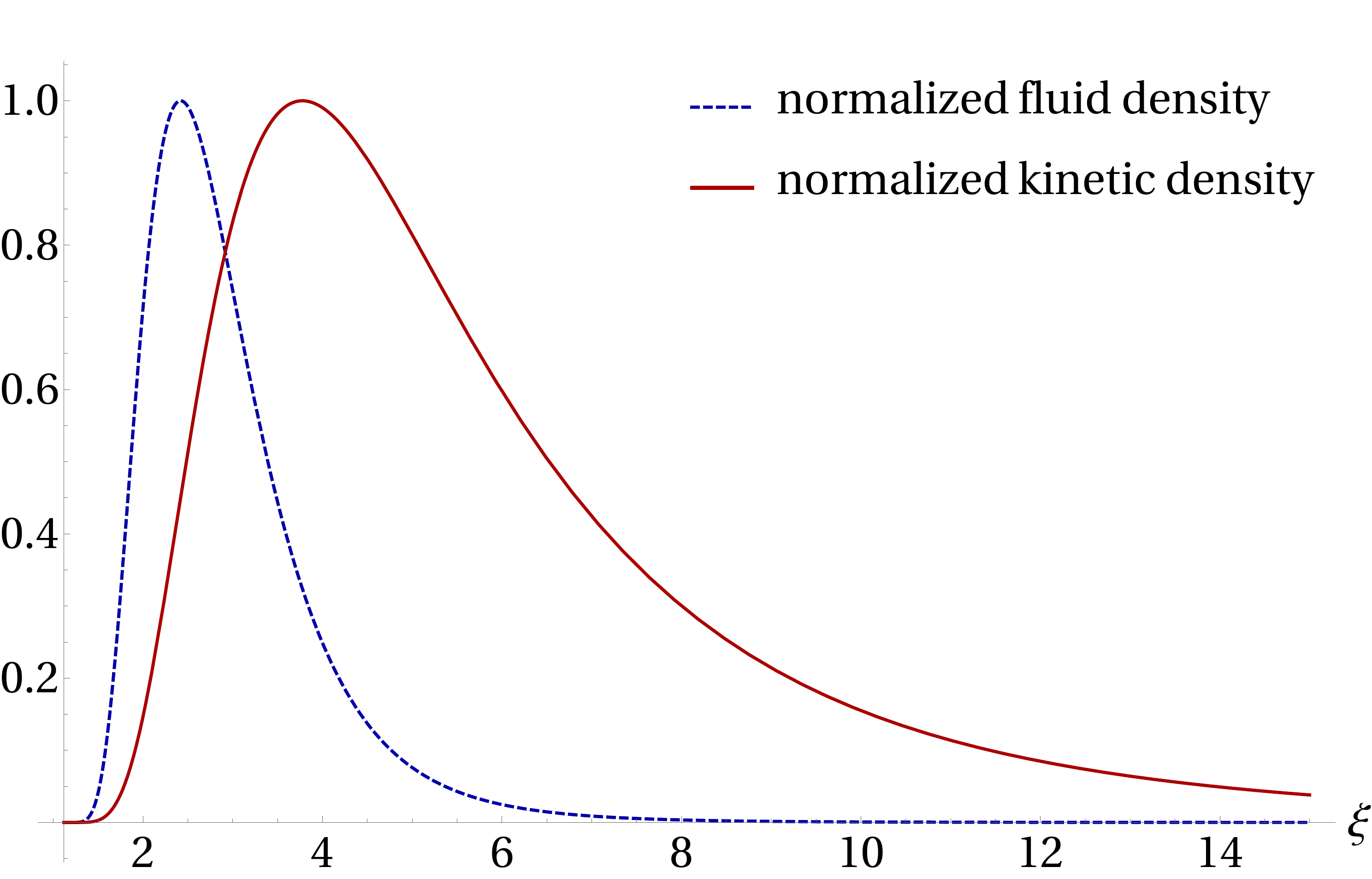}}
\subfigure{\includegraphics[scale=0.28]{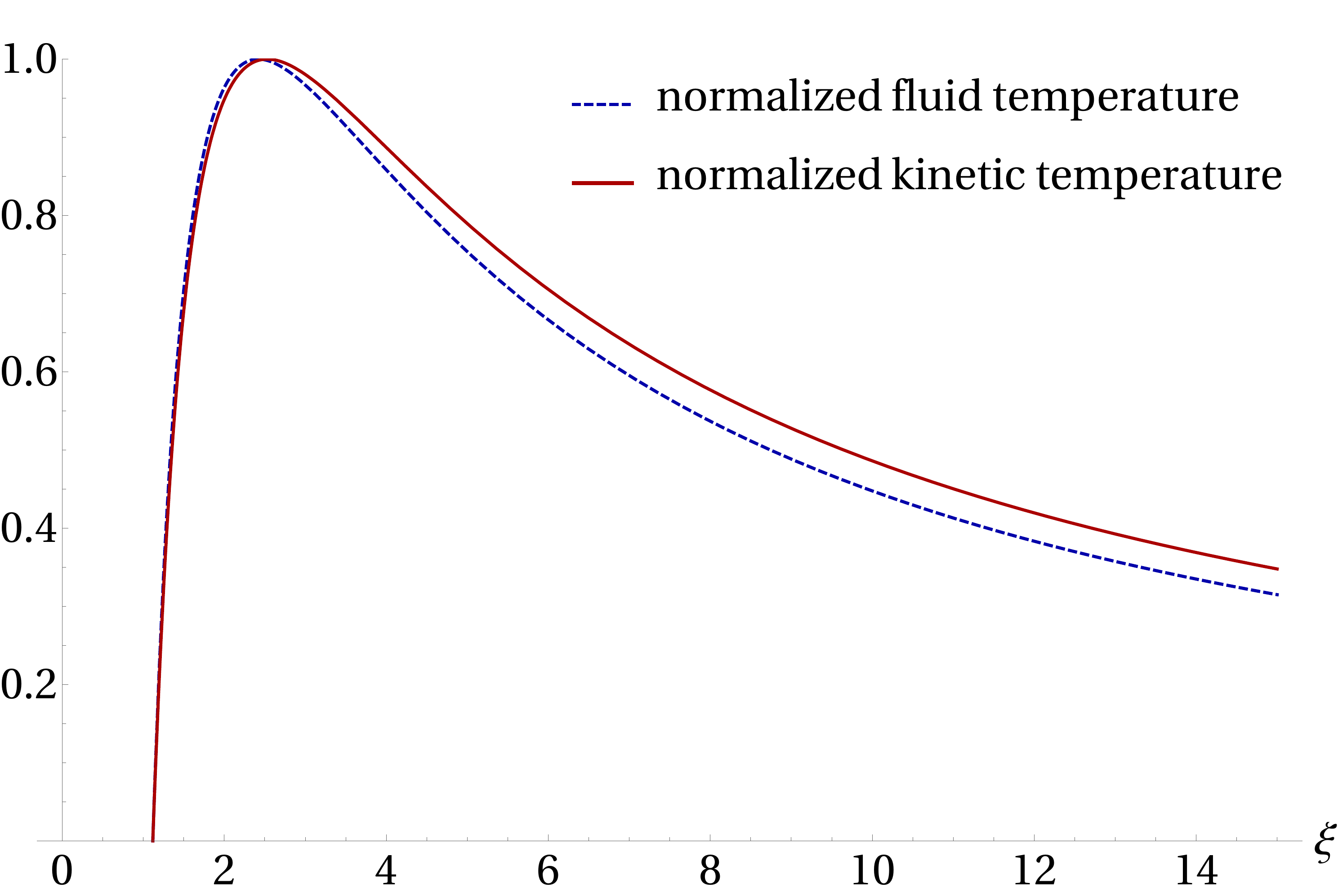}}
\caption{Comparison between the normalized densities (left panel) and between the normalized temperatures (right panel) in the equatorial plane for $\kappa = \lambda_0 = 1$ and $(k,l) =(4,0)$ and the corresponding value $\gamma =1.18397$ for the adiabatic index. In both cases the normalization is chosen such that the maximum is one.}
\label{Fig:NormalizedTfluvsTkin}
\end{figure}

In figure~\ref{Fig:NormalizedTfluvsTkin} we show the comparison between the normalized mass densities and temperature profiles for $(k,l) = (4,0)$ and fixed $\lambda_0 = \kappa = 1$. Here, the normalization is performed in such a way that the maxima of the profiles is one.
As is visible from these plots, the density distribution is different, the fluid configuration being more compact than the kinetic one (a similar behavior is found for other values of $k$ and $l$). In contrast, the temperature profiles look very similar to each other. In order to partially quantify this assertion, we show in tables~\ref{Table:RatioTemperatures} and~\ref{Table:RatioTemperaturesLamdba} the ratios between the radii and the values of the maxima of the temperature profiles for different values of $(k,l)$ and $\lambda_0$. We see from these results that the locations of these maxima are very similar in both matter models (within $5\%$ relative error), at least for $l=0$, which is consistent with the right plot in figure~\ref{Fig:NormalizedTfluvsTkin}. For higher values of $l$, the radius of the temperature maximum in the kinetic model tends to increase with respect to the fluid case, while the example shown in table~\ref{Table:RatioTemperaturesLamdba} indicates that both ratios become independent of $\lambda_0$ for large values of $\lambda_0$. Finally, we see from these tables that the temperature in the fluid case is somehow larger than in the kinetic description.
\begin{table}
\centering
\caption{Top row: Ratio between the radii corresponding to the maxima of the fluid and kinetic temperature profiles. Bottom row: Ratio between the corresponding maxima of the temperature. Here we choose $\kappa = \lambda_0 = 1$ and different values for $k$ and $l$ and set $\gamma = \gamma^{(\text{kinetic})}$ in the fluid case.}
\label{Table:RatioTemperatures}
\begin{tabular}{|c|c|c|c|c|c|c|c|c|c|c|c|c|}
\hline
 & \multicolumn{4}{c|}{$l=0$} & \multicolumn{4}{c|}{$l=1$} & \multicolumn{4}{c|}{$l=2$} \\ \hline
 & $k=4$ & $k=5$ & $k=6$ & $k=10$ & $k=5$ & $k=6$ & $k=7$ & $k=10$ & $k=5$ & $k=6$ & $k=7$ & $k=10$ \\ 
\hline
 & & & & & & & & & & & & \\
$\displaystyle \frac{mh_0}{E_0}\frac{r_\text{max}^{(\text{fluid})}}{r_\text{max}^{(\text{kinetic})}}$ & 0.956 & 0.960 & 0.963 & 0.971 & 0.941 & 0.945 & 0.948 & 0.955 & 0.929 & 0.933 & 0.937 & 0.945 \\ 
 & & & & & & & & & & & & \\
\hline
 & & & & & & & & & & & & \\
$\displaystyle\frac{E_0}{m h_0}\frac{T_\text{max}^{(\text{fluid})}}{T_\text{max}^{(\text{kinetic})}}$ & 1.206 & 1.199 & 1.195 & 1.186 & 1.253 & 1.245 & 1.239 & 1.228 & 1.286 &  1.275 & 1.267 & 1.251 \\ 
 & & & & & & & & & & & & \\
\hline
\end{tabular}
\end{table}
\begin{table}
\centering
\caption{Same as in previous table fixing the parameter values $(\kappa, k, l) = (1,5,1)$ and varying $\lambda_0$. As in the previous table, we set $\gamma = \gamma^{(\text{kinetic})}$ in the fluid case.}
\label{Table:RatioTemperaturesLamdba}
\begin{tabular}{|c|c|c|c|c|}
\hline
 & $\lambda_0 = 1$ & $\lambda_0 = 4$ & $\lambda_0 = 7$ & $\lambda_0 = 10$ \\ 
\hline
 & & & & \\
$\displaystyle \frac{mh_0}{E_0}\frac{r_\text{max}^{(\text{fluid})}}{r_\text{max}^{(\text{kinetic})}}$ & 0.941 & 0.960 & 0.962 & 0.962 \\ 
 & & & & \\
\hline
 & & & & \\
$\displaystyle\frac{E_0}{m h_0}\frac{T_\text{max}^{(\text{fluid})}}{T_\text{max}^{(\text{kinetic})}}$ & 1.253 & 1.29 & 1.30 & 1.30 \\ 
 & & & & \\
\hline
\end{tabular}
\end{table}

\section{Conclusions}
\label{Sec:Conclusions}

We have provided a complete analytic description for stationary axisymmetric and collisionless kinetic gas clouds consisting of identical massive and neutral particles surrounding a massive central object. Our models are based on the assumptions that the gravitational field is dominated by the central object, giving rise to a central Newtonian potential, and that the gas particles follow bound orbits in this potential. Although our nonrelativistic description restricts the validity of our configurations to radii much larger than the central object's gravitational radius, it serves as a basis for the corresponding general relativistic model in our accompanying article~\cite{cGoS2022c} and will allow us to better understand the effects from the relativistic corrections on the properties of the gas clouds. The one-particle DF in our models is a function depending only on the energy $E$ and the azimuthal angular momentum $L_z$ of the particles through the polytropic ans\"atze~(\ref{Eq:FFactorizationAnsatz},\ref{Eq:Polytrope},\ref{Eq:PolytropeLzEven},\ref{Eq:PolytropeLzRot}) describing both rotating and nonrotating stationary and axisymmetric configurations. In the former case, all particles have positive values of $L_z$ larger than a cut-off value $L_0$, giving rise to a net angular momentum while in the latter case the DF is an even function of $L_z$. By choosing the polytropic index $k$ sufficiently large, the resulting configurations have finite total mass, energy and angular momentum, although they have infinite extend. (Finite volume configurations could easily be considered as well by introducing a cut-off in the energy.)

We have computed the most relevant macroscopic observables associated with the DF, including the particle and energy densities, mean particle velocity, pressure tensor, and the kinetic temperature. Interestingly, these quantities can be expressed explicitly in terms of the Gamma function and Gauss' hypergeometric function for any central potential. Using action-angle variables we have been able to reduce the expressions for the total mass, energy and angular momentum to a triple integral which involves the period function $T_r(E,L)$ describing the period of the radial motion for an orbit with energy $E$ and total angular momentum $L$. For the particular class of H\'enon's isochrone potential (which is characterized by a parameter $b$ whose limits $b\to 0$ and $b\to \infty$ correspond to Kepler's and the spherical harmonic potential, respectively) this period function is independent of $L$ and given by exactly the same expression as in the Kepler case, and for this class the total quantities can be computed analytically.

Based on our analytic expressions, we have provided a detailed analysis for the behavior of the macroscopic quantities, including their dependence on the parameters $k,l,L_0,b$ of the model and their asymptotic decay for large radii $r$. In particular, we have analyzed the morphology of our configurations and shown that both slim and thick toroidal-type configurations can be constructed. Further, we have analyzed the temperature profile in the equatorial plane and shown that it decays as $1/r$ for large values of $r$. For the rotating model, the temperature at the inner edge is zero, has a maximum at some radius and then decays while for the even models the temperature at the inner edge is positive and decays monotonically. Next, we have analyzed the pressure anisotropy. In all our models, the principal pressures in the radial and polar directions are always equal to each other (a property that does not necessarily hold in the relativistic case~\cite{cGoS2022c}), while the principal pressure in the azimuthal direction can be quite different from the radial and polar ones, as we have shown.

Finally, we have compared our kinetic configurations to analogous fluid configurations with fixed specific angular momentum $\ell_0$ which we briefly reviewed in Appendix~\ref{App:ClassicalPolishDoughnuts}. This comparison has led to the following discoveries: (i) By suitably identifying the parameters $L_0/m$ and $\ell_0$ we obtain configurations with exactly the same boundary surface in both models. (ii) With this identification, the fluid configurations are slightly more compact. More precisely, the radius at which both the density and temperature have their maximum value in the fluid configurations is smaller than the corresponding radii of the kinetic configurations,\footnote{Unlike the fluid configurations, the radii of maximum density and temperature do not coincide in the kinetic configurations. This is probably due to the pressure anisotropy.} see table~\ref{Table:RatioTemperatures} for a few representative examples. (iii) By matching the asymptotic behavior of the temperature profile, one can introduce an effective adiabatic index $\gamma^{(\textrm{kinetic})}$ in the kinetic case, which, in our models has a maximum value of about $1.1840$. This allows one to compare the density and temperature profiles of the fluid configurations (which depend on the adiabatic index $\gamma$) to those of the kinetic configurations by setting $\gamma = \gamma^{(\textrm{kinetic})}$. This comparison reveals that while the fluid configurations are more compact, the normalized temperature profile is very similar in both cases, see figure~\ref{Fig:NormalizedTfluvsTkin}. Nevertheless, the fluid configurations are slightly hotter than the kinetic ones, as can be seen from tables~\ref{Table:RatioTemperatures} and~\ref{Table:RatioTemperaturesLamdba}. The fact that the behavior of the temperature is at least qualitatively similar in both models is surprising, since in the fluid case one assumes local thermodynamic equilibrium, while in our kinetic model the gas is collisionless and the temperature cannot be associated with a thermal state, as has been exemplified towards the end of section~\ref{SubSec:Polytropes}. 

The fully relativistic generalization of our models, for which the central object is a black hole, is under investigation. In a first step towards this goal we have restricted this generalization to a non-rotating (Schwarzschild) black hole, see~\cite{cGoS2022c} and~\cite{cGoS2022a} for a related model based on a slightly different ansatz for the DF. Another interesting generalization of the models discussed in the present article is to study the effects of binary collisions between the gas particles or to include a magnetic field and study its effect on a gas of charged particles.

\acknowledgments

We thank Francisco Astorga, Ana Laura Garc\'ia, Ulises Nucamendi, Emilio Tejeda, and Thomas Zannias for useful comments and fruitful discussions. C.G. was supported by a PhD CONACyT fellowship. O.S. was partially supported by a CIC Grant to Universidad Michoacana. We also acknowledge support from the CONACyT Network Project No. 376127 ``Sombras, lentes y ondas gravitatorias generadas por objetos compactos astrof\'isicos".

\appendix

\section{Integrals over angular momentum and energy}
\label{App:Integrals}

In this appendix we list a few integrals which are useful when computing the observables in our models. They arise in subsection~\ref{SubSec:TheELzModels} when performing the explicit integrals over the total and azimuthal angular momenta and over the energy. For notational simplicity we use the short-hand notations $\displaystyle L_1:= |L_z| \sin^{-1}\vartheta$ and $L_2 := L_{\text{max}}(E,r)$ and $L_3 := L_2\sin\vartheta$. Note that in our integrals, $L_1 \leq L_2$. 

First, by means of the variable substitution $u = \sqrt{L^2 - L_1^2}$ one easily finds
\begin{eqnarray}
&& \int\limits_{L_1}^{L_2} \frac{LdL}{\sqrt{L^2 - L_1^2} \sqrt{L_2^2 - L^2}} = \frac{\pi}{2},
\label{Eq:A1}\\
&& \int\limits_{L_1}^{L_2} \frac{\sqrt{L_2^2 - L^2}}{\sqrt{L^2-L_1^2}} L dL 
 = \int\limits_{L_1}^{L_2} \frac{\sqrt{L^2 - L_1^2}}{\sqrt{L_2^2-L^2}} L dL
 = \frac{\pi}{4}\left( L_2^2 -L_1^2 \right).
\label{Eq:A2}
\end{eqnarray}
Next, the integrals over the azimuthal angular momentum encountered in subsection~\ref{SubSec:TheELzModels} have the form
\begin{eqnarray}
\int\limits_0^{L_3}\left(\frac{L_z}{L_0}-1\right)_+^l dL_z
 &=& \frac{L_0}{l+1} \left(\frac{L_3}{L_0}-1\right)_+^{l+1}, 
 \\
\int\limits_0^{L_3} \left(\frac{L_z}{L_0}-1\right)_+^l L_z dL_z 
 &=& \frac{L_0^2}{l+1} \left(\frac{L_3}{L_0}-1\right)_+^{l+1} 
  + \frac{L_0^2}{l+2} \left(\frac{L_3}{L_0}-1\right)_+^{l+2},
 \\
\int\limits_0^{L_3}  \left(\frac{L_z}{L_0}-1\right)_+^l L_z^2 dL_z 
 &=& \frac{L_0^3}{l+1} \left(\frac{L_3}{L_0}-1\right)_+^{l+1} 
 + \frac{2L_0^3}{l+2} \left(\frac{L_3}{L_0}-1\right)_+^{l+2} 
 + \frac{L_0^3}{l+3} \left(\frac{L_3}{L_0}-1\right)_+^{l+3}.
\end{eqnarray}

Finally, the energy integrals found in subsection~\ref{SubSec:TheELzModels} have the form
\begin{eqnarray}
\int\limits_{m\Phi(r)}^0 \left( -\frac{E}{E_0} \right)^s
\left(\frac{R}{L_0}\sqrt{2m[E-m\Phi(r)]}-1\right)^{l+1}_+ dE
 &=& \frac{L_0^2}{m R^2}\left[\frac{L_0^2}{2m R^2 E_0}\right]^s
\int\limits_1^\eta \left(\eta^2 - \xi^2 \right)^\alpha \left(\xi-1\right)^{l+1}_+ \xi d\xi
\nonumber\\
 &=& 2E_0\eta^{l+1}\psi(r)^{s+1} J_{s,l}(\eta),
\end{eqnarray}
where we have used the variable substitution $\xi := R\sqrt{2m[E-m\Phi(r)]}/L_0$ and recalled the definitions of the functions $\eta = R\sqrt{-2m^2\Phi(r)}/L_0$, $\psi(r) = -m\Phi(r)/E_0$ and $J_{s,l}(\eta)$ defined in Eqs.~(\ref{Eq:eta},\ref{Eq:psi},\ref{Eq:Jsl}), respectively.

\section{Properties of the effective potential belonging to the isochrone model}
\label{App:EffectivePotential}

The effective potential~(\ref{Eq:EffectivePotentialNewton}) associated with H\'enon's isochrone model in Eq.~(\ref{Eq:IsochronePotential}) is
\begin{equation}
V_L(r) = -\frac{GM m}{b+\sqrt{b^2 + r^2}} + \frac{L^2}{2mr^2},\qquad r > 0,
\label{Eq:B1}
\end{equation}
with the parameter $b\geq 0$ having units of length. The equilibrium points are obtained by analyzing the critical points of $V_L$ which are determined by the equation
\begin{equation}
0 = -\frac{d}{dr}V_L(r)
 = -\frac{GMm r}{\sqrt{b^2 +r^2}\left( b+\sqrt{b^2 +r^2}\right)^2} + \frac{L^2}{mr^3}.
\end{equation}
There is a unique critical point describing a global minimum of $V_L$ which is located at $r = r_0(L)$ with
\begin{eqnarray}
r_0^2(L) &=&  \left( \frac{L^2}{4GMm^2} \right)^2
\left(1+\sqrt{1+\frac{4bGMm^2}{L^2}} \right)^4 - b^2
\nonumber\\
 &=& \left( \frac{L^2}{2GMm^2} \right)^2\sqrt{1+\frac{4bGMm^2}{L^2}}
 \left(1+\sqrt{1+\frac{4bGMm^2}{L^2}} \right)^2.
\label{Eq:B3}
\end{eqnarray}
The corresponding minimal energy is
\begin{equation}
E_0(L) = V_L(r_0) 
 = -\frac{2(GM)^2m^3}{L^2\left(1+\sqrt{1+\frac{4bGMm^2}{L^2}}\right)^2}.
\label{Eq:B4}
\end{equation}
The condition for the upper bound of the total angular momentum $L_{\text{ub}}(E)$ is obtained from this by setting $E = E_0(L)$, which yields
\begin{equation}
L_{\text{ub}}(E) = \sqrt{\frac{m}{2}}\left(\frac{GMm}{\sqrt{-E}}-2b\sqrt{-E} \right).
\label{Eq:B5}
\end{equation}

\begin{figure}[h!]
\centerline{
\includegraphics[scale=0.3]{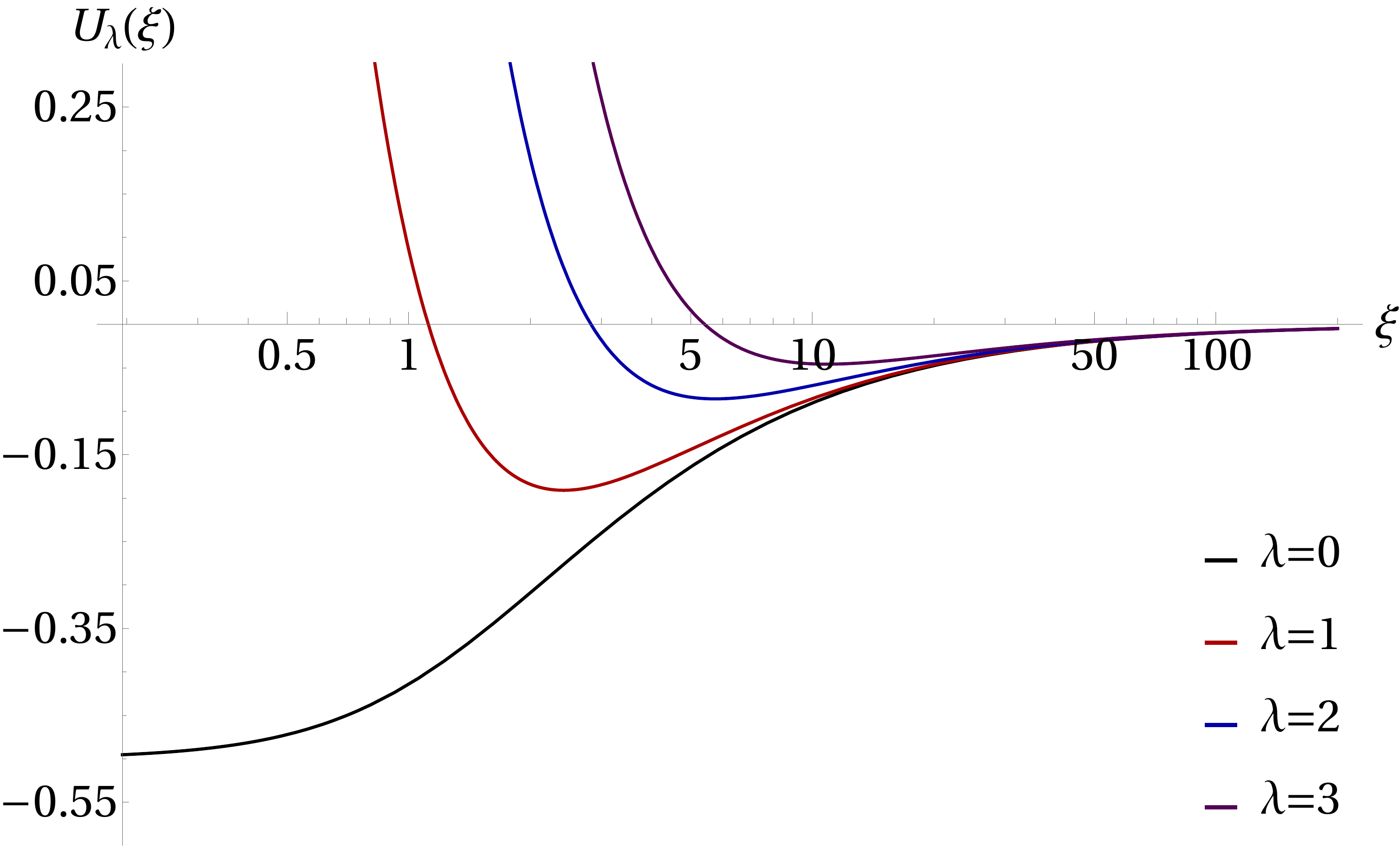}}
\caption{Behavior of the dimensionless effective potential $U_\lambda(\xi)$ for the isochrone model for different values of the parameter $\lambda$ and setting $\kappa=1$. The black line  corresponding to $\lambda=0$ shows the profile of the isochrone potential itself.}
\label{Fig:EffectivePotentialIsochrone}
\end{figure}

Using the dimensionless quantities defined in Eqs.~(\ref{Eq:LengthScaleKinetic},~\ref{Eq:DimensionlessQuantitiesKinetic}), the dimensionless effective potential $U_\lambda(\xi) = V_L(r)/E_0$ reads 
\begin{equation}
U_\lambda(\xi) = -\frac{1}{\kappa + \sqrt{\kappa^2 + \xi^2}} + \frac{\lambda^2}{2 \xi^2},\qquad
\xi > 0,
\end{equation} 
and for illustrative purposes its behavior is shown in figure~\ref{Fig:EffectivePotentialIsochrone}.

\section{Total mass, energy and angular momentum}
\label{App:TotalMass}

In this appendix we provide details for the computation of the integrals determining the total mass, energy and angular momentum of the gas. These integrals have the following form (cf. Eqs.~(\ref{Eq:TotalMass1},\ref{Eq:TotalAngularMomentum1})):
\begin{equation}
I_{M,E,J_z} := (2\pi)^2\int\limits_{E_{\text{min}}(0)}^0 dE \int\limits_0^{L_{\text{ub}}(E)} dL \int\limits_{-L}^{+L} dL_z T_r(E,L) F(E,L,L_z) W_{M,E,J_z},
\label{Eq:A11} 
\end{equation}
where the indices $M,E,J_z$ refer, respectively, to the total mass $M_{\text{gas}}$, energy $E_{\text{gas}}$ and the z-component of the angular momentum $\ve{J}_{\text{gas}}\cdot\ve{e}_z$ and correspondingly, $W_M = m$, $W_E = E$ and $W_{J_z} = L_z$. For the example discussed in section~\ref{SubSec:TotalMassAngularMomentum} the period $T_r$ and the DF are independent of $L$. Using the explicit expressions for $T_r$, $E_{\text{min}}$ and $L_{\text{ub}}$ in Eqs.~(\ref{Eq:RadialPeriod},\ref{Eq:LimitsOfIntegration}) , the length scale defined in~(\ref{Eq:LengthScaleKinetic}), the dimensionless quantities~(\ref{Eq:DimensionlessQuantitiesKinetic}) and the corresponding definitions for the particular $(E,L_z)$-models~(\ref{Eq:Polytrope},~\ref{Eq:FFactorizationAnsatz},~\ref{Eq:PolytropeLzRot}), one obtains from Eq.~(\ref{Eq:A11}):
\begin{equation}
\frac{M_{\text{gas}}^{(\text{rot})}}{\alpha \Lambda^3 \sqrt{m^5 E_0}} = 4\sqrt{2}\pi^3 \int\limits_{-E_0/2\kappa}^0 dE \left(-\frac{E}{E_0}\right)^{k-3}_+ \int\limits_0^{\lambda_{\text{ub}}(E)} d\lambda \int\limits_{-\lambda}^{+\lambda} d\lambda_z \left(\frac{\lambda_z}{\lambda_0}-1\right)^{l}_+,\qquad
\lambda_{\textrm{ub}}(E) = \sqrt{-\frac{E_0}{2E}} - \kappa\sqrt{-\frac{2E}{E_0}}.
\end{equation}
The integrals over $\lambda_z$ and $\lambda$ can be carried out immediately. In order to perform the integral over $E$ we use the variable substitution
\begin{equation}
E := -\frac{E_0}{2\kappa}\sigma^2,
\end{equation}
where $\sigma$ is dimensionless, and obtain
\begin{equation}
\frac{M_{\text{gas}}^{(\text{rot})}}{\alpha \Lambda^3 \sqrt{m^5 E_0^3}} = \frac{\pi^3}{2^{k-\frac{11}{2}}}  \frac{1}{(l+1)(l+2)\lambda_0^{l}\kappa^{k-3-\frac{l}{2}}} \int\limits^1_0 d\sigma \sigma^{2k-l-7} \left(1-\frac{\lambda_0}{\sqrt{\kappa}}\sigma-\sigma^2\right)^{l+2}_+.
\end{equation}
The second-order polynomial in $\sigma$ can be expressed as $(\sigma_+ - \sigma)_+(\sigma - \sigma_-)_+$ where the roots are given by
\begin{equation}
\sigma_{\pm} = -\frac{\lambda_0}{2\sqrt{\kappa}} \pm \sqrt{1+\frac{\lambda_0^2}{4\kappa}}, 
\end{equation}
and satisfy $\sigma_- < -1$ and $0 < \sigma_+ < 1$. Since the integrand only contributes when $\sigma_- < \sigma < \sigma_+$ and the integration limits vary from $0$ to $1$ one can replace the upper limit with $\sigma_+$ and remove the index $+$ in the integrand. Using~\cite[Eq. 3.197.8]{iGiR2007} yields
\begin{equation}
\frac{M_{\text{gas}}^{(\text{rot})}}{\alpha \Lambda^3 \sqrt{m^5 E_0^3}} = \frac{\pi^3}{2^{k-\frac{11}{2}}}  \frac{(-\sigma_-)^{l+2} (\sigma_+)^{2k-4}}{(l+1)(l+2)\lambda_0^{l}\kappa^{k-3-\frac{l}{2}}} B(l+3,2k-l-6)\;{}_{2}F_{1}\left(-l-2, 2k-l-6, 2k-3, \frac{\sigma_+}{\sigma_-} \right).
\end{equation}
To make further progress we introduce the variable $\chi$, defined by $\displaystyle \frac{2\sqrt{\kappa}}{\lambda_0}:=\sinh(2\chi)$, in terms of which the roots can be expressed as $\sigma_+ = \tanh\chi$ and $\sigma_- = -\coth\chi$. After some algebra one finally obtains  
\begin{equation}
\frac{M_{\text{gas}}^{(\text{rot})}}{\alpha \Lambda^3 \sqrt{m^5 E_0^3}} = \frac{\pi^3}{2^{k-\frac{11}{2}}}  \frac{\Gamma(l+1)\Gamma(2k-l-6)}{\Gamma(2k-3)} \frac{{}_{2}F_{1}\left(-(l+2), 2k-l-6, 2k-3, -\tanh^2\chi \right)}{\lambda_0^{2k-6}(\cosh\chi)^{4k-2l-12}}.
\end{equation}
In a similar way, choosing $W_E = E$ in Eq.~(\ref{Eq:A11}) yields
\begin{eqnarray}
\frac{E_{\text{gas}}^{(\text{rot})}}{\alpha \Lambda^3 \sqrt{m^3 E_0^5}} &=& -\frac{\pi^3}{2^{k-\frac{9}{2}}}  \frac{1}{(l+1)(l+2)\lambda_0^{l}\kappa^{k-2-\frac{l}{2}}} \int\limits^{\sigma_+}_0 d\sigma \sigma^{2k-l-5} \left(\sigma_+ -\sigma\right)^{l+2}_+ \left(\sigma - \sigma_- \right)^{l+2}_+ \nonumber\\
&=& -\frac{\pi^3}{2^{k-\frac{9}{2}}} \frac{\Gamma(l+1)\Gamma(2k-l-4)}{\Gamma(2k-1)} \frac{{}_{2}F_{1}\left(-(l+2), 2k-l-4, 2k-1, -\tanh^2\chi \right)}{\lambda_0^{2k-4}(\cosh\chi)^{4k-2l-8}}.
\end{eqnarray}
For the z-component of the total angular momentum one obtains
\begin{eqnarray}
\frac{|\ve{J}_{\text{gas}}^{(\text{rot})}|}{\alpha \Lambda^4 m^2 E_0^2}
&=& \frac{\pi^3}{2^{k-\frac{11}{2}}}\frac{\Gamma(l+1)\Gamma(2k-l-6)}{\Gamma(2k-3)}\frac{{}_{2}F_{1}\left(-(l+2), 2k-l-6, 2k-3, -\tanh^2\chi \right)}{\lambda_0^{2k-7}(\cosh\chi)^{4k-2l-12}} \nonumber\\
& & \times \left[ 1 + \frac{l+1}{2k-l-7}\cosh^2\chi \frac{{}_{2}F_{1}\left(-(l+3), 2k-l-7, 2k-3, -\tanh^2\chi \right)}{{}_{2}F_{1}\left(-(l+2), 2k-l-6, 2k-3, -\tanh^2\chi \right)}\right].
\end{eqnarray}

\section{Short review of circular hydrodynamic flows}
\label{App:ClassicalPolishDoughnuts}

For comparison with the kinetic, collisionless configurations analyzed in this article, in this appendix we discuss their hydrodynamical equivalent, namely the circular flows describing steady-state, axisymmetric rotating perfect fluid configurations around a compact object (see~\cite{oZdP2015} for a recent reference which generalize these configurations to include a magnetic field). These flows can be obtained by integrating the continuity and Euler equations for an external axisymmetric potential $\Phi(\ve{x})$,
\begin{equation}
\frac{\partial\rho}{\partial t} + \nabla\cdot\left(\rho\ve{v}\right) = 0,
\qquad
\frac{\partial\ve{v}}{\partial t} + \left(\ve{v}\cdot\nabla\right)\ve{v} + \frac{1}{\rho}\nabla P + \nabla \Phi = 0,
\label{Eq:C1}
\end{equation}
where the state of the fluid is specified by its mass density $\rho(t,\ve{x})$, pressure $P(t,\ve{x})$, velocity field $\ve{v}(t,\ve{x})$, temperature $T(t,\ve{x})$, and the enthalpy per unit of mass (or specific enthalpy) $h(t,\ve{x})$. For a steady-state axisymmetric configuration, the scalar quantities only depend on the radius $r$ and the polar angle $\vartheta$, and the velocity field has the form $\ve{v} = r\sin\vartheta \Omega(r,\vartheta)\ve{e}_\varphi$, with $\Omega(r,\vartheta)$ the angular velocity. It follows that the continuity equation is automatically satisfied. Furthermore, using the fact that $\displaystyle\ve{v}\cdot\nabla = \Omega\frac{\partial}{\partial \varphi}$, the nontrivial components of the Euler equations yield
\begin{eqnarray}
\label{Eq:C4}
-r\sin^2\vartheta\Omega^2
 + \frac{1}{\rho}\frac{\partial P}{\partial r} + \frac{\partial \Phi}{\partial r} &=& 0, \\
\label{Eq:C5}
- r^2\sin\vartheta\cos\vartheta\Omega^2 + \frac{1}{\rho}\frac{\partial P}{\partial \vartheta} + \frac{\partial \Phi}{\partial \vartheta} &=& 0.
\end{eqnarray}
Considering a baritropic fluid for which the pressure is a function of the mass density only, it follows from the first law of thermodynamics (with constant entropy) that $\displaystyle \frac{dP}{\rho} = d \mathscr{h} h$. Substituting this relation into Eqs.~(\ref{Eq:C4}) and~(\ref{Eq:C5}) one obtains
\begin{equation}
d\left[ h + \Phi\right] 
 = \Omega^2\left[ r \sin^2\vartheta dr + r^2\sin\vartheta\cos\vartheta d\vartheta \right]
 = \frac{1}{2}\Omega^2 d(R^2),
\label{Eq:C6}
\end{equation}
where $R := r\sin\vartheta$ is the cylindrical radius. On the other hand, introducing the specific azimuthal angular momentum $\ell := R^2\Omega$, one can rewrite Eq.~(\ref{Eq:C6}) as
\begin{equation}
d\left[ h + \Phi\right] 
  = \frac{1}{2}\Omega^2d\left(\frac{\ell}{\Omega}\right)
  = \frac{1}{2}\left[ \Omega d\ell - \ell d\Omega \right].
\label{Eq:C7}
\end{equation}
Taking the exterior derivative on both sides and recalling $d^2 = 0$ yields
\begin{equation}
d\Omega \wedge d\ell = 0,
\end{equation}
which implies that $d\Omega$ is proportional to $d\ell$ and hence that $\Omega$ is a function of $\ell$ only.

Therefore, given any function $\Omega_0(\ell)$ of $\ell$, one obtains a fluid configuration by setting $\Omega = \Omega_0$ which yields the implicit relation $\ell = R^2\Omega_0(\ell)$ for $\ell$. This allows one to express $\Omega$ and $\ell$ as functions of the cylindrical coordinate $R$. Subsequently, Eq.~(\ref{Eq:C7}) is integrated,
\begin{equation}
h = h_\infty -\Phi - \frac{1}{2}\ell\Omega_0(\ell) + \int\limits_0^\ell \Omega_0(\bar{\ell}) d\bar{\ell},
\end{equation}
with a free constant $h_\infty$. In particular, the quantities $\Omega$, $\ell$, $h + \Phi$ are constant on the ``von Zeipel" cylinders $R = const$~\cite{vZeipel1924}.

As a simple example, consider the case for which $\ell = \ell_0$ is constant which implies that $\Omega = \ell_0/R^2$ and hence
\begin{equation}
h = h_\infty - \left( \Phi + \frac{\ell_0^2}{2R^2} \right),
\label{Eq:C8}
\end{equation}
with a constant $h_\infty$ describing the asymptotic value of the specific enthalpy. The boundary of the configuration is determined by the equation $h = 0$ which corresponds to a level set of the ``effective potential" $\Phi + \ell_0^2/(2R^2)$.

\begin{figure}[h!]
\centering
\subfigure[\:$\bar{h}_\infty=0, \lambda_0=1$.]{\includegraphics[scale=0.245]{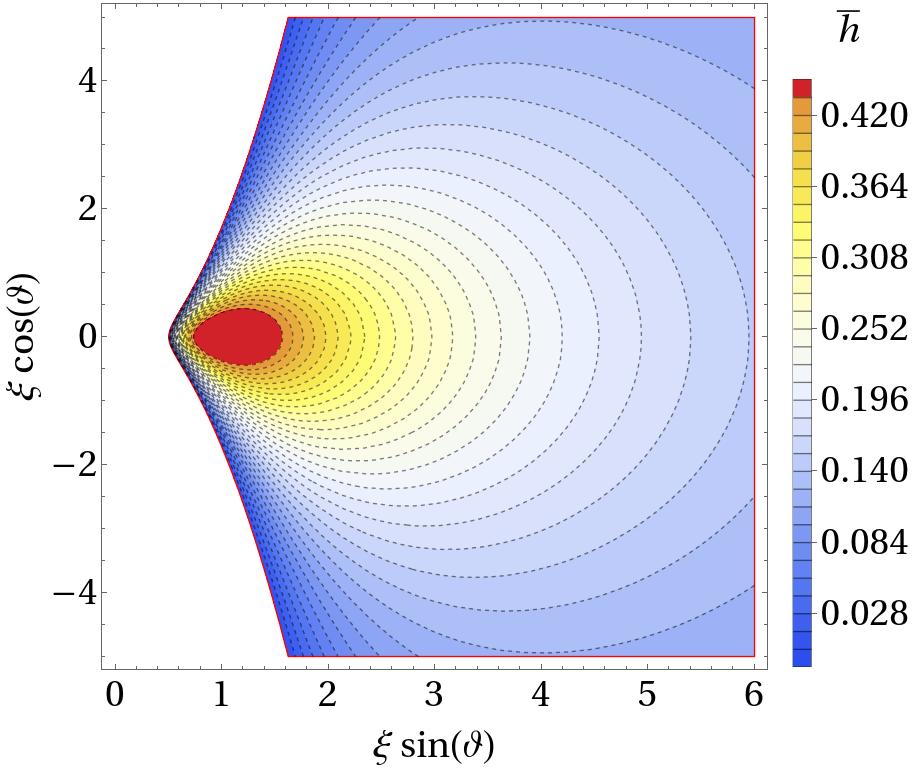}}
\subfigure[\:$\bar{h}_\infty=1, \lambda_0=1$.]{\includegraphics[scale=0.245]{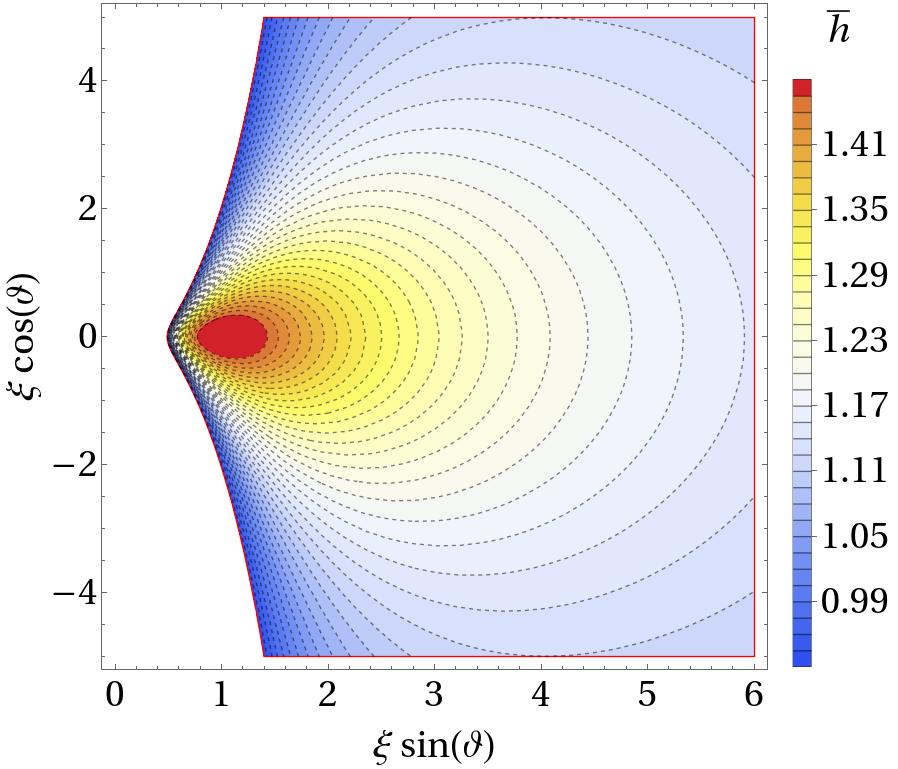}}
\subfigure[\:$\bar{h}_\infty=1, \lambda_0=2$.]{\includegraphics[scale=0.245]{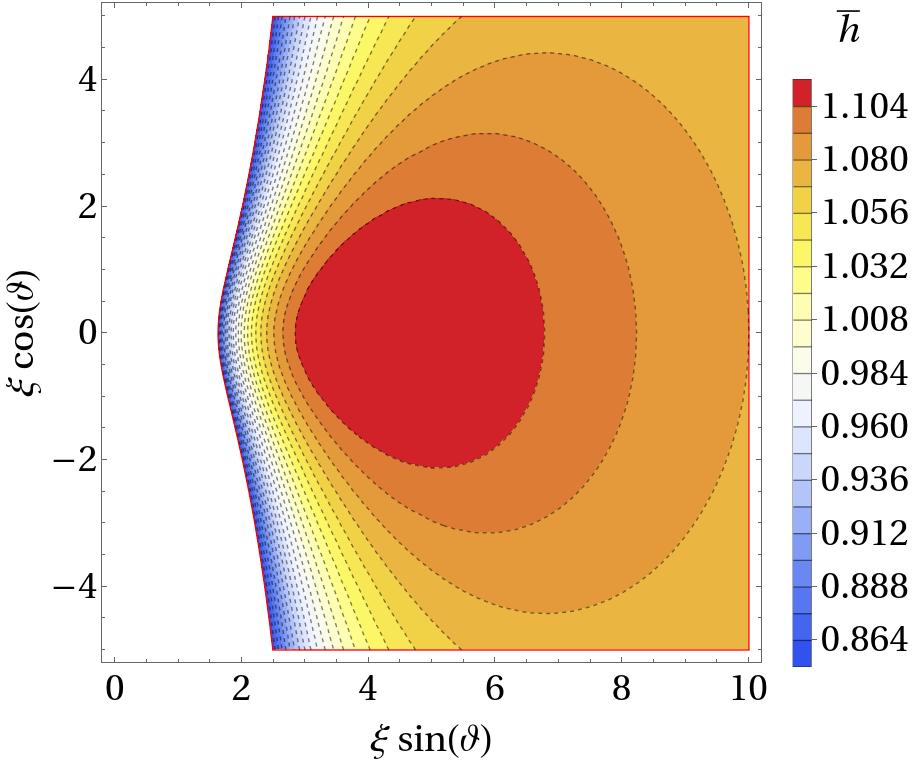}}
\caption{Contour plot for the enthalpy $\bar{h}$ in the $xz$-plane for different values of $\bar{h}_\infty$ and $\lambda_0$, assuming the Kepler potential.}
\label{Fig:BoundaryEnthalpyKepler}
\end{figure}

For the case of the Kepler potential, Eq.~(\ref{Eq:C8}) can be rewritten in terms of the following dimensionless quantities
\begin{equation}
r = \frac{GM}{h_0}\xi,\quad \ell_0 = \frac{GM}{\sqrt{h_0}} \lambda_0, \quad h_\infty = h_0\bar{h}_\infty, \quad \hbox{and} \quad h = h_0\bar{h},
\label{Eq:CDimensionless}
\end{equation}
with a positive constant $h_0$, such that
\begin{equation}
\bar{h}(\xi,\vartheta) = \bar{h}_\infty + \frac{1}{\xi} - \frac{\lambda_0^2}{2 \xi^2 \sin^2\vartheta}.
\label{Eq:C9}
\end{equation}
The condition $\bar{h} > 0$ implies that $\xi > \xi^{\text{min}}(\vartheta)$ with 
\begin{equation}
\xi^{\text{min}}(\vartheta) = 
\left\{
\begin{array}{lll}
\displaystyle \frac{\lambda_0^2}{2\sin^2\vartheta}, & & \bar{h}_\infty = 0, \\
\displaystyle \frac{1}{2\bar{h}_\infty}\left(-1 + \sqrt{1 + \frac{2 \bar{h}_\infty \lambda_0^2}{\sin^2\vartheta}}\right), & & \bar{h}_\infty > 0. 
\end{array}
\right.
\end{equation}
The boundary surface and the contours of the enthalpy function $\bar{h}$ in the $xz$-plane are shown in figure~\ref{Fig:BoundaryEnthalpyKepler} for different values of $\bar{h}_\infty$ and $\lambda_0$.

\begin{figure}[h!]
\centerline{
\subfigure[\:$\bar{h}_\infty = -1/8$.]
{\includegraphics[scale=0.240]{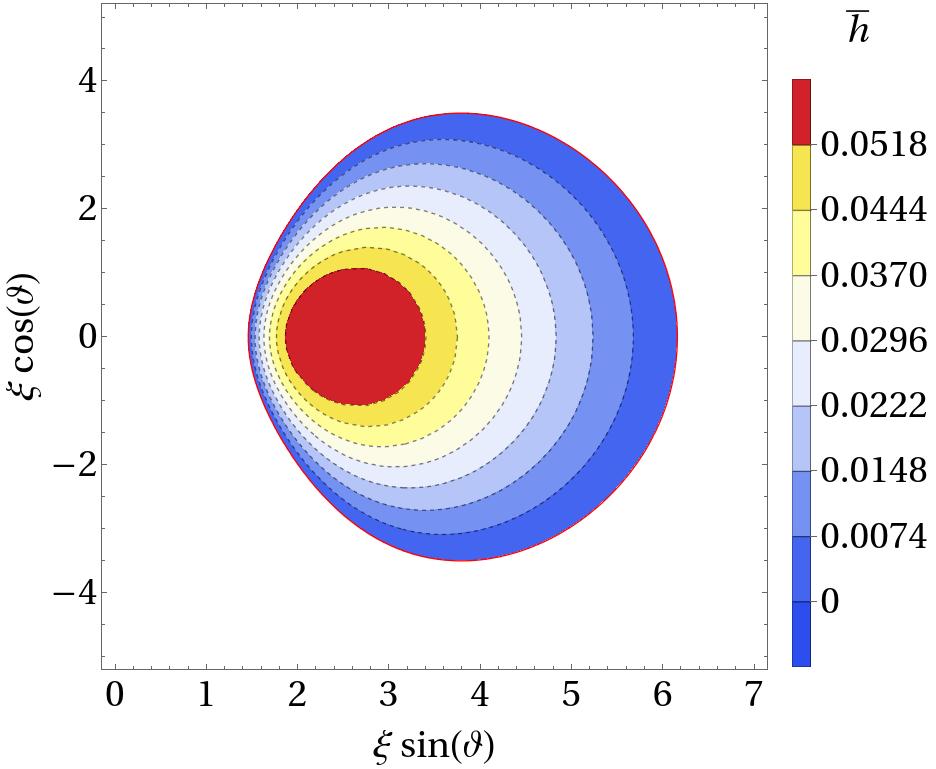}}
\subfigure[\:$\bar{h}_\infty = 0$.]
{\includegraphics[scale=0.240]{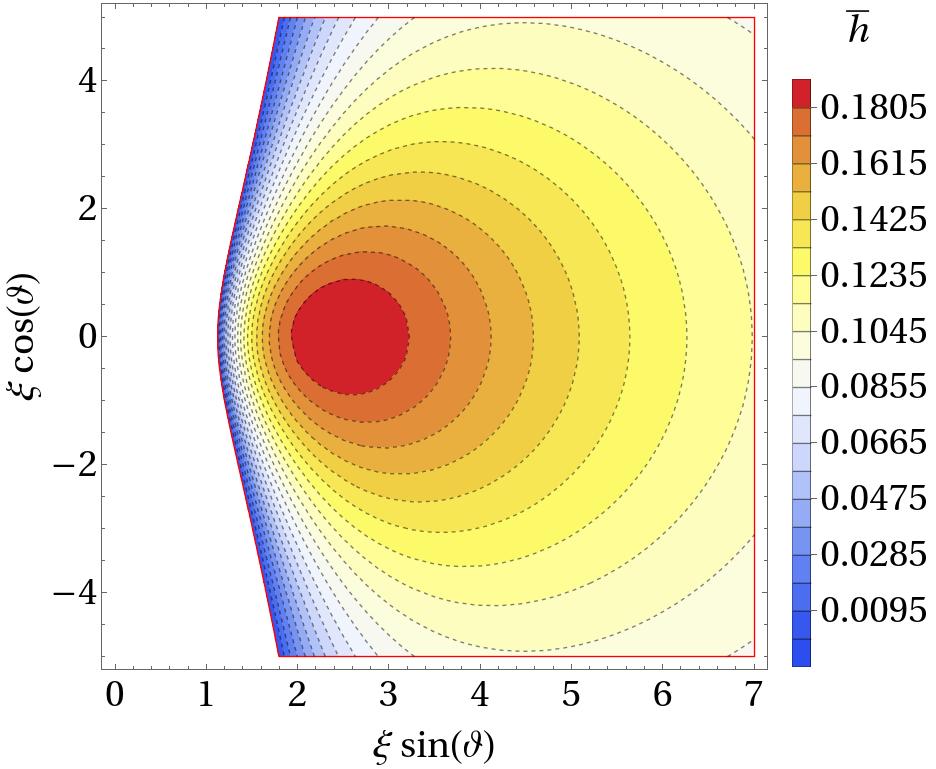}}
\subfigure[\:$\bar{h}_\infty = 1/8$.]
{\includegraphics[scale=0.240]{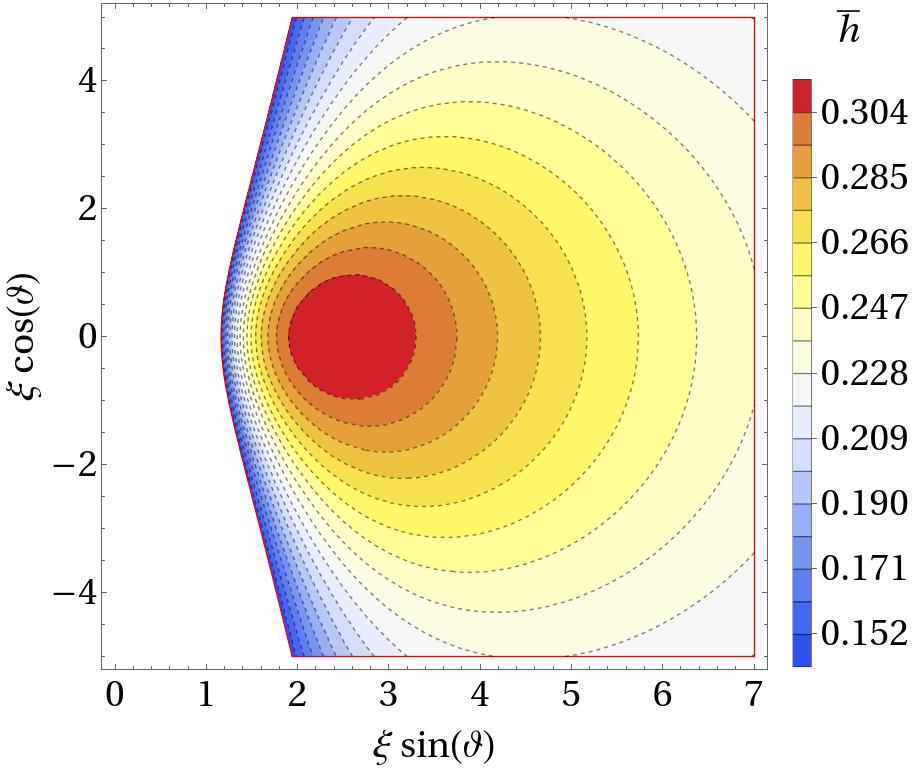}}}
\caption{Contour plot for the enthalpy $\bar{h}$ in the $xz$-plane for different values of $\bar{h}_\infty$ and $\lambda_0 = \kappa = 1$. Left panel: Here a negative value of $\bar{h}_\infty$ is chosen, which leads to a configuration with finite extend. Center panel: the marginal case $\bar{h}_\infty = 0$. Right panel: a case with $\bar{h}_\infty > 0$.}
\label{Fig:BoundaryEnthalpyIsochrone}
\end{figure}

For the more general case of the isochrone potential defined in Eq.~(\ref{Eq:IsochronePotential}), the dimensionless enthalpy is given by
\begin{equation}
\bar{h}(\xi,\vartheta) = \bar{h}_\infty + \frac{1}{\kappa + \sqrt{\kappa^2 + \xi^2}} - \frac{\lambda_0^2}{2 \xi^2 \sin^2\vartheta},
\label{Eq:EnthalpyFluid}
\end{equation}
where $\kappa := \frac{h_0 b}{GM}$. In this case, the boundary condition for the minimum radius of the configuration generally leads to a cubic equation for $\xi^2$ which yields
\begin{equation}
\xi^{\text{min}}(\vartheta) = 
\left\{
\begin{array}{lll}
\displaystyle \frac{\lambda_0^2}{2\sin^2\vartheta}\sqrt{1 + \frac{4\kappa\sin^2\vartheta}{\lambda_0^2}}, & & \bar{h}_\infty = 0, \\
\displaystyle \frac{1}{\sqrt{2}\bar{h}_\infty} \sqrt{1 + \bar{h}_\infty \left( 2\kappa + \frac{\lambda_0^2}{\sin^2\vartheta}\right) - \sqrt{ \left(1 + 2 \bar{h}_\infty \kappa \right)^2 + \frac{2 \bar{h}_\infty \lambda_0^2}{\sin^2\vartheta}}}, & & \bar{h}_\infty > 0. 
\end{array}
\right.
\end{equation}
The boundary surfaces and the contours of $\bar{h}$ in the $xz$-plane for different values of $\bar{h}$ and $\lambda_0 = \kappa = 1$ are shown in figure~\ref{Fig:BoundaryEnthalpyIsochrone}.

\bibliographystyle{unsrt.bst} 
\bibliography{refs_kinetic}
%
\end{document}